\documentclass{article}
\usepackage{amsthm, amssymb, latexsym, amsmath}
\newtheorem{theorem}{Theorem}[subsection]
\newtheorem{lemma}[theorem]{Lemma}
\newtheorem*{lemmaB}{Lemma B.1}
\newtheorem*{lemmaB2}{Lemma B.2}
\newtheorem*{lemmaC}{Lemma C.1}
\newtheorem*{propositionC}{Proposition C.2}
\newtheorem{proposition}[theorem]{Proposition}
\newtheorem{corollary}[theorem]{Corollary}
\newtheorem{Conjecture}{Conjecture}

\newtheorem{remark}{Remark}
\newtheorem{definition}{Definition}
\newenvironment{them}{\textbf{Theorem}\it}{}
\begin{document}
\huge
\title{Asymptotic Stability of Nonlinear Schr\"odinger Equations with
Potential\thanks{This paper is a part of the first author's Ph.D
thesis.}}
\Large 
\author{Zhou Gang$^{1,\ddagger}$\thanks{Supported by NSERC under Grant NA7901.} \ I.M. Sigal$^{1}$\thanks{Supported by
NSF under Grant DMS-0400526.}}\maketitle
\setlength{\leftmargin}{.1in} \setlength{\rightmargin}{.1in}
\centerline{ \small{$^1$Department of Mathematics, University of
Notre Dame, Notre Dame, U.S.A.}}
\centerline{\small{$^{1}$Department of Mathematics, University of
Toronto, Toronto, Canada}} \normalsize \vskip.1in
\setcounter{page}{1}
\section*{Abstract}
We prove asymptotic stability of trapped solitons in the
generalized nonlinear Schr\"odinger equation with a potential in
dimension 1 and for even potential and even initial conditions.
\section{Introduction}
In this paper we study the generalized nonlinear Schr\"odinger
equation with a potential
\begin{equation}\label{NLS}
i\frac{\partial\psi}{\partial
t}=-\psi_{xx}+V_{h}\psi-f(|\psi|^{2})\psi
\end{equation}
in dimension 1. Here $V_{h}: \mathbb{R}\rightarrow \mathbb{R}$ is
a family of external potentials, $\psi_{xx}=\partial_{x}^{2}\psi,$
and $f(s)$ is a nonlinearity to be specified later. Such equations
arise in the theory of Bose-Einstein condensation \footnote[1]{In
this case Equation (~\ref{NLS}) is called the Gross-Pitaevskii
equation.}, nonlinear optics, theory of water waves
\footnote[2]{In these two areas one usually takes $V_{h}=0,$ but
taking into account impurities and/or variations in geometry of
the medium one arrives at (~\ref{NLS}) with, in general, a
time-dependent $V_{h}.$}and in other areas. To fix ideas we assume
the potentials to be of the form $V_{h}(x):=V(hx)$ with $V$ smooth
and decaying at $\infty.$ Thus for $h=0,$ Equation (~\ref{NLS})
becomes the standard generalized nonlinear Schr\"odinger equation
(gNLS)
\begin{equation}\label{gNLS}
i\frac{\partial \psi}{\partial
t}=-\psi_{xx}+\mu\psi-f(|\psi|^{2})\psi,
\end{equation} where $\mu=V(0).$
For a certain class of nonlinearities, $f(|\psi|^{2})$ (see
Section ~\ref{s2}), there is an interval $\mathcal{I}_{0}\subset
\mathbb{R}$ such that for any $\lambda\in \mathcal{I}_{0}$
Equation (~\ref{gNLS}) has solutions of the form
$e^{i(\lambda-\mu)t}\phi_{0}^{\lambda}(x)$ where
$\phi_{0}^{\lambda}\in \mathcal{H}_{2}(\mathbb{R})$ and
$\phi_{0}^{\lambda}>0.$ Such solutions (in general without the
restriction $\phi_{0}^{\lambda}>0$) are called the
\textit{solitary waves} or \textit{solitons} or, to emphasize the
property $\phi_{0}^{\lambda}>0,$ the \textit{ground states}. For
brevity we will use the term \textit{soliton} applying it also to
the function $\phi_{0}^{\lambda}$ without the phase factor
$e^{i(\lambda-\mu)t}.$

Equation (~\ref{gNLS}) is translationally and gauge invariant.
Hence if $e^{i(\lambda-\mu)t}\phi_{0}^{\lambda}(x)$ is a solution
for Equation (~\ref{gNLS}), then so is $
e^{i(\lambda-\mu)t}e^{i\alpha}\phi_{0}^{\lambda}(x+a),$ for any
$\lambda\in \mathcal{I}_{0},\ a\in \mathbb{R},\ \alpha\in
[0,2\pi).$ This situation changes dramatically when the potential
$V_{h}$ is turned on. In general, as was shown in
~\cite{Floer,oh1,ABC} out of the three-parameter family
$e^{i(\lambda-\mu)t}e^{i\alpha}\phi_{0}^{\lambda}(x+a)$ only a
discrete set of two parameter families of solutions to Equation
(~\ref{NLS}) bifurcate: $e^{i\lambda
t}e^{i\alpha}\phi^{\lambda}_{h}(x),$ $\alpha\in [0,2\pi)$ and
$\lambda\in \mathcal{I}$ for some $\mathcal{I}\subseteq
\mathcal{I}_{0}$, with $\phi^{\lambda}_{h}\in
\mathcal{H}_{2}(\mathbb{R})$ and $\phi^{\lambda}_{h}>0$. Each such
family corresponds to a different critical point of the potential
$V_{h}(x).$ It was shown in ~\cite{Oh2} that the solutions
corresponding to minima of $V_{h}(x)$ are orbitally (Lyapunov)
stable and to maxima, orbitally unstable. We call the solitary
wave solutions described above which correspond to the minima of
$V_{h}(x)$ \textit{trapped solitons} or just \textit{solitons} of
Equation (~\ref{NLS}) omitting the last qualifier if it is clear
which equation we are dealing with.

The main result of this paper is a proof that the trapped solitons
of Equation (~\ref{NLS}) are asymptotically stable. The latter
property means that if an initial condition of (~\ref{NLS}) is
sufficiently close to a trapped soliton then the solution
converges in some weighted $\mathcal{L}^{2}$ space to, in general,
another trapped soliton of the same two-parameter family. In this
paper we prove this result under the additional assumption that
the potential and the initial condition are even. This limits the
number of technical difficulties we have to deal with. In the
subsequent paper we remove this restriction and allow the soliton
to 'move'.

In fact, in this paper we prove a result more general than
asymptotic stability of trapped solitons. Namely, we show that if
the initial conditions are of the form
$$\psi_{0}=e^{i\gamma_{0}}(\phi_{h}^{\lambda_{0}}+\chi_{0}),$$
with $\chi_{0}$ being small in the space
$(1+x^{2})\mathcal{H}^{1}$, $\gamma_{0}\in \mathbb{R}$ and
$\lambda_{0}\in \mathcal{I}$ ($\mathcal{I}$ will be defined
later). Then the solution, $\psi(t),$ of Equation (~\ref{NLS}) can
be written as
\begin{equation}\label{decom1}
\psi(t)=e^{i\gamma(t)}(\phi_{h}^{\lambda(t)}+\chi(t)),
\end{equation} where $\gamma(t)\in \mathbb{R},$ $\chi(t)\rightarrow 0$
in some local norm, and $\lambda(t)\rightarrow \lambda_{\infty}$
for some $\lambda_{\infty}$ as $t\rightarrow \infty$.

We observe that (~\ref{NLS}) is a Hamiltonian system with
conserved energy (see Section ~\ref{s2}) and, though orbital
(Lyapunov) stability is expected, the asymptotic stability is a
subtle matter. To have asymptotic stability the system should be
able to dispose of excess of its energy, in our case, by radiating
it to infinity. The infinite dimensionality of a Hamiltonian
system in question plays a crucial role here.

First attack on the asymptotic stability in infinite dimensional
Hamiltonian systems was made in the pioneering work of Soffer and
Weistein ~\cite{SW1} where the asymptotic stability of nonlinear
bound states was proved for the nonlinear Schr\"odinger equation
with a potential and a weak nonlinearity in the dimensions higher
than or equal to $3$. Asymptotic stability of moving solitons in
the (generalized) nonlinear Schr\"odinger equation without
potential and dimension 1 was first proven by Buslaev and Perelman
~\cite{Buslaev}. The above results were significantly extended by
Soffer and Weinstein, Buslaev and Perelman, Tsai and Yau, Buslaev
and Sulem, Cuccagna(see
~\cite{SW2,SW3,SW4,BP2,TY1,TY2,TY3,BuSu,Cu1,Cu2,Cu3}). Related
results in multi-soliton dynamics were obtained by Perelman, and
Rodnianski, Schlag and Soffer (see ~\cite{Pere,RSS1,RSS2}). Deift
and Zhou (see ~\cite{DeZh})used a different approach, inverse
scattering method, to asymptotic behavior of solitons of the
1-dimensional nonlinear Schr\"odinger equations.

Among earlier work we should mention the works of Shatah and
Strauss, Weinstein, Grillakis, Shatah and Strauss on orbital
stability (see ~\cite{SS,We1,We2,GSS1,GSS2}) whose results were
extended by Comech and Pelinosky, Comech, Cuccagna, Pelinovsky and
Vougalter, Cuccagna and Pelinovsky, and Schlag (see
~\cite{CP,CO,Cu1,CPV,CuPe,Schlag}).

There is an extensive physics literature on the subject; some of
the references can be found in Grimshaw and Pelinovsky ~\cite{GP}.

Long-term dynamics of solitons in external potentials is
determined by Bronski and Jerrard, Fr\"ohlich, Tsai and Yau,
Keraani, and Fr\"{o}hlich, Gustafson, Jonsson and Sigal (see
~\cite{BJ,FTY,Keraani,sigal}).

Our approach is built on the beautiful theory of one-dimensional
Schr\"odinger operators developed by Buslaev and Perelman, and
Buslaev and Sulem(see \cite{Buslaev,BP2,BuSu}). One of the key
points in this approach is obtaining suitable (and somewhat
surprising) estimates on the propagator for the linearization of
Equation (~\ref{NLS}) around the soliton family
$e^{i\gamma}\phi_{h}^{\lambda}.$ One of the difficulities here
lies in the fact that the corresponding generator, $L(\lambda),$
is not self-adjoint. To obtain the desired estimates one develops
the spectral representation for the propagator (in terms of the
boundary values of the resolvent; this can be also extended to
other functions of the generator (see Subsection
~\ref{integralkernel})) and then estimates the integral kernel of
the resolvent using estimates on various solutions of the
corresponding spectral problem $(L(\lambda)-\sigma)\xi=0$
(Appendix ~\ref{app}). These estimates are close to the
correponding estimates of ~\cite{Buslaev,BP2,BuSu}. Since these
estimate are somewhat involved we take pain to provide a detailed
and readable account. Note that, independently, W. Schlag
~\cite{Schlag} has developped spectral representation similiar to
ours (see Subsection ~\ref{integralkernel}) and Goldberg and
Schlag ~\cite{GoSc} obtained (by a different technique) estimates
on the propagators of the one-dimensional, self-adjoint (scalar)
Schr\"odinger operators similiar to some of our estimates (see
Subsection ~\ref{ap1}) but under more general assumptions on the
potential than in our (non-self-adjoint, vector) case.

The paper is organized as follows: in Section ~\ref{s2} we
describe the Hamiltonian structure of Equation (~\ref{NLS}), cite
a well-posedness result, formulate our conditions on the
nonlinearity and the potential and our main result. In Section
~\ref{oper} we describe the spectral structure of the linearized
equation around the trapped soliton. In Section ~\ref{asym} we
decompose the solution into a part moving in the 'soliton
manifold' and a simplectically orthogonal fluctuation and find the
equations for the soliton parameters and for the fluctuation. In
the same section we estimate the soliton parameters and the
fluctuation assuming certain estimates on the linearized
propagators (i.e. the solutions of the linearized equation). The
latter estimates are proven in Section ~\ref{Estimatepropagator},
modulo estimates of the generalized eigenfunctions which are
obtained in Appendix ~\ref{app}. In Appendix ~\ref{proofassu} we
analyze the implicit conditions on the nonlinearity and the
potential made in Section ~\ref{oper}.

As customary we often denote derivatives by subindices as in
$\phi^{\lambda}_{\lambda}=\frac{d}{d\lambda}\phi^{\lambda}$ and
$\phi^{\lambda}_{x}=(\frac{d}{dx}\phi^{\lambda})$ for
$\phi^{\lambda}=\phi^{\lambda}(x).$ The Sobolev and $L^{2}$ spaces
are denoted by $\mathcal{H}^{1}$ and $\mathcal{L}^{2}$
respectively.
\section*{Acknowledgment}
We are grateful to S. Cuccagna, V. Vougalter and, especially, V.S.
Buslaev for fruitful discussions. In particular, V.S.Buslaev
taught us beautiful ODE techniques many of which are due to him
and his collaborators.

This paper is a part of the first author's Ph.D requirement. While
the paper has been prepared for publication the authors received
the preprints by Comech and Pelinovsky ~\cite{CP}, Schlag
~\cite{Schlag} and Goldberg and Schlag ~\cite{GoSc} whose results
overlap with some of the results of the present paper, namely,
with the description of the linearized spectrum (Section
~\ref{oper}) and with construction of functions of nonself-adjoint
operators (Subsection ~\ref{integralkernel}) and with an estimate
of one-dimensional propagators (Subsection ~\ref{ap1})
respectively. The authors are grateful to Comech, Pelinovsky and
Schlag for communicating their results prior to publications.
\section{Properties of (~\ref{NLS}), Assumptions and Results}\label{s2}
In this section we discuss some general properties of Equation
(~\ref{NLS}) and formulate our results.
\subsection{Hamiltonian Structure}
Equation (~\ref{NLS}) is a Hamiltonian system on Sobolev space
$\mathcal{H}^{1}(\mathbb{R},\mathbb{C})$ viewed as a real space
$\mathcal{H}^{1}(\mathbb{R},\mathbb{R})\oplus
\mathcal{H}^{1}(\mathbb{R},\mathbb{R})$ with the inner product
$(\psi,\phi)=Re\int_{\mathbb{R}}\bar{\psi}\phi$ and with the
simpletic form
$\omega(\psi,\phi)=Im\int_{\mathbb{R}}\bar{\psi}\phi.$ The
Hamiltonian functional is: $$H(\psi):=\int
[\frac{1}{2}(|\psi_{x}|^{2}+V_{h}|\psi|^{2})-F(|\psi|^{2})],$$
where $F(u):=\frac{1}{2}\int_{0}^{u}f(\xi)d\xi.$

Equation (~\ref{NLS}) has the time-translational and gauge
symmetries which imply the the following conservation laws: for
any $t\geq 0,$ we have
\begin{enumerate}
 \item[(CE)] conservation of energy: $$H(\psi(t))=H(\psi(0));$$
 \item[(CP)]
 conservation of the number of particles: $$N(\psi(t))=N(\psi(0)),$$ where $N(\psi):=\int
 |\psi|^{2}.$
\end{enumerate}
We need the following condition on the nonlinearity $f$ for the
global well-posedness of (~\ref{NLS}).
\begin{enumerate}
 \item[(fA)] The nonlinearity $f$ is locally Lipschitz and $f(\xi)\leq c(1+|\xi|^{q})$ for some
 $c>0$ and $q<2.$
\end{enumerate}

The following theorem is proved in ~\cite{Oh3,Cazenave}.\\
\begin{them}
Assume that the nonlinearity $f$ satisfies the condition (fA), and
that the potential $V$ is bounded. Then Equation (~\ref{NLS}) is
globally well posed in $\mathcal{H}^{1}$, i.e. the Cauchy problem
for Equation (~\ref{NLS}) with initial datum $\psi(0)\in
\mathcal{H}^{1}$ has a unique solution $\psi(t)$ in the space
$\mathcal{H}^{1}$ and this solution depends continuously on
$\psi(0)$.

Moreover $\psi(t)$ satisfies the conservation laws (CE) and (CP).

If $\psi(0)$ has a finite norm $\|(1+|x|)\psi(0)\|_{2},$ then we
have the following estimates:
\begin{equation}\label{weightestimate}
\|(1+|x|)\psi(t)\|_{2}\leq
e(\|\psi(0)\|_{\mathcal{H}^{1}})[\|(1+|x|)\psi(0)\|_{2}+t\|\psi(0)\|_{\mathcal{H}^{1}}],
\end{equation}
where $e:\mathcal{R}_{+}\rightarrow \mathcal{R}_{+}$ is a smooth
function.
\end{them}
\subsection{Existence and Stability of Solitons}
In this subsection we discuss the problem of existence and
stability of solitons.

It is proved in  ~\cite{Buslaev,existence} that if the
nonlinearity $f$ in Equation (~\ref{NLS}) is smooth, real and
satisfies the following condition
\begin{enumerate}
 \item[(fB)] There is an interval $\mathcal{I}_{0}\in \mathbb{R}^{+}$ s.t. for any $\lambda\in \mathcal{I}_{0}$
 $$U(\phi,\lambda):=-\lambda\phi^{2}+\int_{0}^{\phi^{2}}f(\xi)d\xi$$ has a positive root and the smallest positive root
 $\phi_{0}(\lambda)$ satisfies
 $U_{\phi}(\phi_{0}(\lambda),\lambda)>0,$
\end{enumerate}
then for any $\lambda\in \mathcal{I}_{0}$ there exists a unique
solution of Equation (~\ref{gNLS}) of the form $e^{i(\lambda-\mu)
t}\phi_{0}^{\lambda}$ with $\phi_{0}^{\lambda}\in \mathcal{H}^{2}$
and $\phi_{0}^{\lambda}>0$. Such solutions are called the solitary
waves or solitons or to emphasize that $\phi_{0}^{\lambda}>0$, the
ground states. For brevity we use the term soliton and we apply it
to the function $\phi_{0}^{\lambda}$. Note the function
$\phi_{0}^{\lambda}$ satisfies the equation:
\begin{equation}\label{solitonnopotential}
-(\phi_{0}^{\lambda})_{xx}+\lambda\phi_{0}^{\lambda}-f((\phi_{0}^{\lambda})^{2})\phi_{0}^{\lambda}=0.
\end{equation}
\begin{remark}
If $f(\xi)=c\xi^{p}+o(\xi^{p})$ with $c, \ p>0$, then Condition
(fB) is satisfied for $\lambda\in (0,\delta)$ with $\delta$
sufficiently small.
\end{remark}
When the potential $V$ is present, then some of the solitons above
bifurcate into solitons for Equation (~\ref{NLS}). Namely,
similarly as in ~\cite{Floer,oh1} one can show that if $f$
satisfies the following condition,
\begin{enumerate}
\item[(fC)] $f$ is smooth, $f(0)=0$ and there exists $p\geq 1$,
such that $|f^{'}(\xi)|\leq c(1+|\xi|^{p}),$
\end{enumerate} and if $V$ satisfies the condition
\begin{enumerate}
\item[(VA)] $V$ is smooth and $0$ is a non-degenerate local
minimum of $V$,
\end{enumerate}
and if the soliton, $\phi_{0}^{\lambda},$ exists for Equation
(~\ref{solitonnopotential}), then for any $\lambda\in
\mathcal{I}_{0V}:=\{\lambda|\lambda>\displaystyle\inf_{x\in\mathbb{R}}\{V(x)\}\}\cap\{\lambda|\lambda+V(0)\in
\mathcal{I}_{0}\}$ there exists a soliton $\phi^{\lambda}_{h}$
satisfying the equation
$$-\frac{d^{2}}{dx^{2}}\phi^{\lambda}_{h}+(\lambda+V_{h})\phi^{\lambda}_{h}-f((\phi^{\lambda}_{h})^{2})\phi^{\lambda}_{h}=0$$
and which is of the form
$\phi_{h}^{\lambda}=\phi^{\lambda+V(0)}_{0}+O(h^{3/2})$ where
$\phi_{0}^{\lambda}$ is the soliton of Equation
(~\ref{solitonnopotential}). (The subindex should not be confused
with the derivative in $h$.)

Under more restrictive conditions on the nonlinearity $f$ one can
show as in ~\cite{GSS1,sigal,We2} that the soliton
$\phi_{h}^{\lambda}$ is a minimizer of the energy functional
$H(\psi)$ for a fixed number of particles $N(\psi)=constant$ if
and only if
\begin{equation}\label{stabli}
\frac{d}{d\lambda}\|\phi_{h}^{\lambda}\|^{2}_{2}>0.
\end{equation}
The latter condition is also equivalent to the orbital stability
of $\phi^{\lambda}_{h}$. In what follows we set
\begin{equation}\label{stab}
\mathcal{I}=\{\lambda\in \mathcal{I}_{0V}:\frac{\partial}{\partial
\lambda}\|\phi^{\lambda}_{h}\|_{2}>0\}.
\end{equation}

Observe that there exist some constants $c,\ \delta>0$ such that
\begin{equation}\label{expondecay}
|\phi_{h}^{\lambda}(x)|\leq ce^{-\delta|x|}\ \text{and}\
|\frac{d}{d\lambda}\phi_{h}^{\lambda}|\leq ce^{-\delta|x|},
\end{equation}
and similarly for the derivatives of $\phi^{\lambda}_{h}$ and
$\frac{d}{d\lambda}\phi_{h}^{\lambda}$. The first estimate can be
found in ~\cite{GSS1} and the second estimate follows from the
fact that the function $\frac{d}{d\lambda}\phi_{h}^{\lambda}$
satisfies the equation
$$[-\frac{d^{2}}{dx^{2}}+V_{h}+\lambda-f((\phi_{h}^{\lambda})^{2})-2f^{'}((\phi_{h}^{\lambda})^{2})(\phi_{h}^{\lambda})^{2}]\frac{d}{d\lambda}\phi^{h}_{\lambda}=-\phi_{h}^{\lambda}$$
and standard arguments.

For our main result we will also require the following condition
on the potential $V:$
\begin{enumerate}
\item[(VB)] $|V(x)|\leq ce^{-\alpha|x|}$ for some $c,\alpha>0.$
\end{enumerate}
\subsection{Linearized Operator and Spectral Conditions}
In our analysis we use some implicit spectral conditions on the
$\text{Fr}\acute{e}\text{chet}$ derivative $\partial
G(\phi^{\lambda}_{h})$ of the map
\begin{equation}\label{symmetryG}
G(\psi)=-i(-\frac{d^{2}}{dx^{2}}+\lambda+V_{h})\psi+if(|\psi|^{2})\psi
\end{equation}
appearing on the right hand side of Equation (~\ref{NLS}). We
compute
\begin{equation}\label{defineoperator}
\partial
G(\phi_{h}^{\lambda})\chi=-i(-\frac{d^{2}}{dx^{2}}+\lambda+V_{h})\chi+if((\phi^{\lambda}_{h})^{2})\chi+2if^{'}((\phi^{\lambda}_{h})^{2})(\phi^{\lambda}_{h})^{2}Re\chi.
\end{equation}
This is a real linear but not complex linear operator. To convert
it to a linear operator we pass from complex functions to real
vector-functions:
$$\chi\longleftrightarrow \vec{\chi}=\left(
\begin{array}{lll}
\chi_{1}\\
\chi_{2}
\end{array}
\right),$$ where $\chi_{1}=Re\chi$ and $\chi_{2}=Im\chi.$ Then
$$\partial G(\phi_{h}^{\lambda})\chi\longleftrightarrow
L(\lambda)\vec{\chi}$$ where
\begin{equation}\label{operaL}
L(\lambda) :=  \left(
\begin{array}{lll}
0&L_{-}(\lambda)\\
-L_{+}(\lambda)&0
\end{array}
\right),
\end{equation}
with
\begin{equation}\label{firstoperator}
L_{-}(\lambda):=-\frac{d^{2}}{dx^{2}}+V_{h}+\lambda-f((\phi^{\lambda}_{h})^{2}),
\end{equation} and
\begin{equation}\label{secondoperator}
L_{+}(\lambda):=-\frac{d^{2}}{dx^{2}}+V_{h}+\lambda-f((\phi_{h}^{\lambda})^{2})-2f^{'}((\phi^{\lambda}_{h})^{2})(\phi^{\lambda}_{h})^{2}.
\end{equation}
Then we extend the operator $L(\lambda)$ to the complex space
$\mathcal{H}^{2}(\mathbb{R},\mathbb{C})\oplus
\mathcal{H}^{2}(\mathbb{R},\mathbb{C}).$ By a general result (see
e.g. ~\cite{RSIV}),
$\sigma_{ess}(L(\lambda))=(-i\infty,-i\lambda]\cap
[i\lambda,i\infty)$ if the potential $V_{h}$ in Equation
(~\ref{NLS}) decays at $\infty.$

We show in the next section that the operator $L(\lambda)$ has at
least four usual or associated eigenvectors: the zero eigenvector
$\left(
\begin{array}{lll}
0\\
\phi^{\lambda}_{h}
\end{array}
\right)$ and associated zero eigenvector $\left(
\begin{array}{lll}
\frac{d}{d\lambda}\phi_{h}^{\lambda}\\
0
\end{array}
\right)$ related to the gauge symmetry $\psi(x,t)\rightarrow
e^{i\alpha}\psi(x,t)$ of the original equation, and two
eigenvectors with $O(h^{2})$ eigenvalues originating from the zero
eigenvector $\left(
\begin{array}{lll}
\partial_{x}\phi_{0}^{\lambda}\\
0
\end{array}
\right)$ of the $V=0$ equation due to the translational symmetry
of that equation and associated zero eigenvector $\left(
\begin{array}{lll}
0\\
x\phi_{0}^{\lambda}
\end{array}
\right)$ related to the boost transformation $\psi(x,t)\rightarrow
e^{ibx}\psi(x,t)$ coming from the Galilean symmetry of the $V=0$
equation.

Besides of eigenvalues, the operator $L(\lambda)$ may have
resonances at the tips, $\pm i\lambda$, of its essential spectrum
(those tips are called thresholds). The definition of the
resonance is as follows:
\begin{definition}
A function $h\not=0$ is called a resonance of $L(\lambda)$ at
$i\lambda$ if and only if $h$ is $C^{2},$ is bounded and satisfies
the equation $$(L(\lambda)-i\lambda)h=0.$$ Similarly we define a
resonance at $-i\lambda.$
\end{definition}
In what follows we make the following spectral assumptions:
\begin{enumerate}
 \item[(SA)] Dimension of the generalized eigenvector space for
 isolated eigenvectors is 4,
 \item[(SB)]$L(\lambda)$ has no embedded eigenvalues,
 \item[(SC)]$L(\lambda)$ has no resonances at $\pm
i\lambda$.
\end{enumerate}
Condition (SA) is satisfied for a large class of nonlinearities,
but it is not generic. For some open set of nonlinearities the
operator $L(\lambda)$ might have other purely imaginary, isolated
eigenvalues besides those mentioned above. Our technique can be
extended to this case. For the consideration of space this will be
done elsewhere.
\begin{Conjecture}\label{assu}
Conditions (SB) and (SC) are satisfied for generic nonlinearities
$f$ and potentials $V$ provided that $V$ decays exponentially fast
at $\infty.$
\end{Conjecture}
There are standard techniques for proving the (SB) part of this
conjecture. This will be addressed elsewhere.

The following results support the (SC) part of the conjecture.
Introduce the family of operators
$$L_{general}(U):=L_{0}+U,$$ where
$$L_{0}:=\left(
 \begin{array}{lll}
 0&-\frac{d^{2}}{dx^{2}}+\beta\\
 \frac{d^{2}}{dx^{2}}-\beta&0
 \end{array}
 \right)\ \text{and}\ U:=\left(
 \begin{array}{lll}
 0&V_{1}\\
 -V_{2}&0
 \end{array}
 \right)$$ parameterized by $\beta>0$, $s\in \mathbb{C}$ and the
 functions $V_{1}(x)$ and $V_{2}(x)$ satisfying
\begin{equation}\label{alpha2}
|V_{1}(x)|,|V_{2}(x)|\leq ce^{-\alpha|x|}
\end{equation}
for some constants $c,\alpha>0.$ Then we have
\begin{proposition}\label{spectral}
\begin{enumerate}
 \item[(A)] If (SB) and (SC) are satisfied for a given
 $U^{0},$ then (SB) and (SC) are satisfied for any $U$ such that
 $\|e^{\alpha |x|}(U-U^{0})\|_{\mathcal{L}^{\infty}}$
 is sufficiently small, where $\alpha$ is the same as in Equation (~\ref{alpha2}).
 \item[(B)] If for some $U^{0}$ the operator $L_{general}(U^{0})$
 has a resonance at $i\beta$ (or at $-i\beta$) and if $$\int_{-\infty}^{\infty} V^{0}_{1}(x)+V^{0}_{2}(x)dx\not=0,$$ then there exists a small
 neighborhood $\mathcal{A}\subset\mathbb{C}$ of $1$ such that $L_{general}(sU^{0})$ has no
 resonance at $i\beta$ for $s\in\mathcal{A}\backslash\{1\}.$
\end{enumerate}
\begin{remark}
\begin{enumerate}
 \item For $U=0$, the operator $L_{general}(U)=L_{0}$ has
 resonances at $\pm i\beta$. Hence Statement (B) shows that the
 operators $L_{general}(sU)$ with $s\not=0$ and sufficiently
 small have no resonance at $\pm i\beta.$
 \item
 It is proved in ~\cite{Kaup} that if $f(u)=u$ and $V(x)=0$ in Equation (~\ref{NLS}),
 then Conditions (SA) and (SB) hold, but Condition (SC) fails.
 Proposition ~\ref{spectral} (B) implies that Equation
 (~\ref{NLS}) with $f(u)=u$ and $V(x)=sV^{0}(x),$ for a large class
 of $V^{0}(x)$ and for $s\not=0$ sufficiently small, satisfies (SB)
 and (SC). It can be proved that for a large subclass of potentials $V^{0}(x)$ Condition (SA)
 remains to be satisfied. For consideration of space it will be done elsewhere.
 \item Equation (~\ref{NLS}) with $f(u)=u^{2}$ and $V(x)=0$ is integrable. It can be shown that the operator $L(\lambda)$ in
 this case satisfies the conditions (SB) and (SC). However, this
 equation fails Condition (~\ref{stabli}) (it is a critical NLS) and
 (SA) (its generalized zero eigenvector space is of dimension 6.)
 It is easy to stabilize this equation by changing the
 nonlinearity slightly, say, taking $f(u)=u^{2-\epsilon}$ or $f(u)=u^{2}-\epsilon
 u^{4}.$ The resulting equations satisfy (~\ref{stabli}), (SB) and (SC) but not
 (SA). Specifically, if the nonlinearity $f(u)=u^{2-\epsilon}$ and the potential $V=0$,
 then Equation (~\ref{NLS}) has a standing wave solution $\psi(x,t)=e^{it}\phi(x)$ with
 $$\phi(x)=(12-4\epsilon)^{-\frac{1}{4-2\epsilon}}[e^{(2-\epsilon)x}+e^{-(2-\epsilon)x}]^{-\frac{1}{2-\epsilon}}.$$
 Then by Proposition ~\ref{spectral} Statement (A) and an explicit form of the soliton $\phi$ the corresponding linearized operator,
 $$\left(
 \begin{array}{lll}
 0&-\frac{d^{2}}{dx^{2}}+1-\phi^{4-2\epsilon}\\
 \frac{d^{2}}{dx^{2}}-1+(5-2\epsilon)\phi^{4-2\epsilon}&0
 \end{array}
 \right),$$ has no resonances at $\pm
 i$ provided that $\epsilon>0$ is sufficiently small.
\end{enumerate}
\end{remark}
\end{proposition}
\subsection{Main theorem}\label{main}
We state the main theorem of this paper.
\begin{theorem}\label{maintheorem}
Assume Conditions (VA), (VB), (fA)-(fC) and (SA)-(SC) and assume
that the nonlinearity $f$ is a polynomial of degree $p\geq 4.$
Assume the external potential $V$ is even, and $\lambda\in
\mathcal{I}$ with $\mathcal{I}$ defined in Equation (~\ref{stab}).
There exists a constant $\delta>0$ such that if $\psi(0)$ is even
and satisfies
$$\inf_{\gamma\in \mathcal{R}}\{\|x^{2}(e^{i\gamma}\psi(0)-\phi^{\lambda}_{h})\|_{2}+\|e^{i\gamma}\psi(0)-\phi^{\lambda}_{h}\|_{\mathcal{H}^{1}}\}\leq \delta,$$ then there exists a constant $\lambda_{\infty}\in \mathcal{I}$, such that
$$\displaystyle\inf_{\gamma\in \mathcal{R}}\|
(1+|x|)^{-\nu}(\psi(t)-e^{i\gamma}\phi^{\lambda_{\infty}})\|_{2}\rightarrow
0$$ as $t\rightarrow \infty$ where $\nu>3.5,$ in other words, the
trapped soliton is asymptotically stable.
\end{theorem}
\section{Properties of Operator $L(\lambda)$}\label{oper}
In this section we find eigenvectors and define the essential
spectrum subspace of $L(\lambda)$. Here we do not assume that the
potential $V$ is even. Our main theorem is:
\begin{theorem}\label{mainpo}
If $V$ satisfies Conditions (VA) and (VB) and if $\lambda\in
\mathcal{I}$, then $L(\lambda)$ has $3$ independent eigenvectors
and one associated eigenvector with small eigenvalues: one
eigenvector $\left(
\begin{array}{lll}
0\\
\phi^{\lambda}_{h}
\end{array}
\right)$ and one associated eigenvector $\left(
\begin{array}{lll}
\frac{d}{d\lambda}\phi^{\lambda}_{h}\\
0
\end{array}
\right)$ with eigenvalue 0, both of which are even if $V$ is even;
2 independent eigenvectors with small non-zero imaginary
eigenvalues, which are odd if $V$ is even.
\end{theorem}
\begin{proof}
The proof is based on the following facts: the operator
$L(\lambda)$ has the eigenvector $\left(
\begin{array}{lll}
0\\
\phi^{\lambda}_{h}
\end{array} \right):$
$$
L(\lambda)\left(
\begin{array}{lll}
0\\
\phi^{\lambda}_{h}
\end{array} \right)=0,
$$
related to the gauge symmetry of the map $G(\psi)$ (see Equation
(~\ref{symmetryG})), and associated zero eigenvector $\left(
\begin{array}{lll}
\frac{d}{d\lambda}\phi^{\lambda}_{h}\\
0
\end{array} \right)$:
$$
L(\lambda)\left(
\begin{array}{lll}
\frac{d}{d\lambda}\phi^{\lambda}_{h}\\
0
\end{array} \right)=\left(
\begin{array}{lll}
0\\
\phi^{\lambda}_{h}
\end{array}
\right).
$$
Moreover, for $h=0,$ the operator has the zero eigenvector
$\left(
\begin{array}{lll}
\partial_{x}\phi_{0}^{\lambda+V(0)}\\
0
\end{array}
\right):$
$$
L^{h=0}(\lambda)\left(
\begin{array}{lll}
\partial_{x}\phi_{0}^{\lambda+V(0)}\\
0
\end{array}
\right)=0,
$$
coming from the translational symmetry of the map $G(\psi)$ and
the associated zero eigenvector $\left(
\begin{array}{lll}
0\\
x\phi_{0}^{\lambda+V(0)}
\end{array}
\right)$:
$$
 L^{h=0}(\lambda)\left(
\begin{array}{lll}
0\\
x\phi_{0}^{\lambda+V(0)}
\end{array}
\right)=\left(
\begin{array}{lll}
2\partial_{x}\phi_{0}^{\lambda+V(0)}\\
0
\end{array}
\right),
$$
coming from the boost transformation.

The first two properties above yield the first part of the
theorem. The last two properties and elementary perturbation
theory will yield the second part of this theorem.

To prove the second part of the theorem we first observe that
since the operator $L(\lambda)$ is of the form
$L(\lambda)=JH(\lambda)$ where $J=\left(
\begin{array}{lll}
0&1\\
-1&0
\end{array}
\right)$ is the anti-self-adjoint matrix and
$H(\lambda)=\left(
\begin{array}{lll}
L_{+}(\lambda)&0\\
0&L_{-}(\lambda)
\end{array} \right)$ is a real self-adjoint operator, the spectrum
of $L(\lambda)$ is symmetric with respect to the real and
imaginary axis. Hence the eigenvectors $\left(
\begin{array}{lll}
\partial_{x}\phi_{0}^{\lambda+V(0)}\\
0
\end{array}
\right)$ and $\left(
\begin{array}{lll}
0\\
x\phi_{0}^{\lambda+V(0)}
\end{array}
\right)$ for $h=0$ give rise to either two pure imaginary or two
real eigenvalues. We claim for $V^{''}(0)>0$ the former case takes
place; and for $V^{''}(0)<0,$ the latter one. To prove this we use
the Feshbach projection method (see ~\cite{GuSi}) with the
projections $\bar{P}:=I-P$ and $$P:=\text{Projection on
Span}\{\left(
\begin{array}{lll}
0\\
\phi_{h}^{\lambda}
\end{array}
\right),\ \left(
\begin{array}{lll}
\partial_{\lambda}\phi_{h}^{\lambda}\\
0
\end{array}
\right),\ \left(
\begin{array}{lll}
\partial_{x} \phi_{h}^{\lambda}\\
0
\end{array}
\right),\ \left(
\begin{array}{lll}
0\\
x\phi_{h}^{\lambda}
\end{array}
\right)\}.$$ Then the eigenvalue equation $L(\lambda)\psi=\mu\psi$
is equivalent to the nonlinear eigenvalue problem
$$(PL(\lambda)P-W)\phi=\mu\phi$$ where $\phi\in \text{Span}P$ and
$W:=PL(\lambda)\bar{P}(\bar{P}L(\lambda)\bar{P}-\mu)^{-1}\bar{P}L(\lambda)P.$
It is easy to see that there exists some constant $\delta_{1},\
\delta_{2}>0$ such that if $h$ is sufficiently small and if
$|\mu|\leq \delta_{1}$ then for $n=0,1,2$
\begin{equation}\label{estimatemu}
\|\partial_{\mu}^{n}(\bar{P}L(\lambda)\bar{P}-\mu)^{-1}\|_{\mathcal{L}^{2}\cap
\text{Range}\bar{P}\rightarrow \mathcal{L}^{2}\cap
\text{Range}\bar{P}}\leq \delta_{2}.
\end{equation}
We claim that
$$\|W\|=O(h^{3}).$$ Indeed, similarly as in ~\cite{oh1}
we can get that $L(\lambda)=L^{h=0}(\lambda)+O(h^{3/2})$ and
$P=P^{h=0}+O(h^{3/2}).$ Therefore
$$PL(\lambda)\bar{P}=P^{h=0}L^{h=0}(\lambda)\bar{P}^{h=0}+O(h^{3/2})=O(h^{3/2})$$
and similarly $\bar{P}L(\lambda)P=O(h^{3/2}).$ Since we look for
small eigenvalues $\mu$ we could use Estimate (~\ref{estimatemu})
to prove $\partial_{\mu}^{n}W=O(h^{3})$, $n=0,1,2$. We have the
following observations for the term $PL(\lambda)P:$
$$
\begin{array}{lll}
PL(\lambda)P\left(
\begin{array}{lll}
0\\
\phi^{\lambda}_{h}
\end{array}
\right)=0,\  PL(\lambda)P\left(
\begin{array}{lll}
\partial_{x}\phi^{\lambda}_{h}\\
0
\end{array}
\right)=\left(
\begin{array}{lll}
0\\
\phi^{\lambda}_{h}
\end{array} \right)
\end{array}
$$
$$
PL(\lambda)P\left(
\begin{array}{lll}
0\\
x\phi^{\lambda}_{h}
\end{array}
\right)=\left(
\begin{array}{lll}
2\partial_{x}\phi_{h}^{\lambda}\\
0
\end{array}
\right),
$$
$$
PL(\lambda)P\left(
\begin{array}{lll}
\partial_{x}\phi_{h}^{\lambda}\\
0
\end{array}
\right)=\left(
\begin{array}{lll}
0\\
h^{2}V^{''}(0)x\phi_{h}^{\lambda}
\end{array}
\right)+O(h^{3})
$$
The operator $PL(\lambda)P+W$ restricted to the 4-dimensional
space $Ran P$ has the $4\times 4$ matrix:
$$
\left(
\begin{array}{cccc}
0 & 1 & 0 & 0\\
0 & 0 & 0 & 0\\
0 & 0 & 0 & h^{2}V^{''}(0)\\
0 & 0 & 2 & 0
\end{array}
\right)+O(h^{3}).
$$
By a standard contraction argument we could prove that there are
four eigenvalues, i.e. four values of $\mu$:
$0+O(h^{3/2}),0+O(h^{3/2}),\pm\sqrt{-2h^{2}V^{''}(0)}+O(h^{3/2}).$
Since we already know that $L(\lambda)$ has an eigenvalue 0 with
multiplicity 2, the other two eigenvalues are
$\pm\sqrt{-2h^{2}V^{''}(0)}+O(h^{3/2}).$
\end{proof}
\begin{corollary}
There exist a real function $\xi_{1}$ and an imaginary function
$\eta_{1}$ such that $\left(
\begin{array}{lll}
\xi_{1}\\
\eta_{1}
\end{array}
\right)$ is the eigenvector of $L(\lambda)$ with small, nonzero
and imaginary eigenvalue $i\epsilon_{1}$. Therefore
\begin{equation}\label{imaginary}
L(\lambda)\left(
\begin{array}{lll}
\xi_{1}\\
\eta_{1}
\end{array}
\right)=i\epsilon_{1}\left(
\begin{array}{lll}
\xi_{1}\\
\eta_{1}
\end{array}
\right),\ L(\lambda)\left(
\begin{array}{lll}
\xi_{1}\\
-\eta_{1}
\end{array}
\right)=-i\epsilon_{1}\left(
\begin{array}{lll}
\xi_{1}\\
-\eta_{1}
\end{array}
\right)
\end{equation}
\end{corollary}
For the operator $L(\lambda)$ is not self-adjoint we define the
projection onto the pure point spectrum subspace of $L(\lambda)$
as:
$$P^{L(\lambda)}_{d}:=\frac{1}{2i\pi}\int_{\Gamma}(L(\lambda)-z)^{-1}dz,$$
where curve $\Gamma$ is a small circle around $0:$
$$\Gamma:=\{z||z|=\min(\lambda,2\epsilon_{1})\},$$ where, recall
$\epsilon_{1}$ from Equation (~\ref{imaginary}).
\begin{proposition}\label{projection}In the Dirac notation
$$
\begin{array}{lll}
P_{d}^{L(\lambda)}&=&\frac{1}{\langle
\phi_{h}^{\lambda},\frac{d}{d\lambda}\phi_{h}^{\lambda}\rangle}(\left|
\begin{array}{lll}
0\\
\phi_{h}^{\lambda}
\end{array}
\right\rangle \left\langle
\begin{array}{lll}
\frac{d}{d\lambda}\phi_{h}^{\lambda}\\
0
\end{array}
\right|+\left|
\begin{array}{lll}
\frac{d}{d\lambda}\phi_{h}^{\lambda}\\
0
\end{array}
\right\rangle \left\langle
\begin{array}{lll}
0\\
\phi_{h}^{\lambda}
\end{array}
\right|)\\
& &+\frac{1}{2\langle \xi_{1},\eta_{1}\rangle}(\left|
\begin{array}{lll}
\xi_{1}\\
\eta_{1}
\end{array}
\right\rangle\left\langle
\begin{array}{lll}
-\eta_{1}\\
\xi_{1}
\end{array}
\right| - \left|
\begin{array}{lll}
\xi_{1}\\
-\eta_{1}
\end{array}
\right\rangle\left\langle
\begin{array}{lll}
\eta_{1}\\
\xi_{1}
\end{array}
\right|).
\end{array}
$$
\end{proposition}
The proof of this proposition is straightforward but tedious, and
given in Appendix ~\ref{appendixB} where it is proved in a more
general setting.
\begin{definition}\label{continuousspace}
We define the essential spectrum subspace of $L(\lambda)$ as
$\text{Range}(1-P_{d}^{L(\lambda)}),$ where $P_{d}^{\lambda}$ is
defined before Proposition ~\ref{projection}. And we define the
operator
\begin{equation}\label{defineessential}
P_{ess}^{L(\lambda)}:=1-P_{d}^{L(\lambda)}.
\end{equation}
\end{definition}
\section{Re-parametrization of $\psi(t)$}\label{asym}
In this section we introduce a convenient decomposition of the
solution $\psi(t)$ to Equation (~\ref{NLS}) into a solitonic
component and a simplectically fluctuation.
\subsection{Decomposition of $\psi(t)$}
In this subsection we decompose $\psi(t)$, and derive equations of
each component. From now on we fix one sufficiently small $h$, and
we will drop the subindex $h$ and denote $\phi_{h}^{\lambda}$ by
$\phi^{\lambda}$, and $\frac{d}{d\lambda}\phi^{\lambda}_{h}$ by
$\phi_{\lambda}^{\lambda}.$
\begin{theorem} Assume $V$ and $\psi(0)$ are even.
There exists a constant $\delta>0$, so that if the initial datum
$\psi(0)$ satisfies $\displaystyle\inf_{\gamma\in
\mathcal{R}}\|\psi(0)-e^{i\gamma}\phi^{\lambda}\|_{\mathcal{H}^{1}}<\delta$,
then there exist differentiable functions $ \lambda,\
\gamma:\mathbb{R^{+}}\rightarrow \mathbb{R}$, such that
\begin{equation}\label{decomposition}
\psi(t)=e^{i\int_{0}^{t}\lambda(t)dt+i
\gamma(t)}(\phi^{\lambda(t)}+R),
\end{equation}
where $R$ is in the essential spectrum subspace, i.e.
\begin{equation}\label{orthogonal}
Im \langle R,i\phi^{\lambda}\rangle=Im\langle
R,\phi^{\lambda}_{\lambda}\rangle=0.
\end{equation}
\end{theorem}
\begin{proof}
By the Lyapunov stability (see ~\cite{Oh2,GSS1}), $\forall\
\epsilon>0,$ there exists a constant $\delta,$ such that if
$\displaystyle\inf_{\gamma\in
R}\|\psi(0)-e^{i\gamma}\phi^{\lambda}\|_{\mathcal{H}^{1}}<\delta,$
then $\forall \ t>0$, $\displaystyle\inf_{\gamma}
\|\psi(t)-e^{i\gamma}\phi^{\lambda}\|_{\mathcal{H}^{1}}<\epsilon.$
The decompositions (~\ref{decomposition}) (~\ref{orthogonal})
follow from Splitting Theorem in ~\cite{sigal} and the fact that
$\psi(t)$ are even while all the eigenvectors, besides even
eigenvectors $\left(
\begin{array}{lll}
0\\
\phi^{\lambda}
\end{array}
\right)$ and $\left(
\begin{array}{lll}
\phi^{\lambda}_{\lambda}\\
0
\end{array}
\right),$ are odd.
\end{proof}
Plug Equation (~\ref{decomposition}) into Equation (~\ref{NLS}) to
obtain:
\begin{equation}\label{plug}
\begin{array}{lll}
& &
-\dot{\gamma}(\phi^{\lambda}+R)+i(\dot{\lambda}\phi_{\lambda}^{\lambda}+R_{t})\\
& &=- R_{xx}+\lambda
R+V_{h}R-f(|\phi^{\lambda}|^{2})R-f^{'}(|\phi^{\lambda}|^{2})(\phi^{\lambda})^{2}(R+\bar{R})+N(R),
\end{array}
\end{equation}
where
$$N(R)=-f(|\psi|^{2})(\phi^{\lambda}+R)+f(|\phi^{\lambda}|^{2})(\phi^{\lambda}+R)+f^{'}(|\phi^{\lambda}|^{2})(\phi^{\lambda})^{2}(R+\bar{R}).$$
Passing from complex functions $R=R_{1}+iR_{2}$ to real
vector-functions $\left(
\begin{array}{lll}
R_{1}\\
R_{2}
\end{array}
\right)$ we obtain
\begin{equation}\label{continu}
\frac{d}{dt}\left(
\begin{array}{lll}
R_{1}\\
R_{2}
\end{array}
\right)=\left(
\begin{array}{lll}
0&L_{-}(\lambda)\\
-L_{+}(\lambda)&0
\end{array}
\right)\left(
\begin{array}{lll}
R_{1}\\
R_{2}
\end{array}
\right)+\left(
\begin{array}{lll}
Im N(R)\\
-Re N(R)
\end{array}
\right)+\left(
\begin{array}{lll}
-\dot{\lambda}\phi^{\lambda}_{\lambda}\\
-\dot{\gamma}\phi^{\lambda}
\end{array}
\right).
\end{equation}

Differentiating $Im\langle R,i\phi^{\lambda}\rangle=0$ (see
Decomposition ~\ref{orthogonal}) with respect to $t$, we get
\begin{equation}\label{Dorth}
Im\langle R_{t},\phi^{\lambda}\rangle+\dot{\lambda}Im\langle R,
i\phi^{\lambda}_{\lambda}\rangle=0.
\end{equation}
Multiply Equation (~\ref{plug}) by $i\phi^{\lambda}$ and use
Equation (~\ref{Dorth}) to obtain:
$$
\dot{\lambda}\langle
\phi_{\lambda}^{\lambda},\phi^{\lambda}\rangle-\dot{\lambda}Re\langle
R,\phi^{\lambda}_{\lambda}\rangle -\dot{\gamma}Im\langle
R,\phi^{\lambda}\rangle=Im\langle N(R),\phi^{\lambda}\rangle.
$$
By similar reasoning the relation $Im\langle
R,\phi_{\lambda}^{\lambda}\rangle=0$ implies that
$$
-\dot{\gamma}\langle
\phi^{\lambda},\phi^{\lambda}_{\lambda}\rangle-\dot{\gamma}Re\langle
R, \phi_{\lambda}^{\lambda}\rangle+\dot{\lambda}Im\langle R,
\phi_{\lambda\lambda}^{\lambda}\rangle =Re\langle
N(R),\phi_{\lambda}^{\lambda}\rangle.
$$
Combine the last two equations into a matrix form:
\begin{lemma}
The parameters $\lambda$ and $\gamma$ fixed by Equations
(~\ref{Dorth}) (~\ref{orthogonal}) satisfy the equations:
\begin{equation}\label{originlambdagamma}
\left[ \begin{array}{ccc} \langle
\phi_{\lambda}^{\lambda},\phi^{\lambda}\rangle-Re\langle
R,\phi_{\lambda}^{\lambda}\rangle & -Im\langle
R,\phi^{\lambda}\rangle\\ -Im\langle R,
\phi_{\lambda\lambda}^{\lambda}\rangle & \langle
\phi_{\lambda}^{\lambda},\phi^{\lambda}\rangle+Re\langle R,
\phi_{\lambda}^{\lambda}\rangle
\end{array} \right]
\left[\begin{array}{cc} \dot{\lambda}\\ \dot{\gamma}
\end{array}\right]=
\left[\begin{array}{cc} Im\langle N(R),\phi^{\lambda}\rangle \\
-Re\langle N(R),\phi^{\lambda}_{\lambda}\rangle
\end{array}\right] .
\end{equation}
\end{lemma}
By our requirement and orbital stability of the solitons, $\langle
\phi_{\lambda}^{\lambda},\phi^{\lambda}\rangle
>\epsilon_{0}>0$ for some constant $\epsilon_{0},$ and $\langle R,\phi_{\lambda}^{\lambda}\rangle$ and $\langle
R,\phi_{\lambda\lambda}^{\lambda}\rangle$ are small. Thus the
matrix on the left hand side is invertible and
\begin{equation}\label{invert}
 \| \left[
\begin{array}{ccc} \langle
\phi_{\lambda}^{\lambda},\phi^{\lambda}\rangle-Re\langle
R,\phi_{\lambda}^{\lambda}\rangle & -Im\langle
R,\phi^{\lambda}\rangle\\
-Im\langle R, \phi_{\lambda\lambda}^{\lambda}\rangle & \langle
\phi_{\lambda}^{\lambda},\phi^{\lambda}\rangle+Re\langle R,
\phi_{\lambda}^{\lambda}\rangle
\end{array}\right ]^{-1}  \|\leq c
\end{equation} for some $c>0$ independent of time $t.$
\subsection{Change of Variables}
In this subsection we study key Equation (~\ref{continu}). The
study is complicated by the fact that the linearized operator
$L(\lambda(t))$ depends on time $t$. To circumvent this difficulty
we rearrange Equation (~\ref{continu}) as follows. We fix time
$T>0$, and define function $g^{T}$ by
\begin{equation}\label{rg}
e^{i\int_{0}^{t}\lambda(s)ds+i\gamma(t)}R=:e^{i\lambda_{1}
 t+i\gamma_{1}}g^{T}
\end{equation}  where $\gamma_{1}=\gamma(T)$ and
$\lambda_{1}=\lambda(T)$. Denote
\begin{equation}\label{Delta}
\Delta_{1}:=-\int_{0}^{t}\lambda(t)dt-\gamma(t)+\lambda_{1}t+\gamma_{1}.
\end{equation}
From Equations (~\ref{rg}) and (~\ref{Delta}), we derive the
equation for $g^{T}$. Let $g^{T}=g^{T}_{1}+ig^{T}_{2},$ then
Equation (~\ref{continu}) implies
\begin{equation}\label{rewrit}
\frac{d}{dt}\left(
\begin{array}{lll}
g^{T}_{1}\\
g^{T}_{2}
\end{array}
\right)=L(\lambda_{1})\left(
\begin{array}{lll}
g^{T}_{1}\\
g^{T}_{2}
\end{array}
\right)+\left(
\begin{array}{lll}
Im D\\
-Re D
\end{array}
\right),
\end{equation}
where $$D=D_{1}+D_{2}+D_{3},$$
$$D_{1}=\dot{\gamma}\phi^{\lambda}e^{-i\Delta_{1}}-i\dot{\lambda}\phi_{\lambda}^{\lambda}e^{-i\Delta_{1}},$$
$$
\begin{array}{lll}
D_{2}&=&[f(|\phi^{\lambda_{1}}|^{2})+f^{'}(|\phi^{\lambda_{1}}|^{2})(\phi^{\lambda_{1}})^{2}-f(|\phi^{\lambda}|^{2})-f^{'}(|\phi^{\lambda}|^{2})(\phi^{\lambda})^{2}]g^{T}\\
&+&[f^{'}(|\phi^{\lambda_{1}}|^{2})(\phi^{\lambda_{1}})^{2}-f^{'}(|\phi^{\lambda}|^{2})(\phi^{\lambda})^{2}]\bar{g^{T}}+f^{'}(|\phi^{\lambda}|^{2})(\phi^{\lambda})^{2}[1-e^{-2i\Delta_{1}}]\bar{g^{T}},
\end{array}
$$
$$D_{3}=e^{-i\Delta_{1}}N(R).$$
We need to decompose $g^{T}$ along the point spectrum and
essential spectrum subspaces of the operator $L(\lambda_{1})$.
Since $g^{T}$ is even and $\langle
\phi^{\lambda_{1}},\phi^{\lambda_{1}}_{\lambda_{1}}\rangle >0$,
there are differentiable real functions $k_{1}^{T},\ k_{2}^{T}:\
[0,T]\rightarrow \mathbb{R}$ such that
\begin{equation}\label{decomposeg}
g^{T}=ik_{1}^{T}\phi^{\lambda_{1}}+k_{2}^{T}\phi^{\lambda_{1}}_{\lambda_{1}}+h^{T},
\end{equation}
and $h^{T}$ is in the essential spectrum subspace of
$L(\lambda_{1})$, where, recall $P_{ess}$ from Equations
(~\ref{defineessential}).
\begin{lemma}
The functions $k_{1}^{T},$ $k_{2}^{T}$ and
$h^{T}=h^{T}_{1}+ih^{T}_{2}$ satisfy the following equations:
\begin{equation}\label{k1k22}
\left[
\begin{array}{cc}
-\sin(\Delta_{1})\langle \phi^{\lambda_{1}},\phi^{\lambda}\rangle,
& \cos(\Delta_{1})\langle
\phi^{\lambda_{1}}_{\lambda_{1}},\phi^{\lambda}\rangle\\
\cos(\Delta_{1})\langle
\phi^{\lambda}_{\lambda},\phi^{\lambda_{1}}\rangle,&
\sin(\Delta_{1})\langle
\phi^{\lambda_{1}}_{\lambda_{1}},\phi^{\lambda}_{\lambda}\rangle
\end{array}\right]\left[
\begin{array}{lll}
k_{1}^{T}\\
k_{2}^{T}
\end{array}
\right]=-\left[
\begin{array}{lll}
Re\langle e^{i\Delta_{1}}h^{T},\phi^{\lambda} \rangle\\
Im\langle e^{i\Delta_{1}}h^{T},\phi^{\lambda}_{\lambda}\rangle
\end{array} \right],
\end{equation}
\begin{equation}\label{equationh}
\frac{d}{dt}\left(
\begin{array}{lll}
h^{T}_{1}\\
h^{T}_{2}
\end{array}
\right)=L(\lambda_{1})\left(
\begin{array}{lll}
h^{T}_{1}\\
h^{T}_{2}
\end{array}
\right)+P_{ess}\left(
\begin{array}{lll}
Im D\\
-Re D
\end{array}
\right).
\end{equation}
\end{lemma}
\begin{proof}
By Equation (~\ref{orthogonal}), we have the following two
equations:
$$
\begin{array}{lll}
0&=&Im\langle R,i\phi^{\lambda}\rangle=Re\langle
e^{i\Delta_{1}}g^{T},\phi^{\lambda}\rangle\\
&=&k_{2}^{T}\cos(\Delta_{1})\langle
\phi^{\lambda_{1}}_{\lambda_{1}},\phi^{\lambda}\rangle-k_{1}^{T}\sin(\Delta_{1})\langle
\phi^{\lambda_{1}},\phi^{\lambda}\rangle+Re\langle
e^{i\Delta_{1}}h^{T},\phi^{\lambda}\rangle;
\end{array}
$$
$$
\begin{array}{lll}
0&=&Im\langle R,\phi^{\lambda}_{\lambda}\rangle\\
&=&k_{2}^{T}\sin(\Delta_{1})\langle
\phi^{\lambda_{1}}_{\lambda_{1}},\phi^{\lambda}_{\lambda}\rangle+k_{1}^{T}\cos(\Delta_{1})\langle
\phi^{\lambda_{1}},\phi^{\lambda}_{\lambda}\rangle+Im\langle
e^{i\Delta_{1}}h^{T},\phi^{\lambda}_{\lambda}\rangle.
\end{array}
$$
Since $\left(
\begin{array}{lll}
0\\
\phi^{\lambda_{1}}
\end{array}
\right)$ and $\left(
\begin{array}{lll}
\phi^{\lambda_{1}}_{\lambda_{1}}\\
0
\end{array}
\right)$ are eigenvectors of $L(\lambda_{1}),$ Equation
(~\ref{rewrit}) implies Equation (~\ref{equationh}).
\end{proof}
When $|\lambda-\lambda_{1}|$ is small, $\langle
\phi_{\lambda_{1}}^{\lambda_{1}},\phi^{\lambda}\rangle,\ \langle
\phi^{\lambda}_{\lambda},\phi^{\lambda_{1}}\rangle>\epsilon_{0}>0$
for some constant $\epsilon_{0}$. Thus in this case the matrix
\[ \left[
\begin{array}{cc}
-\sin(\Delta_{1})\langle \phi^{\lambda_{1}},\phi^{\lambda}\rangle,
& \cos(\Delta_{1})\langle
\phi^{\lambda_{1}}_{\lambda_{1}},\phi^{\lambda}\rangle\\
\cos(\Delta_{1})\langle
\phi^{\lambda}_{\lambda},\phi^{\lambda_{1}}\rangle,&
\sin(\Delta_{1})\langle
\phi^{\lambda_{1}}_{\lambda_{1}},\phi^{\lambda}_{\lambda}\rangle
\end{array}\right]
\]
has an inverse uniformly bounded in $t$ and $T$.
\subsection{Estimates of the Parameters $\lambda,\ \gamma$ and the
Function $R$}\label{estimateparameters} In this subsection we will
estimate the parameters $\lambda(t),\ \gamma(t)$ and the function
$R(t).$
\begin{proposition}\label{lambdagammak1k2}
Let $\nu>7/2$ and $\rho_{\nu} :=  (1+|x|)^{-\nu}.$ We have for
time $t\geq 0,$
$$|\dot{\lambda}(t)|+|\dot{\gamma}(t)|\leq c(1+t)^{-3},$$
$$\|\rho_{\nu}R\|_{2}\leq c(1+t)^{-3/2}$$
where the constant $c$ is independent of $t$.
\end{proposition}
The proof of this proposition is based on estimates of the
evolution operator
$$U(t)=e^{tL(\lambda_{1})}$$ which we formulate now. Note that $U(t)$ is defined in a standard way
(see Lemma ~\ref{definepropagator} for detailed definition).
Recall that the operator $P_{ess}$, defined in Equation
(~\ref{defineessential}), is the projection onto the essential
spectrum subspace $\mathcal{H}_{pp}(L^{*}(\lambda))^{\perp}$. We
prove in Section
 ~\ref{Estimatepropagator} that in the 1-dimensional case $U(t)$ satisfies the
following estimates:
\begin{equation}\label{second}
\|\rho_{\nu} U(t)P_{ess} h\|_{2}\leq
c(1+t)^{-\frac{3}{2}}\|\rho_{-2}h\|_{2};
\end{equation}
\begin{equation}\label{fourth}
\|\rho_{\nu}U(t)P_{ess}h\|_{2}\leq
c(1+t)^{-3/2}(\|\rho_{-2}h\|_{1}+\|h\|_{2});
\end{equation}
\begin{equation}\label{third}
\|U(t)P_{ess} h\|_{\mathcal{L}^{\infty}}\leq
ct^{-1/2}(\|\rho_{-2}h\|_{1}+\|h\|_{2});
\end{equation}
\begin{equation}\label{first}
\|U(t)P_{ess} h\|_{\mathcal{L}^{\infty}}\leq
c(1+t)^{-\frac{1}{2}}\|\rho_{-2}h\|_{\mathcal{H}^{1}};
\end{equation}
where $\nu>7/2.$ Recall $\rho_{\nu}(x) = (1+|x|)^{-\nu}. $\\
\textit{Proof of Proposition ~\ref{lambdagammak1k2}} We will
estimate the following quantities:
$$m^{T}_{1}(t)=\|\rho_{\nu} h^{T}\|_{2},\ \ \ \ m_{2}^{T}(t)=\|g^{T}\|_{\mathcal{L}^{\infty}},$$
$$M_{1}(T)=\sup_{\tau\leq T}(1+\tau)^{3/2}m^{T}_{1}(\tau),\ \ \ \ M_{2}(T)=\sup_{\tau\leq T}(1+\tau)^{1/2}m^{T}_{2}(\tau),$$
where $\nu$ is a constant greater than $3.5$.

Note various constants $c$ used below do not depend on $t$ or $T$.

The matrix on the left hand side of Equation (~\ref{k1k22}) has a
uniformly bounded inverse. Hence using the definition of $M_{1}$
we obtain
\begin{equation}\label{k1k2}
|k_{1}^{T}|+|k_{2}^{T}| \leq c M_{1}(1+t)^{-\frac{3}{2}}.
\end{equation}
From Equations (~\ref{originlambdagamma}), (~\ref{invert}),
(~\ref{rg}) and (~\ref{decomposeg}), we have $$|
\dot{\lambda}|+|\dot{\gamma}|\leq c \|\rho_{\nu}R\|_{2}^{2} \leq
c(|k_{1}^{T}|+|k_{2}^{T}|+\|\rho_{\nu}h^{T}\|_{2})^{2}.
$$
Hence by the definition of $M_{1},$
\begin{equation}\label{lambdagamma}
|\dot{\lambda}|+ |\dot{\gamma}|\leq cM_{1}^{2}(1+t)^{-3}.
\end{equation}

By the definition of $g^{T}$ in (~\ref{rg}),
$\|\rho_{\nu}R\|_{2}=\|\rho_{\nu}g^{T}\|_{2}.$ By the
decomposition (~\ref{decomposeg}), we have
\begin{equation}\label{Estimater}
\|\rho_{\nu}R\|_{2}\leq
c(|k_{1}^{T}|+|k_{2}^{T}|)+\|\rho_{\nu}h^{T}\|_{2}\leq
c(1+t)^{-3/2}M_{1}.
\end{equation}

We need to estimate $\Delta_{1}$ given in Equation (~\ref{Delta}).
By the observations that
$$\lambda(t)-\lambda_{1}=-\int_{t}^{T}\dot{\lambda}(\tau)d\tau\
\text{and}\
\gamma(t)-\gamma_{1}=-\int_{t}^{T}\dot{\gamma}(\tau)d\tau$$ and
Estimate (~\ref{lambdagamma}), we have
$$
\begin{array}{lll}
|e^{i\Delta_{1}}-1|&=&|e^{-\int_{0}^{t}i(\lambda(t)-\lambda_{1})dt-
\int_{0}^{t}i\dot{\gamma}(t)dt}-1|\\
&=&|e^{-i\int_{0}^{t}\int^{T}_{t}\dot{\lambda}(\tau)d\tau dt-i\int_{t}^{T}\dot{\gamma}(t)dt}-1|\\
&\leq &cM_{1}^{2}.
\end{array}
$$

Now we estimate $h^{T}$. Using the Duhamel principle, we rewrite
Equation (~\ref{equationh}) as
$$\left(
\begin{array}{lll}
h_{1}^{T}\\
h_{2}^{T}
\end{array} \right)=U(t)\left(
\begin{array}{lll}
h_{1}^{T}(0)\\
h_{2}^{T}(0)
\end{array} \right)+\int_{0}^{t}U(t-\tau)P_{ess}\left(
\begin{array}{lll}
Im D\\
-Re D
\end{array}
\right) d\tau,$$ where, recall, $U(t)=e^{tL(\lambda_{1})}.$ Using
Estimate (~\ref{third}), we obtain
\begin{equation}\label{hinfty}
\begin{array}{lll}
\|h^{T}\|_{\mathcal{L}^{\infty}}&\leq
&\|U(t)h^{T}(0)\|_{\mathcal{L}^{\infty}}+\int_{0}^{t}\sum_{n=1}^{3}\|U(t-\tau)P_{ess}\left(
\begin{array}{lll}
Im D_{n}\\
-Re D_{n}
\end{array}
\right)\|_{\mathcal{L}^{\infty}}d\tau\\
&\leq
&\|U(t)h^{T}(0)\|_{\mathcal{L}^{\infty}}\\
&
&+c\int_{0}^{t}(1+|t-\tau|)^{-1/2}\sum_{n=1}^{3}(\|\rho_{-2}D_{n}\|_{1}+\|D_{n}\|_{2})d\tau,
\end{array}
\end{equation}
and using Estimate (~\ref{fourth}) we derive
\begin{equation}\label{h2}
\begin{array}{lll}
\|\rho_{\nu}h^{T}\|_{2}&\leq
&\|\rho_{\nu}U(t)h^{T}(0)\|_{2}+\int_{0}^{t}\sum_{n=1}^{3}\|\rho_{\nu}U(t-\tau)P_{ess}\left(
\begin{array}{lll}
Im D_{n}\\
-Re D_{n}
\end{array}
\right)\|_{2}d\tau\\
&\leq
&\|\rho_{\nu}U(t)h^{T}(0)\|_{2}\\
&
&+c\int_{0}^{t}(1+|t-\tau|)^{-3/2}\sum_{n=1}^{3}(\|\rho_{-2}D_{n}\|_{1}+\|D_{n}\|_{2})d\tau.
\end{array}
\end{equation}

Next we estimate $\|\rho_{-2}D_{n}\|_{1}+\|D_{n}\|_{2},$
$n=1,2,3.$

By Estimate (~\ref{lambdagamma}) we can estimate $D_{1}:$
$$\|\rho_{-2}D_{1}\|_{1}+\|D_{1}\|_{2}\leq c( |\dot{\gamma}|+|\dot{\lambda}|)\leq c
M_{1}^{2}(1+t)^{-3}.$$

For $D_{2},$
$$\|\rho_{-2}D_{2}\|_{1}+\|D_{2}\|_{2}\leq cM_{1}^{2}(1+t)^{-\frac{3}{2}}.$$

For $D_{3}$, recall $f$ is a polynomial, so we can take out the
terms containing at least one power of $\phi_{\lambda}$, denote it
by $D_{I},$ and $D_{II} :=  D_{3}-D_{I}$. Since the leading term
of $D_{I}$ is $c|g^{T}|^{2}\phi^{\lambda}$, we have
\begin{equation}\label{dI}
\|D_{I}\|_{2}+\|\rho_{-2}D_{I}\|_{1}\leq cM_{2}M_{1}(1+t)^{-2}.
\end{equation} Here we have ignored the
higher order terms which are estimated by
$P(M_{2})M_{1}(1+t)^{-3},$ where the function $P$ is a polynomial
such that $P(0)=0.$

Since $c|g^{T}|^{2p+1}$ is the leading-order term of $D_{II}$,
$$\|D_{II}\|_{2}+\|(1+|x|)^{2}D_{II}\|_{1}\leq c(\|g^{T}\|_{\mathcal{L}^{\infty}}^{2p}\|g^{T}\|_{2}+\|g^{T}\|^{2p-1}_{\infty}\|(1+|x|)g^{T}\|_{2}^{2}).$$
Since $\|g^{T}\|_{2}\leq c$, $\|(1+|x|)g^{T}\|_{2}\leq c(1+t)$ by
Equation (~\ref{weightestimate}), and
$\|g^{T}\|_{\mathcal{L}^{\infty}}^{2p-1}\leq
M_{2}^{2p-1}(1+t)^{-p+1/2},$ and using that $p\geq 4$ we have
\begin{equation}\label{dII}
\begin{array}{lll}
& &\|D_{II}\|_{2}+\|(1+|x|)^{2}D_{II}\|_{1}\\
&\leq &c(M_{2}^{2p}(1+t)^{-p}+M_{2}^{2p-1}(1+t)^{-p+1/2}(1+t)^{2})\\
&\leq &c(M_{2}^{2p}(1+t)^{-4}+M_{2}^{2p-1}(1+t)^{-\frac{3}{2}}).
\end{array}
\end{equation}

By Estimates (~\ref{dI}) and (~\ref{dII}) and the fact that
$D_{3}=D_{I}+D_{II},$ we obtain
$$\|D_{3}\|_{2}+\|\rho_{-2}D_{3}\|_{1}\leq c(1+t)^{-3/2}(M_{2}M_{1}+M_{2}^{2p}+M_{2}^{2p-1}).$$

This finishes the estimates of the $D_{i}$'s. Now we return to the
estimation of the $h^{T}.$

By Estimate (~\ref{h2}), we have
$$
\begin{array}{lll}
& &\|\rho_{\nu}h^{T}\|_{2}\\
&\leq
&c(1+t)^{-3/2}(\|g^{T}(0)\|_{2}+\|\rho_{-2}g^{T}(0)\|_{1})+\\ & &
c\int_{0}^{t}(M_{2}M_{1}+M_{1}^{2}+M_{2}^{2p}+M_{2}^{2p-1})\frac{1}{(1+t-\tau)^{3/2}(1+\tau)^{3/2}}d\tau,
\end{array}
$$
so
$$\begin{array}{lll}
(1+t)^{3/2}\|\rho_{\nu}h^{T}\|_{2}&\leq&
c(\|g^{T}(0)\|_{2}+\|\rho_{-2}g^{T}(0)\|_{1})\\
& &+c(M_{2}M_{1}+M_{1}^{2}+M_{2}^{2p}+M_{2}^{2p-1}).
\end{array}
$$

Remember the definition of $M_{1}$ and
$$S=\|\rho_{-2}g^{T}(0)\|_{1}+\|g^{T}(0)\|_{\mathcal{H}^{1}}.$$
Then we have
\begin{equation}\label{M2M3}
M_{1}\leq cS+c(M_{2}M_{1}+M_{1}^{2}+M_{2}^{2p}+M_{2}^{2p-1}).
\end{equation}

By Estimate (~\ref{hinfty}), we have
$$
\begin{array}{lll}
\|h^{T}\|_{\mathcal{L}^{\infty}} &\leq
&c(1+t)^{-\frac{1}{2}}(\|\rho_{-2}g^{T}(0)\|_{1}+\|g^{T}(0)\|_{\mathcal{H}^{1}})\\
&
&+\int_{0}^{t}\frac{d\tau}{(t-\tau)^{1/2}}(\|D\|_{2}+\|\rho_{-2}D\|_{1}).
\end{array}
$$

By the definition of $M_{2}$ we have
\begin{equation}\label{M4}
M_{2}\leq cS+c(M_{2}M_{1}+M_{1}^{2}+M_{2}^{2p}+M_{2}^{2p-1}).
\end{equation}

By Equation (~\ref{rg}) we have
$$S=\|\rho_{-2}g^{T}(0)\|_{1}+\|g^{T}(0)\|_{\mathcal{H}^{1}}=\|\rho_{-2}R(0)\|_{1}+\|R(0)\|_{\mathcal{H}^{1}}.$$
Thus $S$ only depends on $R(0)$.

If $M_{n}(0),\ (n=1,2)$ are sufficiently close to zero, by
Estimates (~\ref{M2M3}) and (~\ref{M4}) we have shown that
$M_{1}(T), M_{2}(T)\leq\mu(S)S$ for any time $T$, where $\mu(S)$
is a function that is bounded for sufficiently small $S$. Thus we
have shown that $$M_{1}(T)+M_{2}(T)\leq
c(\|\rho_{-2}R(0)\|_{1}+\|R(0)\|_{\mathcal{H}^{1}}).$$ The last
estimate together with Estimates (~\ref{lambdagamma}) and
(~\ref{Estimater}) implies Proposition ~\ref{lambdagammak1k2}.
\begin{flushright}
$\square$
\end{flushright}
\subsection{Proof of Theorem ~\ref{maintheorem}}
In this subsection we prove our main Theorem ~\ref{maintheorem}.
To this end we use Proposition ~\ref{lambdagammak1k2}. Since
$$
|\dot{\lambda}| +|\dot{\gamma}|\leq c(1+t)^{-3}
$$
for some $c>0$, there exist $ \lambda_{\infty},\ \gamma_{\infty}$
such that
\begin{equation}\label{finallambda}
|\lambda(t)-\lambda_{\infty}|+|\gamma(t)-\gamma_{\infty}|\leq
c(1+t)^{-2}.
\end{equation}

Recall that the solutions $\psi(t)$ can be written as in Equation
(~\ref{decomposition})
$$\psi(t)=e^{-i\int_{0}^{t}\lambda(t)dt+i
\gamma(t)}(\phi^{\lambda(t)}+R).$$ By Proposition
~\ref{lambdagammak1k2} and Equation (~\ref{finallambda}) we have
for $\nu>3.5$
$$
\begin{array}{lll}
& &\|\rho_{\nu}(\phi^{\lambda(t)}+R-\phi^{\lambda_{\infty}})\|_{2}\\
&\leq&\|\rho_{\nu}(\phi^{\lambda(t)}-\phi^{\lambda_{\infty}})\|_{2}+\|\rho_{\nu}R\|_{2}\\
&\leq&c[(1+t)^{-2}+(1+t)^{-3/2}]\\
&\leq&c(1+t)^{-3/2},
\end{array}
$$
which implies Theorem ~\ref{maintheorem}.
\section{Estimates On Propagators for Matrix Schr\"odinger
Operators}\label{Estimatepropagator}
\subsection{Formulation of the Main Result}\label{ap1}
In this subsection we prove Estimates
(~\ref{second})-(~\ref{first}) on the propagator
$U(t)=e^{tL(\lambda_{1})},$ where $L(\lambda)$ is defined in
Definition ~\ref{operaL}. Actually we study more general
propagators generated by the operators $$ L_{general}=\left(
\begin{array}{lll}
0&L_{2}\\
-L_{1}&0
\end{array}
\right),
$$
where
$$L_{1}:=-\frac{d^{2}}{dx^{2}}+V_{1}+\beta,\
\text{and}\ L_{2}:=-\frac{d^{2}}{dx^{2}}+V_{2}+\beta.$$ Here the
constant $\beta>0$, and the functions $V_{1}$ and $V_{2}$ are
even, real and satisfy the estimates:
$$
|V_{1}(x)|, \ |V_{2}(x)|\leq c e^{-\alpha|x|}
$$ for some constants $c,\ \alpha>0$.

By standard arguments (see, e.g. ~\cite{RSIV}) we have that
$$\sigma_{ess}(L_{general})=i(-\infty,-\beta]\cup i[\beta,\infty).$$
The points $-i\beta$ and $i\beta$ are called thresholds. They
affect the long time behavior of the semigroup $e^{tL_{general}}$
in a crucial way.

The following notion will play an important role:
\begin{definition}
A function $h\not=0$ is called the resonance of $L_{general}$ at
$i\beta\ (\text{or}\ -i\beta)$, if and only if $h$ is bounded and
satisfies
$$(L_{general}-i\beta I)h=0\ \ (\text{or}\ (L_{general}+i\beta I)h=0).$$
\end{definition}
\begin{lemma}\label{definepropagator}
The operator $L_{general}$ generates a semigroup,
$e^{tL_{general}},\ t\geq 0.$
\end{lemma}
\begin{proof}
We write the operator $L_{general}$ as
$$L_{general}=L_{0}+U,$$ where
$$L_{0}:=\left(
\begin{array}{lll}
0&-\frac{d^{2}}{dx^{2}}+\beta\\
\frac{d^{2}}{dx^{2}}-\beta&0
\end{array} \right),\ U:=\left(
\begin{array}{lll}
0&V_{1}\\
-V_{2}&0
\end{array} \right).$$ It is easy to verify that the operator $L_{0}$ is
a generator of a $(C_{0})$ contraction semigroup (see, e.g.
~\cite{Gold,RSII}). Also the operator $U:
\mathcal{L}^{2}\rightarrow \mathcal{L}^{2}$ is bounded. By
[~\cite{Gold}, Theorem 6.4 and ~\cite{RSII}] $L_{general}=L_{0}+U$
generates a $(C_{0})$ semigroup.
\end{proof}

Let $P_{ess}$ be the projection onto the essential spectrum
subspace of $L_{general}$, where, recall the definition of
$P_{ess}$ in Equation (~\ref{defineessential}).
\begin{theorem}\label{maintheorem1}
Assume that $L_{general}$ has no resonances at $\pm i\beta,$ no
eigenvalues embedded in the essential spectrum, and has no
eigenvalues with non-zero real parts. Then $\forall \mu>3.5$ there
exists a constant $c=c(\mu)>0$ such that
\begin{equation}\label{estimate1}
\|\rho_{\mu}e^{tL_{general}}P_{ess}h\|_{2}\leq
c(1+t)^{-3/2}\|\rho_{-2}h\|_{2};
\end{equation}
\begin{equation}\label{estimate2}
\|\rho_{\mu}e^{tL_{general}}P_{ess}h\|_{2}\leq
c(1+t)^{-3/2}(\|\rho_{-2}h\|_{1}+\|h\|_{2})
\end{equation}
\begin{equation}\label{estimate3}
\|e^{tL_{general}}P_{ess}h\|_{\mathcal{L}^{\infty}}\leq
c(1+t)^{-1/2}\|\rho_{-2}h\|_{\mathcal{H}^{1}},
\end{equation}
\begin{equation}\label{estimate4}
\|e^{tL_{general}}P_{ess}h\|_{\mathcal{L}^{\infty}}\leq
ct^{-1/2}(\|h\|_{2}+\|\rho_{-2}h\|_{1}),
\end{equation}
where, recall $\rho_{\nu}(x)=(1+|x|)^{-\nu}.$
\end{theorem}
The proof of an equivalent theorem of Theorem ~\ref{maintheorem1}
is given in Subsection ~\ref{proofmainth}.

It is more convenient to transform first the operator
$L_{general}$ as
\begin{equation}\label{trans}
H := -iT^{*}L_{general}T,\end{equation} where the $2\times 2$
matrix $$ T:=\frac{1}{\sqrt{2}}\left(
\begin{array}{lll}
1&i\\
i&1
\end{array}
\right). $$ Compute the matrix operator $H$:
\begin{equation}\label{defineH}
H=H_{0}+W,
\end{equation}
where
\begin{equation}\label{h0w}
H_{0}:= \left(
\begin{array}{lll}
-\frac{d^{2}}{dx^{2}}+\beta&0\\
0&\frac{d^{2}}{dx^{2}}-\beta
\end{array}
\right), \ W:=1/2\left(
\begin{array}{lll}
V_{3}&-iV_{4}\\
-iV_{4}&-V_{3}
\end{array}\right),
\end{equation}
with the functions $V_{4} :=  V_{1}-V_{2},$ $V_{3} :=
V_{2}+V_{1}.$ From the properties of functions $V_{1},$ $V_{2}$ we
have
\begin{equation}\label{v3v4}
|V_{4}(x)|,\ |V_{3}(x)|\leq ce^{-\alpha|x|}
\end{equation}
for some constants $c,\alpha>0.$ Hence
$\sigma_{ess}(H)=(-\infty,-\beta]\cup [\beta,\infty).$ The
assumptions on $L_{general}$ are transported to $H$ as following:
$H$ has no resonances at $\pm\beta,$ and has only finitely many
eigenvalues which are in the interval $(-\beta,\beta).$

Clearly the operator $H$ also generates a semigroup $e^{-itH}.$ To
prove the theorem above we relate the propagator $e^{-itH}$ to the
resolvent $(H-\lambda \pm i0)^{-1}$ of the generator $H$ on the
essential spectrum.

We introduce some notations that will be used below: Let
$F=[f_{ij}]$ be an $n\times m$ matrix with entry $f_{ij}\in B,$
where $B$ is a normed space, then
$$\|F\|_{B} :=\sum_{i,j}\|f_{ij}\|_{B}.$$
The H\"older inequality for such vector-valued functions reads: if
the constants $p,q\geq 1$, $\frac{1}{p}+\frac{1}{q}=1,$ the
vectors $F_{1}\in \mathcal{L}^{p}$ and $F_{2}\in \mathcal{L}^{q},$
then
$$\|F_{1}F_{2}\|_{\mathcal{L}^{1}}\leq \|F_{1}\|_{\mathcal{L}^{p}}\|F_{2}\|_{\mathcal{L}^{q}}.$$
\subsection{The Spectral Representation and the Integral Kernel of the Propagator
$e^{-itH}P_{ess}$}\label{integralkernel} In this subsection we
compute the spectral representation and the integral kernel of
$e^{-itH}P_{ess}$, where $P_{ess}$ is the projection onto the
essential spectrum subspace of operator $H$ which is unbounded and
non-self-adjoint. The main theorem is:
\begin{theorem}\label{kernelpc}
Let $\epsilon_{0}$ be a small positive number. Then
\begin{equation}\label{kernelpc2}
\begin{array}{lll}
& &e^{-itH}P_{ess}=\displaystyle\lim_{K\rightarrow
\infty}\displaystyle\lim_{\epsilon\rightarrow 0^{+}}\\
& &
\frac{1}{2i\pi}\{\int_{\beta-\epsilon_{0}}^{K}+\int^{-\beta+\epsilon_{0}}_{-K}\}e^{-it\lambda}[(H-\lambda-i\epsilon)^{-1}-(H-\lambda+i\epsilon)^{-1}]d\lambda,
\end{array}
\end{equation}
where the limits on the right hand side are strong limits. The
limits are independent of $\epsilon_{0}$ because $$(H\pm
z+i0)^{-1}=(H\pm z-i0)^{-1}$$ for any $z$ in the interval
$[\beta-\epsilon_{0},\beta)$.
\end{theorem}
\begin{remark}
Clearly Equation (~\ref{kernelpc2}) can be extended to functions
$f(\lambda)$ such that $\int|\hat{f}(t)|dt\leq \infty$ by
$$f(H)P_{ess}=\frac{1}{2i\pi}\{\int_{\beta}^{\infty}+\int^{-\beta}_{-\infty}\}f(\lambda)[(H-\lambda-i0)^{-1}-(H-\lambda+i0)^{-1}]d\lambda,$$
where the function $\hat{f}$ is the Fourier transform of $f$. It
can also be extended to other classes of functions using that if
$\lambda\in (-\infty,-\beta]\cup [\beta,\infty)$ and $g\in
\mathcal{C}_{0}^{\infty}$ then
\begin{equation}\label{positive}
\langle g, \frac{1}{2\pi
i}\sigma_{3}[(H-\lambda-i0)^{-1}-(H-\lambda+i0)^{-1}]g\rangle\geq
0,
\end{equation} where $\sigma_{3}:=\left(
\begin{array}{lll}
1&0\\
0&-1
\end{array}
\right).$

Since the operator $H$ plays an important role in the nonlinear
theory we discuss this extension elsewhere. It is not used in this
paper.
\end{remark}
We divide the proof of Theorem ~\ref{kernelpc} into two parts: in
Lemmas ~\ref{weaklimit} and ~\ref{decay} we prove that the limits
on the right hand side exist; then we prove that the left and
right hand sides are equal.

For $\theta\in \mathbb{R},$ we define the space
$\mathcal{L}^{2,\theta}:=(1+|x|)^{\theta}\mathcal{L}^{2}:$
$$\|g\|_{\mathcal{L}^{2,\theta}}=\|(1+|x|)^{\theta}g\|_{2}.$$
\begin{lemma}\label{weaklimit}
Let $\beta^{'}<\beta$ with $\beta-\beta^{'}$ sufficiently small.
If $K$ is a sufficiently large constant, then for any function
$g\in \mathcal{L}^{2,2}$
\begin{equation}\label{stronglimit}
\lim_{\epsilon\rightarrow 0^{+}}\int_{\beta^{'}}^{K}
e^{-it(\lambda-i\epsilon)}(H-\lambda+i\epsilon)^{-1}g d\lambda
\end{equation} exists in the space $\mathcal{L}^{2,-2}.$
\end{lemma}
\begin{proof}
Let $f, g\in \mathcal{L}^{2,2}.$ The function
$$u(z):=\langle f,e^{-itz}(H-z)^{-1}g\rangle$$ is
analytic on the set $$\Gamma:=\{z|Im z>0,\ Rez\geq \beta^{'}\}.$$
Therefore
\begin{equation}\label{analyticfunction}
\begin{array}{lll}
& &\int_{\beta^{'}}^{K}u(z-i0)dz\\
&=&\int_{\beta^{'}-i\epsilon}^{K-i\epsilon}u(z)dz+\int_{K-i\epsilon}^{K}u(z)dz+\int_{\beta^{'}}^{\beta^{'}-i\epsilon}u(z)dz.
\end{array}
\end{equation}
Moreover, since $\sigma_{ess}(H)=(-\infty,-\beta]\cup
[\beta,\infty)$, we have for a $\epsilon_{0}\in(0,\beta)$ that
$$|u(z)|\leq c\|f\|_{2}\|g\|_{2}$$ in the interval
$[\beta^{'}-i\epsilon,\beta^{'}]$. Hence
\begin{equation}\label{inversebound}
|\int_{\beta^{'}}^{\beta^{'}-i\epsilon}u(z)dz|\leq
c\epsilon\|f\|_{2}\|g\|_{2}.
\end{equation}

Consider $u(z)$ in the interval $[K-i\epsilon,K]$. We claim that
in this interval
$$|u(z)|\leq c\|f\|_{\mathcal{L}^{2,2}}\|g\|_{\mathcal{L}^{2,2}}.$$
Indeed, the integral kernel of $(H_{0}-z)^{-1}$ is
\begin{equation}\label{g0xyz}
G_{0}(x,y,z)=\frac{1}{2i\pi}\left(
\begin{array}{lll}
\frac{e^{-\sqrt{z-\beta}|x-y|}}{\sqrt{z-\beta}}&0\\
0&-\frac{e^{-\sqrt{z+\beta}|x-y|}}{\sqrt{z+\beta}}
\end{array}
\right),
\end{equation}
where $\sqrt{z-\beta}$ and $\sqrt{\beta-z}$ are defined in such a
way that their real parts are nonnegative. Hence for $z\in
K-i[0,\epsilon]$ the operators $(H_{0}-z)^{-1}:\
\mathcal{L}^{2,2}\rightarrow \mathcal{L}^{2,-2}$ are uniformly
bounded in $|Im z|$, and converge to zero as $K\rightarrow
\infty.$ Thus the operator $1+(H_{0}-z)^{-1}W:
\mathcal{L}^{2,-2}\rightarrow \mathcal{L}^{2,-2}$ has a bounded
inverse for sufficiently large $Re z$. Hence the equation
$$(H-z)^{-1}=(1+(H_{0}-z)^{-1}W)^{-1}(H_{0}-z)^{-1}$$
implies that the operators $(H-z)^{-1}:
\mathcal{L}^{2,2}\rightarrow \mathcal{L}^{2,-2}$ are uniformly
bounded for $|Re z|$ large and $Imz\not=0.$ Since $f,g\in
\mathcal{L}^{2,2},$ our claim follows.

Since $f,g\in \mathcal{L}^{2,2},$ $u(z)$ is bounded in the
interval from $K-i\epsilon$ to $K.$ Hence
\begin{equation}\label{uninversebound}
|\int_{K-i\epsilon}^{K}u(z)dz|\leq
c\epsilon\|f\|_{\mathcal{L}^{2,2}}\|g\|_{\mathcal{L}^{2,2}}.
\end{equation}
Equations (~\ref{analyticfunction}), (~\ref{inversebound}),
(~\ref{uninversebound}) imply Equation (~\ref{stronglimit}).
\end{proof}
\begin{lemma}\label{decay}
For any $g\in \mathcal{L}^{2,2},$
\begin{equation}\label{convergence}
\lim_{K_{1}\rightarrow \infty}\int_{\beta}^{K_{1}} e^{-it\lambda}
[(H-\lambda+i0)^{-1}-(H-\lambda-i0)^{-1}]g d\lambda
\end{equation} exists in the norm $\mathcal{L}^{2,-2}.$
\end{lemma}
\begin{proof}
It is sufficient to prove the following statement: for a fixed
$g\in \mathcal{L}^{2,2}$ and large constant $K_{2}$, the integral
\begin{equation}\label{changeform}
\int_{K_{2}}^{K_{1}}|\langle f,
e^{-it\lambda}[(H-\lambda+i0)^{-1}-(H-\lambda-i0)^{-1}]g\rangle|
d\lambda
\end{equation}
converges uniformly as $K_{1}\rightarrow \infty$ in any $f\in
\mathcal{L}^{2,2}$ such that $\|f\|_{\mathcal{L}^{2,2}}=1.$

Since we are only concerned with the convergence of
(~\ref{changeform}), we always assume the constant $K_{2}$ is
sufficiently large so that if $\lambda\geq K_{2}$ then
$$1+(H_{0}-\lambda\pm i0)^{-1}W: \ \mathcal{L}^{2,-2}\rightarrow
\mathcal{L}^{2,-2}$$ is invertible.

First using the second resolvent equation and formula
$$(H-\lambda\pm i0)^{-1}=[1+W(H_{0}-\lambda\pm i0)^{-1}]^{-1}(H_{0}-\lambda\pm
i0)^{-1}$$ we obtain
$$
\begin{array}{lll}
& &(H-\lambda+i0)^{-1}-(H-\lambda-i0)^{-1}\\
&=&(1+(H_{0}-\lambda+i0)^{-1}W)^{-1}\\
& &[(H_{0}-\lambda+i0)^{-1}-(H_{0}-\lambda-i0)^{-1}]\\
& &(1+W(H_{0}-\lambda-i0)^{-1})^{-1}.
\end{array}
$$
Next, by a standard argument we derive
$$
\begin{array}{lll}
& &\frac{1}{2\pi i}[(H_{0}-\lambda+i0)^{-1}-(H_{0}-\lambda-i0)^{-1}]\\
&=&\left(
\begin{array}{lll}
\frac{\cos{k(x-y)}}{k}&0\\
0&0
\end{array}
\right)\\
&=&\frac{1}{2k}[\left(
\begin{array}{lll}
e^{ikx}\\
0
\end{array}
\right)\left(
\begin{array}{lll}
e^{-iky},&0
\end{array}
\right)+\left(
\begin{array}{lll}
e^{-ikx}\\
0
\end{array}
\right)\left(
\begin{array}{lll}
e^{iky},&0
\end{array}
\right)].
\end{array}
$$
Since $f,\ g\in \mathcal{L}^{2,2},$ the following functions are
well defined
$$f^{*}_{\lambda}:=[1+W^{*}(H_{0}-\lambda-i0)^{-1}]^{-1}f,$$
$$g_{\lambda}:=[1+W(H_{0}-\lambda-i0)^{-1}]^{-1}g,$$
and $f_{\lambda}^{*}, g_{\lambda}\in \mathcal{L}^{2,2}.$
Furthermore by Equation (~\ref{g0xyz}) we obtain that for large
$\lambda$
\begin{equation}\label{decayestimate}
\begin{array}{lll}
\|f-f_{\lambda}^{*}\|_{\mathcal{L}^{2,2}}&\leq&
\frac{c}{|k|}\|f\|_{\mathcal{L}^{2,2}},\\
\|g-g_{\lambda}\|_{\mathcal{L}^{2,2}}&\leq&
\frac{c}{|k|}\|g\|_{\mathcal{L}^{2,2}},
\end{array}
\end{equation}
where, recall $k=\sqrt{\lambda-\beta}.$ Therefore
$$
\begin{array}{lll}
& &\int_{K_{1}}^{K_{2}}|\langle f,
(H-\lambda+i0)^{-1}-(H-\lambda-i0)^{-1}]g\rangle| dk^{2}\\
&\leq&\int_{K_{1}}^{K_{2}}|\langle f_{\lambda}^{*},\left(
\begin{array}{lll}
e^{ikx}\\
0
\end{array}
\right)\rangle \langle \left(
\begin{array}{lll}
e^{ikx}\\
0
\end{array}
\right),g_{\lambda}\rangle|\\
& &+|\langle f_{\lambda}^{*},\left(
\begin{array}{lll}
e^{-ikx}\\
0
\end{array}
\right)\rangle \langle \left(
\begin{array}{lll}
e^{-ikx}\\
0
\end{array}
\right),g_{\lambda}\rangle|dk.
\end{array}
$$
We only consider the first term of the right hand side, we claim
that
\begin{equation}\label{zeroestimate}
\begin{array}{lll}
& &\int_{K_{1}}^{K_{2}}|\langle f_{\lambda}^{*},\left(
\begin{array}{lll}
e^{ikx}\\
0
\end{array}
\right)\rangle \langle \left(
\begin{array}{lll}
e^{ikx}\\
0
\end{array}
\right),g_{\lambda}\rangle|dk\\
&\leq&
c\int_{K_{1}}^{K_{2}}a_{K_{1}}^{-2}\frac{\|g\|^{2}_{\mathcal{L}^{2,2}}}{k^{2}}+(a_{K_{1}}^{2}+b_{K_{1}}^{2})\frac{\|f\|^{2}_{\mathcal{L}^{2,2}}}{k^{2}}\\
& &+b_{K_{1}}^{-2}|\langle g, \left(
\begin{array}{lll}
e^{ikx}\\
0
\end{array}
\right)\rangle|^{2}+(a_{K_{1}}^{2}+b_{K_{1}}^{2})|\langle\left(
\begin{array}{lll}
e^{ikx}\\
0
\end{array}
\right),f\rangle|^{2}dk,
\end{array}
\end{equation}
where
$$a_{K}:=(\int_{K}^{\infty}\frac{\|g\|_{\mathcal{L}^{2,2}}^{2}}{k^{2}}dk)^{1/10},$$
$$b_{K}:=(\int_{K}^{\infty}|\langle g,\left(
\begin{array}{lll}
e^{ikx}\\
0
\end{array} \right)\rangle|^{2}d k)^{1/10}.$$ Thus it is easy to see that as
$K_{1}, K_{2}\rightarrow \infty,$ $a_{K_{1}}, b_{K_{1}}\rightarrow
0.$ For the four terms on the right side of Estimate
(~\ref{zeroestimate}),
$$a_{K_{1}}^{-2}\int_{K_{1}}^{K_{2}}\frac{\|g\|_{\mathcal{L}^{2,2}}^{2}}{k^{2}}dk\leq a_{K_{1}}^{8};$$
$$(a_{K_{1}}^{2}+b_{K_{1}}^{2})\int_{K_{1}}^{K_{2}}\frac{\|f\|^{2}_{\mathcal{L}^{2,2}}}{k^{2}}dk\leq (a_{K_{1}}^{2}+b_{K_{1}}^{2})\|f\|^{2}_{\mathcal{L}^{2,2}}\int_{K_{1}}^{K_{2}}\frac{1}{k^{2}}dk;$$
$$b_{K_{1}}^{-2}\int_{K_{1}}^{K_{2}}|\langle g,\left(
\begin{array}{lll}
e^{ikx}\\
0
\end{array}
\right)\rangle|^{2}dk\leq b_{K_{1}}^{8};$$
$$(a_{K_{1}}^{2}+b_{K_{1}}^{2})\int_{K_{1}}^{K_{2}}|\langle\left(
\begin{array}{lll}
e^{ikx}\\
0
\end{array}
\right),f\rangle|^{2}dk\leq
(a_{K_{1}}^{2}+b_{K_{1}}^{2})\|f\|_{2}^{2}.$$ Therefore
$$\int_{K_{1}}^{K_{2}}|\langle f_{\lambda}^{*},\left(
\begin{array}{lll}
e^{ikx}\\
0
\end{array}
\right)\rangle \langle \left(
\begin{array}{lll}
e^{ikx}\\
0
\end{array}
\right),g_{\lambda}\rangle|dk\rightarrow 0$$ for fixed $g\in
\mathcal{L}^{2,2},$ and the decay is independent of $f.$

What left is to prove Estimate (~\ref{zeroestimate}): Indeed,
$$
\begin{array}{lll}
& &|\langle f_{\lambda}^{*},\left(
\begin{array}{lll}
e^{ikx}\\
0
\end{array}
\right)\rangle\langle \left(
\begin{array}{lll}
e^{ikx}\\
0
\end{array}
\right),g_{\lambda}\rangle|\\
&=&|\langle f_{\lambda}^{*}-f+f,\left(
\begin{array}{lll}
e^{ikx}\\
0
\end{array}
\right)\rangle\langle \left(
\begin{array}{lll}
e^{ikx}\\
0
\end{array}
\right),g_{\lambda}-g+g\rangle|\\
&\leq &a_{K_{1}}|\langle f_{\lambda}^{*}-f,\left(
\begin{array}{lll}
e^{ikx}\\
0
\end{array}
\right)\rangle a^{-1}_{K_{1}}\langle \left(
\begin{array}{lll}
e^{ikx}\\
0
\end{array}
\right),g_{\lambda}-g\rangle|\\
& &+b_{K_{1}}|\langle f,\left(
\begin{array}{lll}
e^{ikx}\\
0
\end{array}
\right)\rangle b^{-1}_{K_{1}}\langle \left(
\begin{array}{lll}
e^{ikx}\\
0
\end{array}
\right),g\rangle|\\
& &+a_{K_{1}}|\langle f_{\lambda}^{*}-f,\left(
\begin{array}{lll}
e^{ikx}\\
0
\end{array}
\right)\rangle a^{-1}_{K_{1}}\langle \left(
\begin{array}{lll}
e^{ikx}\\
0
\end{array}
\right),g\rangle|\\
& &+b_{K_{1}}|\langle f,\left(
\begin{array}{lll}
e^{ikx}\\
0
\end{array}
\right)\rangle b^{-1}_{K_{1}}\langle \left(
\begin{array}{lll}
e^{ikx}\\
0
\end{array}
\right),g_{\lambda}-g\rangle|.
\end{array}
$$
By using H\"older Inequality we obtain Estimate
(~\ref{zeroestimate}).
\end{proof}

Lemmas ~\ref{weaklimit} and ~\ref{decay} show the existence of the
limits on the right hand side of Equation (~\ref{kernelpc2}). Now
we prove Equation (~\ref{kernelpc2}):

Instead of the unbounded, non-self adjoint operator $H,$ we
consider the bounded, non-self adjoint operators
$$K_{\epsilon,\kappa} :=  (1-i\epsilon(H+i\kappa))^{-1}$$ for any $\epsilon,\ \kappa\in \mathbb{R},$ where $K_{\epsilon,\kappa}$ is well defined because
$H$ has no complex spectrum.

For the operator $K_{\epsilon,\kappa}$, it is not invertible at
the set (its spectrum)
$$\{(1-i\epsilon(a_{n}+i\kappa))^{-1}|\ a_{n}\ \text{is eigenvalue of}\ H\}\ \cup\ \gamma_{2}\ \cup\ \gamma_{3},$$
where $\gamma_{2}$ and $\gamma_{3}$ are the curves:
$$\gamma_{2}:=\{(1-i\epsilon(\mu+i\kappa))^{-1}| \ \mu\geq \beta\},\ \ \gamma_{3}:=\{(1-i\epsilon(\mu+i\kappa))^{-1}|\ \lambda\leq -\beta.\}.$$ Also it is easy to see that
$P_{d}^{K_{\epsilon,\kappa}}=P_{d}^{H},$ where, recall,
$P_{d}^{K_{\epsilon,\kappa}}$ from Definition
~\ref{continuousspace}. By Definition ~\ref{continuousspace},
$K_{\epsilon,\kappa}$ has the same essential spectrum subspace to
$H$. Denote the projection onto essential spectrum subspace of
$K_{\epsilon,\kappa}$ by $P_{ess}^{K_{\epsilon,\kappa}},$ then
$$P_{ess}^{K_{\epsilon,\kappa}}=P_{ess}.$$

Let $f$ be an entire function on $\mathbb{C}$. Then we can define
the operator $f(K_{\epsilon,\kappa})$ by the Taylor series (see
e.g. ~\cite{JMS}), moreover one has
\begin{equation}\label{initial}
\begin{array}{lll}
f(K_{\epsilon,\kappa})P_{ess}^{K_{\epsilon,\kappa}}
=\frac{1}{2i\pi}\oint_{\gamma_{1}}f(\lambda)(K_{\epsilon,\kappa}-\lambda)^{-1}d\lambda,
\end{array}
\end{equation}
where $\gamma_{1}$ is a contour around the curves
$\gamma_{2},\gamma_{3}$, but leaving
$\{(1-i\epsilon(a_{n}+i\kappa))^{-1}|\ a_{n}\ \text{is eigenvalue
of}\ H\}$ outside.

In order to get a similar formula as in Equation
(~\ref{kernelpc2}) we want to transform the right hand side of
Equation (~\ref{initial}): First we notice that
$$(K_{\epsilon,\kappa}-\lambda)^{-1}=-\frac{1}{\lambda}(H+i\kappa-\frac{1}{i\epsilon}+\frac{1}{i\lambda\epsilon})^{-1}(H+i\kappa-\frac{1}{i\epsilon}).$$
There exists an $\epsilon_{0}>0$ such that if $|\lambda|\leq
\epsilon_{0}$ and $\lambda\in \gamma_{2}\cup \gamma_{3},$ then
\begin{equation}\label{l2l2}
(H+i\kappa-\frac{1}{i\epsilon}+\frac{1}{i(\lambda\pm
0)\epsilon})^{-1}: \mathcal{L}^{2,2}\rightarrow \mathcal{L}^{2,-2}
\end{equation}
are well defined and
$$\|\frac{1}{\lambda}(H+i\kappa-\frac{1}{i\epsilon}+\frac{1}{i(\lambda\pm
0)\epsilon})^{-1}\|_{\mathcal{L}^{2,2}\rightarrow
\mathcal{L}^{2,-2}}\leq \frac{c}{\sqrt{|\lambda|}}$$ for some
constant $c$. Therefore the integral $$\int_{
\begin{subarray}{lll}
\lambda\in \gamma_{2}\cup \gamma_{3},\\
|\lambda|\leq \epsilon_{0}
\end{subarray}}\frac{f(\lambda)}{\lambda}(H+i\kappa-\frac{1}{i\epsilon}+\frac{1}{i(\lambda\pm
0)\epsilon})^{-1}d\lambda$$ exists. Based on the arguments above
we can deform the contour $\gamma_{1}$ as
\begin{equation}\label{firsttransform}
\begin{array}{lll}
& &f(K_{\epsilon,\kappa})P_{ess}^{K_{\epsilon,\kappa}}=\frac{1}{2i\pi}\int_{\gamma_{4}+\gamma_{5}}f(\lambda)(K_{\epsilon,\kappa}-\lambda)^{-1}d\lambda\\
 & &\ \ +\frac{1}{2i\pi}\int_{
\begin{subarray}{lll}
\lambda\in \gamma_{2}\cup \gamma_{3},\\ |\lambda|\leq \epsilon_{0}
\end{subarray}}f(\lambda)[(K_{\epsilon,\kappa}-\lambda+i0)^{-1}-(K_{\epsilon,\kappa}-\lambda-i0)^{-1}]d\lambda
\end{array}
\end{equation}
where $\gamma_{4},\gamma_{5}$ are the contours around the spectral
points $\gamma_{2}\cap \{\lambda||\lambda|> \epsilon_{0}\},$
$\gamma_{3}\cap \{\lambda||\lambda|> \epsilon_{0}\}$ respectively,
and all other spectral points of $K_{\epsilon,\kappa}$ are kept
outside. Since we proved that when $|\lambda|\leq \epsilon_{0}$
the operators $(K_{\epsilon,\kappa}-\lambda\pm
i0)^{-1}:(1-\frac{d^{2}}{dx^{2}})^{-1}\mathcal{L}^{2,2}\rightarrow
\mathcal{L}^{2,-2}$ which justifies the following calculation
$$
\begin{array}{lll}
&
&(K_{\epsilon,\kappa}-\lambda-i0)^{-1}-(K_{\epsilon,\kappa}-\lambda+i0)^{-1}\\
&=&-\frac{1}{i\epsilon\lambda^{2}}[(H+i\kappa-\frac{1}{i\epsilon}+\frac{1}{i\epsilon(\lambda-i0)})^{-1}-(H+i\kappa-\frac{1}{i\epsilon}+\frac{1}{i\epsilon(\lambda+i0)})^{-1}].
\end{array}
$$
By the change of variable
$z=\frac{1}{i\epsilon\lambda}+i\kappa+\frac{i}{\epsilon}$, we have
\begin{equation}\label{chaning}
\begin{array}{lll}
& &\frac{1}{2i\pi}\int_{ \begin{subarray} {lll}
\lambda\in\gamma_{2}\cup \gamma_{3}\\
|\lambda|\leq \epsilon_{0}
\end{subarray}}f(\lambda)[(K_{\epsilon,\kappa}-\lambda+i0)^{-1}-(K_{\epsilon,\kappa}-\lambda-i0)^{-1}]d\lambda\\
&=&\frac{1}{2i\pi}(\int^{\infty}_{\kappa_{1}}+\int_{-\infty}^{\kappa_{2}})f((1-i\epsilon(z+i\kappa))^{-1})((H-z-i0)^{-1}-(H-z+i0)^{-1})dz.
\end{array}
\end{equation}
where $\kappa_{1},\kappa_{2}$ are the points such that
$|\frac{1}{i\epsilon
\kappa_{n}}+i\kappa+\frac{i}{\epsilon}|=\epsilon_{0}\ (n=1,2).$

On the other hand
\begin{equation}\label{secondterm}
\begin{array}{lll}
& &\frac{1}{2i\pi}\int_{\gamma_{4}+\gamma_{5}}f(\lambda)(K_{\epsilon,\kappa}-\lambda)^{-1}d\lambda\\
&=&-\frac{1}{2i\pi}\int_{\gamma_{4}+\gamma_{5}}\frac{f(\lambda)}{\lambda}[H+i\kappa-\frac{1}{i\epsilon}+\frac{1}{i\lambda\epsilon}]^{-1}(H+i\kappa-\frac{1}{i\epsilon})d\lambda\\
&=&\frac{1}{2i\pi}\int_{\gamma_{4}+\gamma_{5}}\frac{f(\lambda)}{i\lambda^{2}\epsilon}[H+i\kappa-\frac{1}{i\epsilon}+\frac{1}{i\lambda\epsilon}]^{-1}d\lambda.
\end{array}
\end{equation}
Let $z=i\kappa-\frac{1}{i\epsilon}+\frac{1}{i\lambda\epsilon}$,
the equation equals to
\begin{equation}\label{secondtermremind}
\begin{array}{lll}
&
&\frac{1}{2i\pi}\int_{\gamma_{6}+\gamma_{7}}f(\frac{1}{i\epsilon(z-i\kappa+\frac{1}{i\epsilon})})(H-z)^{-1}dz\\
&=&\frac{1}{2i\pi}(\int_{\beta}^{\kappa_{1}}+\int^{-\beta}_{\kappa_{2}})f((1-i\epsilon(z+i\kappa))^{-1})[(H-z-i0)^{-1}-(H-z+i0)^{-1}]dz.
\end{array}
\end{equation}
where $\gamma_{6}$ and $\gamma_{7}$ are corresponding to the
curves $\gamma_{4}$ and $\gamma_{5}$, the constants $\kappa_{1},\
\kappa_{2}$ are the same as that in Equation (~\ref{chaning}).

By Equations (~\ref{firsttransform}) (~\ref{chaning})
(~\ref{secondterm}) and (~\ref{secondtermremind}) we get that for
any entire function $f$,
\begin{equation}\label{finalform}
f(K_{\epsilon,\kappa})P_{ess}=\frac{1}{2i\pi}(\int_{\beta}^{\infty}+\int^{-\beta}_{-\infty})f((1-i\epsilon(z+i\kappa))^{-1})[(H-z-i0)^{-1}-(H-z+i0)^{-1}]dz.
\end{equation}
By the results in ~\cite{Gold, RSII} we have that for some
$\kappa_{0}\in \mathbb{R}$
\begin{equation}\label{generatorlimit}
\begin{array}{lll}
e^{-itH}&=&s-\displaystyle\lim_{\epsilon\rightarrow
0^{+}}e^{-itH(1-i\epsilon(i\kappa_{0}+H))^{-1}}\\
&=&s-\displaystyle\lim_{\epsilon\rightarrow
0^{+}}e^{\frac{t}{\epsilon}+(t\kappa_{0}+\frac{t}{\epsilon})(1-i\epsilon(H+i\kappa_{0}))^{-1}}.
\end{array}
\end{equation}
Since the function
$f_{\epsilon}(z):=e^{\frac{t}{\epsilon}+(-it\kappa_{0}+\frac{t}{\epsilon})z}$
is analytic, we could use Equation (~\ref{chaning}):
$$
\begin{array}{lll}
& &e^{-itH(1-i\epsilon(i\kappa_{0}+H))^{-1}}P_{ess}\\
&=&\frac{1}{2i\pi}(\int_{\beta}^{+\infty}+\int^{-\beta}_{-\infty})e^{-it\frac{z}{1-i\epsilon(z+i\kappa_{0})}}((H-z-i0)^{-1}-(H-z+i0)^{-1})dz.
\end{array}
$$
Now let $\epsilon\rightarrow 0^{+},$ the left hand side is
$e^{-itH}$ by Equation (~\ref{generatorlimit}), and the right hand
side is
$$
\begin{array}{lll}
\frac{1}{2i\pi}(\int_{\beta}^{+\infty}+\int^{-\beta}_{-\infty})e^{-itz}((H-z-i0)^{-1}-(H-z+i0)^{-1})dz
\end{array}
$$ by Equation (~\ref{finalform}) and some arguments similar to the proof of Lemmas ~\ref{weaklimit} and
~\ref{decay}. The theorem follows.

Let $L^{\gamma}:=e^{\frac{\gamma}{4}|x|}\mathcal{L}^{2}$ with the
norm
\begin{equation}\label{weightedl2space}
\|g\|_{L^{\gamma}}:=\|e^{\frac{\gamma}{4}|x|}g\|_{\mathcal{L}^{2}}.
\end{equation}
It will be proved in the discussion after Lemma ~\ref{exlambda}
Appendix ~\ref{app}, that for any $\lambda>\beta,$ the operator
$1+(H_{0}-\lambda+i0)^{-1}W:\ L^{-\alpha}\rightarrow L^{-\alpha}$
has a bounded inverse, where, recall the constant $\alpha$ from
Equation (~\ref{alpha}). Define
\begin{equation}\label{definexk}
e(\cdot,k) := [1+(H_{0}-\lambda+i0)^{-1}W]^{-1} \left(
\begin{array}{lll}
e^{ikx}\\
0
\end{array}
\right),
\end{equation}
and $$e^{\%}(x,k):=\sigma_{3}e(x,k)$$ where
$$\sigma_{3}=\left(
\begin{array}{lll}
1&0\\
0&-1
\end{array} \right).$$
There are two terms on the right hand side of Equation
(~\ref{kernelpc2}), we denote the first term by
$e^{-itH}P_{ess}^{+}.$
\begin{lemma}\label{lemmaexpan}
\begin{equation}\label{expanpc}
e^{-iHt}P^{+}_{ess}f =\frac{1}{2\pi}\int_{k\geq 0}
e^{-ik^{2}t}[\langle e^{\%}(x,k), f\rangle e(\cdot,k) +\langle
e^{\%}(-x,k) f\rangle e(-\cdot,k)]dk.
\end{equation}
\end{lemma}
\begin{proof}
Let $f,\ g\in L^{\alpha},$ and define $$f_{\lambda}^{*} :=
(1+W^{*}(H_{0}-\lambda-i0)^{-1})^{-1}f,$$
$$g_{\lambda} := (1+W(H_{0}-\lambda-i0)^{-1})^{-1}g.$$
Thus $f^{*}_{\lambda},\ g_{\lambda}\in L^{\alpha}.$ Therefore all
the following computations make sense:
$$
\begin{array}{lll}
& &\langle f,-i((H-\lambda-i0)^{-1}-(H-\lambda+i0)^{-1})g\rangle\\
&=&\langle
f_{1},-i[(H_{0}-\lambda-i0)^{-1}-(H_{0}-\lambda+i0)^{-1}]g_{1}\rangle\\
&=&\frac{1}{2k}\langle f_{1}, \left(
\begin{array}{ccc}
e^{ikx}\\
0
\end{array}
\right)\rangle \langle \left(
\begin{array}{ccc}
e^{ikx}\\
0
\end{array}
\right),g_{1}\rangle+\frac{1}{2k}\langle f_{1}, \left(
\begin{array}{ccc}
e^{-ikx}\\
0
\end{array}
\right)\rangle \langle \left(
\begin{array}{ccc}
e^{-ikx}\\
0
\end{array}
\right),g_{1}\rangle\\
&=&\frac{1}{2k}\langle f,(1+(H_{0}-\lambda+i0)^{-1}W)^{-1}\left(
\begin{array}{lll}
e^{ikx}\\
0
\end{array}
\right)\rangle\\ &
&\langle(1+(H_{0}-\lambda+i0)^{-1}W^{*})^{-1}\left(
\begin{array}{lll}
e^{ikx}\\
0
\end{array}
\right),g\rangle\\
& &+\frac{1}{2k}\langle f,(1+(H_{0}-\lambda+i0)^{-1}W)^{-1}\left(
\begin{array}{lll}
e^{-ikx}\\
0
\end{array}
\right)\rangle\\
& &\langle(1+(H_{0}-\lambda+i0)^{-1}W^{*})^{-1}\left(
\begin{array}{lll}
e^{-ikx}\\
0
\end{array}
\right),g\rangle\\
&=&\frac{1}{2k}(\langle f, e(\cdot,k)\rangle\langle
e^{\%}(\cdot,k),g\rangle+\langle f, e(-\cdot,k)\rangle\langle
e^{\%}(-\cdot,k),g\rangle).
\end{array}
$$
Then Equation (~\ref{expanpc}) follows.
\end{proof}
Taking $t=0$ in Equation (~\ref{kernelpc2}) we obtain
$$P_{ess}=P_{ess}^{+}+P_{ess}^{-},$$ where the operator $P_{ess}^{+}$ is given by
\begin{equation}\label{positivebrance}
\begin{array}{lll}
P^{+}_{ess}f&=&\displaystyle\lim_{K\rightarrow
\infty}\displaystyle\lim_{\epsilon\rightarrow 0^{+}}
\frac{1}{2i\pi}\int_{\beta-\epsilon_{0}}^{K}[(H-\lambda-i\epsilon)^{-1}-(H-\lambda+i\epsilon)^{-1}]d\lambda,\\
&=&\frac{1}{2\pi}\int_{k\geq 0} \langle e^{\%}(x,k), f\rangle
e(\cdot,k) +\langle e^{\%}(-x,k) f\rangle e(-\cdot,k)dk
\end{array}
\end{equation} for any sufficiently small $\epsilon_{0}>0,$ and similarly for
$P_{ess}^{-}.$ In fact, the operators $P_{ess}^{\pm}$ are spectral
projections corresponding to the branches $[\beta,\infty)$ and
$(-\infty,-\beta]$ of the essential spectrum of the operator $H.$
\subsection{Proof of Theorem ~\ref{maintheorem1}}\label{proofmainth}
In this subsection we prove key Theorem ~\ref{maintheorem1}.
Recall the definition of functions $e(x,k)$ in Equation
(~\ref{definexk}). To this end we use the following technical
results proven in Appendix ~\ref{sectionexk}.
\begin{theorem}\label{estimatexk}
Assume that there are no embedded eigenvalues in the essential
spectrum and there are no resonances at the tips, $\pm\beta$, of
the essential spectrum. Then Equation (~\ref{definexk}) defines
smooth functions $e(x,k)$ which are generalized eigenfunctions of
the operator $H: He(\cdot,k)=\lambda e(\cdot,k)$ with
$k=\sqrt{\lambda-\beta}.$ If $k\geq 0,$ we have the following
estimates:
\begin{equation}\label{supexk0}
\sup_{k\geq 0}|\frac{d^{n}}{dk^{n}}e(x,k)|\leq c\rho_{-n};
\end{equation}
\begin{equation}\label{supexk}
\sup_{k\geq 0}|\frac{d^{n}}{dk^{n}}(e(x,k)/k)|\leq c\rho_{-n-1};
\end{equation}
\begin{equation}\label{supexk2}
[\int_{k\geq 0}|\frac{d^{n}}{dk^{n}}(e(x,k)/k)|^{2}dk]^{1/2}\leq
c\rho_{-n-1};
\end{equation}
\begin{equation}\label{ex0}
e(\cdot,0)=0;
\end{equation}
\begin{equation}\label{el2}
\|\langle e^{\%}(\cdot,k),h\rangle\|_{\mathcal{H}^{2}}\leq
c\|\rho_{-2}h\|_{2},
\end{equation}
where $n=0,1,2,$ all $c$'s are constants independent of $x,$ and
recall $\rho_{\nu}(x)=(1+x^{2})^{-\nu}.$
\end{theorem}
Before starting proving Theorem ~\ref{maintheorem1}, we state the
following standard estimate:
$$\|e^{it\frac{d^{2}}{dx^{2}}}f\|_{\mathcal{L}^{\infty}}\leq
ct^{-1/2}\|f\|_{\mathcal{L}^{1}}$$ valid for some constant $c$ and
for any $f\in \mathcal{L}^{1}.$ This estimate implies that if a
function $g$ satisfies $\int e^{ikx}g(k)dk\in \mathcal{L}^{1}$ and
is even, then
\begin{equation}\label{hfourier}
|\int e^{-ik^{2}t}g(k)dk|\leq ct^{-1/2}\|\int \cos(kx)
g(k)dk\|_{\mathcal{L}^{1}}.
\end{equation}
\textit{Proof of Estimate (~\ref{estimate1})}. We will only
consider the first term on the right hand side of
$e^{-iHt}P_{ess}$ in Equation (~\ref{kernelpc2}) and the first
term of the right hand side of Equation (~\ref{expanpc}). To
simplify the notation we denote
$$h^{\#}(k):=\langle e^{\%}(\cdot,k), h\rangle.$$ Also for a function $g(x,k)$ of two variables such that if for each fixed $x$ (or $k$), $g(x,\cdot)\in
B$ (or $g(\cdot,k)\in B$), then we define
$\|g\|_{B^{k}}=\|g(x,\cdot)\|_{B}$ (or
$\|g\|_{B^{x}}=\|g(\cdot,k)\|_{B}$).

Since $e(x,0)=0$ by Estimate (~\ref{ex0}), we can integrate by
part and use Estimate (~\ref{hfourier}) to obtain:
\begin{equation}\label{forthird}
\begin{array}{lll}
& &|\frac{1}{2\pi}\int
2ik\frac{e(x,k)}{2ik}e^{-ik^{2}t}h^{\#}(k)dk|\\
&=&|\frac{1}{2\pi}t^{-1}\int
\frac{d}{dk}(\frac{e(x,k)h^{\#}(k)}{2ik})e^{-ik^{2}t}dk|\\
&\leq&
ct^{-3/2}\|\frac{d}{dk}\widehat{(\frac{e(x,k)h^{\#}(k)}{2ik})}\|_{\mathcal{L}^{1}(dk)},
\end{array}
\end{equation}
where $\hat{g}(x) :=  \int_{0}^{\infty}\cos(kx)g(k)dk.$ Since
$\|\hat{g}_{2}\|_{1}\leq \|g_{2}\|_{\mathcal{H}^{1}},$ we have
$$
\begin{array}{lll}
& & \|\frac{d}{dk}\frac{\widehat{(e(x,k)h^{\#}(k))}}{2ik}\|_{(\mathcal{L}^{1})^{k}}\\
&\leq&c\displaystyle\sum_{n=0}^{2}\|\frac{d^{n}}{dk^{n}}\frac{e(x,k)}{2ik}h^{\#}(k)\|_{\mathcal{L}^{2}(dk)}\\
&\leq&c\|h^{\#}\|_{\mathcal{H}^{2}}\displaystyle\sum_{n=0}^{2}\|\frac{d^{n}}{dk^{n}}(e/k)\|_{\mathcal{L}^{\infty}(k)}.
\end{array}
$$
By Estimate (~\ref{supexk}),
$$\sum_{n=0}^{2}\|\frac{d^{n}}{dk^{n}}(e/k)\|_{\mathcal{L}^{\infty}(dk)}\leq
c(1+|x|)^{3},$$ and by Estimate (~\ref{el2})
$$\|h^{\#}\|_{\mathcal{H}^{2}}\leq c\|\rho_{-2}h\|_{2}.$$
The last four estimates imply that $$\int\langle
e^{\%}(\cdot,k),h\rangle e(x,k)e^{-ik^{2}t}dk|\leq c
t^{-3/2}(1+|x|)^{3}\|\rho_{-2}h\|_{2}.$$ Similarly we estimate the
other terms in Equations (~\ref{expanpc}), (~\ref{kernelpc2}) to
get
$$|e^{-itH}P_{ess}h|\leq ct^{-3/2}(1+|x|)^{3}\|\rho_{-2}h\|_{2}.$$
Therefore if $\mu>3.5,$ we have
$$\|\rho_{\mu}e^{-iHt}P_{ess}h\|_{2}\leq c
t^{-3/2}\|\rho_{-2}h\|_{2}.$$ By Lemma ~\ref{convergence} we have
$\|\rho_{\mu}e^{-iHt}P_{ess}h\|_{2}\leq c\|\rho_{-2}h\|_{2}.$ The
last two equations give
$$\|\rho_{\mu}e^{-iHt}P_{ess}h\|_{2}\leq c
(1+t)^{-3/2}\|\rho_{-2}h\|_{2}$$ for some $c$. This estimate is
equivalent to Estimate (~\ref{estimate1}).\\
\textit{Proof of
Estimate (~\ref{estimate2})}: We start from the last line of
Equation (~\ref{forthird}),
$$
\begin{array}{lll}
& &
\|\frac{d}{dk}\frac{\widehat{(e(x,k)h^{\#}(k))}}{2ik}\|_{\mathcal{L}^{1}(dk)}\\
&\leq&c\sum_{n=0}^{2}\|\frac{d^{n}}{dk^{n}}\frac{e(x,k)}{2ik}h^{\#}(k)\|_{\mathcal{L}^{2}(dk)}\\
&\leq&c\sum_{n=0}^{2}\|\frac{d^{n}}{dk^{n}}h^{\#}\|_{\mathcal{L}^{\infty}}\sum_{n=0}^{2}\|\frac{d^{n}}{dk^{n}}(e/k)\|_{2}.
\end{array}
$$
By Estimate (~\ref{el2}),
$$\sum_{n=0}^{2}\|\frac{d^{n}}{dk^{n}}h^{\#}\|_{\mathcal{L}^{\infty}}\leq c\|\rho_{-2}h\|_{1},$$ and by
Estimate (~\ref{supexk2}),
$$\sum_{n=0}^{2}\|\frac{d^{n}}{dk^{n}}(e/k)\|_{2}\leq c(1+|x|)^{3}.$$
Therefore we have as before
$$\|\rho_{\nu}e^{-iHt}P_{ess}h\|_{2}\leq
ct^{-3/2}\|\rho_{-2}h\|_{1}.$$ Since
$\|\rho_{\nu}e^{-iHt}P_{ess}h\|_{2}\leq
c\|\rho_{-2}h\|_{2},$ Estimate (~\ref{estimate2}) is proved.\\
\textit{Proof of Estimate (~\ref{estimate3})}. By the Duhamel
principle,
\begin{equation}\label{sery1}
\begin{array}{lll}
&
&\|e^{-itH}P_{ess}\rho_{2}h\|_{\mathcal{L}^{\infty}}\\
&\leq
&\|e^{-itH_{0}}P_{ess}\rho_{2}h\|_{\mathcal{L}^{\infty}}+\|\int_{0}^{t}e^{-i(t-s)H_{0}}P_{ess}We^{-isH}P_{ess}\rho_{2}h
ds\|_{\mathcal{L}^{\infty}}.
\end{array}
\end{equation} Based on the estimates
$$\|h\|_{\mathcal{L}^{\infty}}\leq c\|h\|_{\mathcal{H}^{1}}\ \text{and} \|e^{it\frac{d^{2}}{dx^{2}}}\|_{\mathcal{L}^{1}\rightarrow \mathcal{L}^{\infty}}\leq
ct^{-1/2},$$ we have
\begin{equation}\label{sery2}
\|e^{-itH_{0}}P_{ess}\rho_{2}h\|_{\mathcal{L}^{\infty}}\leq
c(1+t)^{-1/2}\|h\|_{\mathcal{H}^{1}}
\end{equation} and
$$
\begin{array}{lll}
& &\|\int_{0}^{t}e^{-i(t-s)H_{0}}P_{ess}We^{-isH}P_{ess}\rho_{2}h
ds\|_{\mathcal{L}^{\infty}}\\
&\leq&
c\int_{0}^{t}|t-s|^{-1/2}\|We^{-isH}P_{ess}\rho_{2}h\|_{\mathcal{L}^{1}}ds.
\end{array}
$$
Furthermore, by Estimate (~\ref{alpha}) we have $|W|\leq
c\rho_{6}.$ Hence
$$\|We^{-isH}P_{ess}\rho_{2}h\|_{\mathcal{L}^{1}}\leq
c\|\rho_{4}e^{-isH}P_{ess}\rho_{2}h\|_{2}.$$ By Estimate
(~\ref{estimate1}) we have
$$\|\rho_{4}e^{-isH}P_{ess}\rho_{2}h\|_{2}\leq
c(1+s)^{3/2}\|h\|_{2}.$$ Thus
\begin{equation}\label{sery3}
\begin{array}{lll}
&
&\int_{0}^{t}\|e^{-i(t-s)H_{0}}P_{ess}\|_{\mathcal{L}^{1}\rightarrow
\mathcal{L}^{\infty}}\|We^{-isH}P_{ess}\rho_{2}h\|_{\mathcal{L}^{1}}ds\\
&\leq &
c\int_{0}^{t}\frac{1}{|t-s|^{1/2}}\frac{1}{(1+|s|)^{-3/2}}ds\|h\|_{2}\\
&\leq &c(1+t)^{-1/2}\|h\|_{2}.
\end{array}
\end{equation}
Estimates (~\ref{sery1}), (~\ref{sery2}) and (~\ref{sery3}) imply
the inequality
$$\|e^{-itH}P_{ess}\rho_{2} h\|_{\mathcal{L}^{\infty}}\leq
c(1+t)^{-1/2}\|h\|_{\mathcal{H}^{1}}$$ which is equivalent to
Estimate (~\ref{estimate3}).\\
\textit{Proof of Estimate (~\ref{estimate4})}. By the Duhamel
Principle
$$\|e^{-itH}P_{ess}h\|_{\mathcal{L}^{\infty}}\leq \|e^{-itH_{0}}P_{ess}h\|_{\mathcal{L}^{\infty}}+\|\int_{0}^{t}e^{-i(t-s)H_{0}}P_{ess}We^{-isH}P_{ess}h
ds\|_{\mathcal{L}^{\infty}}.$$ For the first term we have
$$\|e^{-itH_{0}}P_{ess}h\|_{\mathcal{L}^{\infty}}\leq
ct^{-1/2}\|h\|_{1};$$ and for the second term we have
$$
\begin{array}{lll}
& &\|\int_{0}^{t}e^{-i(t-s)H_{0}}P_{ess}We^{-isH}P_{ess}h
ds\|_{\mathcal{L}^{\infty}}\\
&\leq&
c\int_{0}^{t}\frac{1}{|t-s|^{1/2}}\|\rho_{-2}We^{-isH}P_{ess}h\|_{2}.
\end{array}
$$
By Estimate(~\ref{estimate2}),
$$\|\rho_{-2}We^{-isH}P_{ess}h\|_{2}\leq c(1+s)^{-3/2}(\|\rho_{-2}h\|_{1}+\|h\|_{2}).$$
Therefore
$$
\begin{array}{lll}
& &\|e^{-itH}P_{ess}h\|_{\mathcal{L}^{\infty}}\\
&\leq &c(t^{-1/2}+\int^{t}_{0}\frac{1}{|t-s|^{1/2}}(1+s)^{-3/2}ds)(\|h\|_{2}+\|\rho_{-2}h\|_{1})\\
&\leq &ct^{-1/2}(\|h\|_{2}+\|\rho_{-2}h\|_{1}),
\end{array}
$$ which gives the last Estimate (~\ref{estimate4}).
\begin{flushright}
$\square$
\end{flushright}
\appendix
\section{Proof of Theorem ~\ref{estimatexk}}\label{app}
In this appendix we study the functions
$$e(x,k):=[1+(H_{0}-\lambda+i0)^{-1}W]^{-1}\left(
\begin{array}{lll}
e^{ikx}\\
0
\end{array}
\right)$$ with $k=\sqrt{\lambda-\beta}$ introduced in Subsection
~\ref{proofmainth}. A simple manipulation shows that $H
e(\cdot,k)=\lambda e(\cdot,k),$ i.e. $e(\cdot,k)$ are generalized
eigenfunctions of the operator $H$ corresponding to spectral
points $\lambda=k^{2}+\beta.$ We begin with some auxiliary results
on solutions to the equation $H\xi=\lambda\xi.$
\subsection{Generalized Eigenfunctions of
$H$}\label{sectionexk} In this subsection we study solutions of
the equation $H\xi=\lambda\xi,$ considered as a differential
equation, with $\lambda$ in an appropriate domain of the complex
plane $\mathbb{C}$.

From now on we will only consider the positive branch of the
essential spectrum subspace. So we always assume $Re\lambda\geq
\beta>0$ and $Im\lambda$ is sufficiently small. The negative
branch is treated exactly the same. If $Re\lambda\geq \beta$ we
define two functions $\sqrt{\lambda-\beta}$ and
$\sqrt{\lambda+\beta}$ such that they are analytic and if
$\lambda-\beta>0$ (or $\lambda+\beta>0$) then
$\sqrt{\lambda-\beta}>0$ (or $\sqrt{\lambda+\beta}>0$).

Define the domain
\begin{equation}\label{omega1}
\Omega:=\{\lambda|Re\lambda\geq \beta, |Im
\sqrt{\lambda-\beta}|+|Im \sqrt{\lambda+\beta}|\leq
\frac{\alpha}{4}\},
\end{equation} where, recall $\alpha$ from (~\ref{alpha}). We always denote
\begin{equation}\label{notationkmu}
k :=  \sqrt{\lambda-\beta}\ \text{and}\ \mu :=
\sqrt{\lambda+\beta}.
\end{equation}
Hence the function $\mu=\sqrt{2\beta-k^{2}}$ is analytic in
$k=\sqrt{\lambda-\beta},\ \lambda\in \Omega.$

Below we will use the space
$\mathcal{L}^{\infty,\beta}:=e^{-\beta|x|}\mathcal{L}^{2}$ with
the norm
$$\|W\|_{\mathcal{L}^{\infty,\beta}}=\|e^{-\beta|x|}W\|_{\mathcal{L}^{\infty}}.$$

We formulate the main result of this appendix. Recall the
definition of the operator $H:=H_{0}+W$, where
$$H_{0}:=\left(
\begin{array}{lll}
-\frac{d^{2}}{dx^{2}}+\beta&0\\
0&\frac{d^{2}}{dx^{2}}-\beta
\end{array}
\right),\ \ W:=1/2\left(
\begin{array}{lll}
V_{3}&-iV_{4}\\
-iV_{4}&-V_{3}
\end{array}\right),$$ with the constant $\beta>0,$ the functions $V_{3},V_{4}$ even, smooth, real and decaying exponentially fast at $\infty:$
\begin{equation}\label{alpha}
|V_{4}(x)|,\ |V_{3}(x)|\leq ce^{-\alpha|x|}
\end{equation} for some constants $c,\ \alpha>0.$ Without loss of
generality we assume $\alpha<\beta.$ The following is the main
theorem:
\begin{theorem}\label{mainsolution}
If $\lambda\in \Omega,$ then the equation
\begin{equation}\label{solutionspace}
(H-\lambda)\phi=0
\end{equation} has $C^{3}$ solutions
$\phi_{1}(\cdot,\mu,W),$ $\psi_{1}(\cdot,k,W),$
$\psi_{2}(\cdot,k,W)$ and $\xi_{1}(\cdot,\mu,W)$ which are
analytic in $k$ and satisfy the following estimates: there exist
constants $R_{1}, c,\epsilon_{0}>0$ such that for $\forall \
x>R_{1},\ \lambda\geq\beta$
\begin{equation}\label{phi1R}
|\frac{d^{n}}{dk^{n}}(\phi_{1}(x,\mu,W)e^{\mu x}-\left(
\begin{array}{lll}
0\\
1
\end{array}
\right))|\leq c e^{-\epsilon_{0}x},
\end{equation}
\begin{equation}\label{psi1R}
|\frac{d^{n}}{dk^{n}}(\psi_{1}(x,k,W)e^{-ikx}-\left(
\begin{array}{lll}
1\\
0
\end{array}
\right))|\leq ce^{-\epsilon_{0}x},
\end{equation}
\begin{equation}\label{psi2R}
 |\frac{d^{n}}{dk^{n}}(\psi_{2}(x,k,W)e^{ikx}-\left(
\begin{array}{lll}
1\\
0
\end{array}
\right))|\leq ce^{-\epsilon_{0}x}
\end{equation} where $n=0,1,2.$

Also
\begin{equation}\label{xi1infty}
\lim_{x\rightarrow +\infty}\xi_{1}(x,\mu,W)e^{-\mu x}=\left(
\begin{array}{lll}
0\\
1
\end{array}
\right).
\end{equation}
For any constant $R_{2}$, there exists a constant $c_{2}>0$, such
that if $\beta\leq \lambda\leq \beta+1$ and $x\geq R_{2},$ then
\begin{equation}\label{phik}
|\frac{d^{n}}{dk^{n}}\phi_{1}(x,\mu,W)|\leq c_{2},
\end{equation} and
\begin{equation}\label{psi1k}
|\frac{d^{n}}{dk^{n}}\psi_{1}(x,k,W)|\leq c_{2}(1+|x|)^{n},
\end{equation}
\begin{equation}\label{psi2k}
|\frac{d^{n}}{dk^{n}}\psi_{2}(x,k,W)|\leq c_{2}(1+|x|)^{n},
\end{equation} where $n=0,1,2.$

Moreover the following maps are continuous
\begin{equation}\label{continu2}
\mathcal{L}^{\infty,\alpha} \ni W\rightarrow
\tilde{\phi}_{1}(W)\in \mathcal{C}^{2}\ \text{and}\
\mathcal{L}^{\infty,\alpha}\ni W\rightarrow
\tilde{\psi}_{1}(W)\in\mathcal{C}^{2}
\end{equation}
are continuous: where, recall the constant $\alpha$ from Equation
(~\ref{alpha}),
$$\tilde{\phi}_{1}(W):=(\frac{d}{dx}\phi_{1}(x,\sqrt{2\beta},W)|_{x=0},\phi_{1}(x,\sqrt{2\beta},W)|_{x=0}),$$
$$\tilde{\psi}_{1}(W):=(\frac{d}{dx}\psi_{1}(x,0,W)|_{x=0},\psi_{1}(x,0,W)|_{x=0}).$$
\end{theorem}
A proof of this theorem follows from Propositions ~\ref{phi1},
~\ref{solutionpsi1} and ~\ref{solutionxi1}.

When the potential $W$ is fixed, for brevity we use for the
solutions above the notations
$\phi_{1}(x,\mu),\psi_{1}(x,k),\psi_{2}(x,k), \xi_{1}(x,\mu)$
respectively if there is no confusion.

Since we need to study the analyticity of all these functions, and
derive them by convergent sequences, we use frequently the
following lemma:
\begin{lemma}\label{analy}
If $\{f_{n}\}_{n=0}^{\infty}$ is a sequence of analytic functions,
and
$$\displaystyle\sum_{n}\|f_{n}\|_{\mathcal{L}^{\infty}}<\infty,$$
then $\displaystyle\sum_{n}f_{n}$ is an analytic function.
\end{lemma}
First let's look at $\phi_{1}$:
\begin{proposition}\label{phi1}
Recall $\mu=\sqrt{\lambda+\beta}$. There is a solution
$\phi_{1}(\cdot,\mu)$ to Equation (~\ref{solutionspace}) which
satisfies the following integral equation:
\begin{equation}\label{a1}
\phi_{1}(x,\mu)=\left(
\begin{array}{lll}
0\\
e^{-\mu x}
\end{array}
\right)-\int_{x}^{+\infty}\left(
\begin{array}{lll}
\frac{\sin{k}(x-y)}{k}&0\\
0&-\frac{e^{-\mu|x-y|}-e^{\mu|x-y|}}{2\mu}
\end{array}
\right)W(y)\phi_{1}(y,\mu)dy,
\end{equation}
Moreover $\phi_{1}(x,\cdot)$ is analytic in $k,$ and satisfies
Estimates (~\ref{phi1R}) (~\ref{phik}), and (~\ref{continu2}).
\end{proposition}
\begin{proof}
First we need to prove the existence of solutions
$\phi_{1}(x,\mu)$ to Equation (~\ref{a1}), which can be rewritten
as
\begin{equation}\label{secondstep}
\psi=\left(
\begin{array}{lll}
0\\
1
\end{array}
\right)+A_{\lambda}\psi,
\end{equation}
where $\psi(x,\mu):=e^{\mu x}\phi_{1}(x,\mu),$ and the operator
$A_{\lambda}:\ \mathcal{L}^{\infty}([T,\infty))\rightarrow
\mathcal{L}^{\infty}([T,\infty))$ is defined as
$$(A_{\lambda}f)(x):=e^{\mu x}\int_{x}^{\infty}\left(
\begin{array}{lll}
\frac{\sin k(x-y)}{k}&0\\
0&-\frac{e^{-\mu(|x-y|)}-e^{\mu|x-y|}}{2\mu}
\end{array}
\right) e^{-\mu y}W(y)f(y)dy$$ with $T\in (-\infty,\infty)$ an
arbitrary constant. We show for a sufficiently large $n$,
$\|A^{n}_{\lambda}\|<1,$ therefore Equation (~\ref{secondstep})
has a unique solution in $\mathcal{L}^{\infty}([T,\infty))$.

Observe that $Re\mu>0$ and $|\frac{\sin k(x-y)}{k}|\leq
c_{1}(1+|x|+|y|)e^{\mu|x-y|}.$ Thus if $x\in [T,\infty),$ then
\begin{equation}\label{contraction}
|A_{\lambda}\psi(x)|\leq
c(T)(1+|x|)\int_{x}^{\infty}(1+|y|)e^{-\alpha
|y|}dy\|\psi\|_{\mathcal{L}^{\infty}([T,\infty))}
\end{equation}
for some $c(T)$ independent of $\lambda$, where, recall $\alpha$
from Inequality (~\ref{alpha}).

Therefore
$$
\begin{array}{lll}
|A_{\lambda}^{n}\psi(x)|&\leq&
c^{n}(T)(1+|x|)\int_{x}^{\infty}(1+|x_{1}|)
e^{-\alpha|x_{1}|}dx_{1}\int_{x_{1}}^{\infty}
(1+|x_{2}|)e^{-\alpha|x_{2}|}dx_{2}\\
& &\cdot\cdot\cdot\int_{x_{n}}^{\infty}(1+|y|)e^{-\alpha|y|}dy\|\psi\|_{\mathcal{L}^{\infty}([T,\infty))}\\
&=&\frac{(1+|x|)c^{n}(T)}{n!}(\int_{x}^{\infty}(1+|x|)e^{-\alpha|x|}dx)^{n}\|\psi\|_{\mathcal{L}^{\infty}([T,\infty))}.
\end{array}
$$
Hence if $n\in \mathbb{N}$ is sufficiently large, then
$$\|A_{\lambda}^{n}\|_{\mathcal{L}^{\infty}([T,\infty))\rightarrow
\mathcal{L}^{\infty}([T,\infty))}<1.$$ By the Neumann series,
there exists a unique $\psi\in \mathcal{L}^{\infty}([T,\infty))$.
Thus there exists a function $\phi_{1}(x,\mu)$ defined in the
interval $x\in [T,\infty)$ which is solution to Equation
(~\ref{a1}). Since $T$ is an arbitrary constant, $\phi_{1}(x,\mu)$
is well defined for $x\in (-\infty,\infty)$.

For the analyticity of $\phi_{1}(x,\cdot):$ Observe that $e^{\mu
x}\phi_{1}(x,\mu)=\sum_{n=0}^{+\infty}A_{\lambda}^{n}\left(
\begin{array}{lll}
1\\
0
\end{array}
\right)$, and the sequence converges absolutely in the
$\mathcal{L}^{\infty}$ norms. Moreover each function
$A_{\lambda}^{n}\left(
\begin{array}{lll}
0\\
1
\end{array}
\right)$ is analytic in $k$, so by Lemma ~\ref{analy}
$\phi_{1}(x,\cdot)$ is analytic.

For Estimate (~\ref{phi1R}): if the constant $T$ is sufficiently
large, then by Estimate (~\ref{contraction}) we have that $\forall
\ x>T,$
$$\|A_{\lambda}\|_{\mathcal{L}^{\infty}((x,\infty))\rightarrow \mathcal{L}^{\infty}((x,\infty))}\leq
ce^{-\epsilon_{0}x}$$ for some constants $c,\ \epsilon_{0}$
independent of $x$ and $\lambda$.

By a direct calculation we can prove that
$(H-\lambda)\phi_{1}(\cdot,\mu)=0.$

To prove (~\ref{continu2}), we only need to consider the case
$\lambda=\beta.$
\end{proof}
Using $\phi_{1}(\cdot,\mu)$, we define another solution
$\phi_{2}(\cdot,\mu)$ to Equation (~\ref{solutionspace}) by
$$\phi_{2}(x,\mu) :=  \phi_{1}(-x,\mu).$$

To prove the existence of solutions $\psi_{1},\ \psi_{2},\
\xi_{1}$ in Theorem ~\ref{mainsolution}, we prove first their
existence on the domain $[R,+\infty),$ where $R$ is a large
constant. Then we continue the solutions to the interval
$(-\infty,\infty)$ by ODE theories. To this end the following
lemma will be used:
\begin{lemma}\label{ODE}
Let $a\geq b$ be constants and $\Omega_{2}\subset \mathbb{C}$ be a
bounded closed set on the complex plane. Define an $4\times 4$
matrix $T(x,k) := [T_{ij}(x,k)]$ such that each entry $T_{ij}$ is
$C^{2}$ continuous in the variable $x\in [a,b],$ and analytic in
the variable $k\in \Omega_{2}.$

Let $X(x,k): [a,b]\times \Omega_{2}\rightarrow\mathbb{C}^{4}$ be a
solution to the ODE system
\begin{equation}\label{systermODE}
\frac{dX(\cdot,k)}{dx}=T(\cdot,k)X(\cdot,k),
\end{equation} with an initial datum $X(a,k).$

If $X(a,\cdot)$ is an analytic function of $k\in \Omega_{2},$ then
$X$ is $C^{2}$ in $x\in [a,b]$, and analytic in $k\in \Omega_{2}.$
\end{lemma}
\begin{proof}
We can rewrite Equation (~\ref{systermODE}) as
$$X(\cdot,k)=X(a,k)+A_{k}X(\cdot,k),$$ where
$A_{k}: \ \mathcal{L}^{\infty}([a,b])\rightarrow\
\mathcal{L}^{\infty}([a,b])$ is an operator defined by
$$A_{k}(X)(x)=\int_{a}^{x}T(y,k)X(y)dy.$$
We can get easily that $$|A_{k}(X)(x)|\leq
c\int_{a}^{x}dx\|X\|_{\mathcal{L}^{\infty}},$$ where $c$ is
independent of $x$ and $k.$

Thus there exists an integer $m\in N$, such that
$\|A^{m}_{k}\|_{\mathcal{L}^{\infty}([a,b])\rightarrow
\mathcal{L}^{\infty}([a,b])}< 1.$ By the same strategy as in
Proposition ~\ref{phi1}, we can get the existence, smoothness and
analyticity of $X(x,k)$.
\end{proof}
\begin{proposition}\label{solutionpsi1}
There exist solutions $\psi_{1}$ and $\psi_{2}$ to Equation
(~\ref{solutionspace}) which are analytic in $k$ and satisfy
Estimates (~\ref{psi1R}), (~\ref{psi2R}), (~\ref{psi1k}),
(~\ref{psi2k}) and (~\ref{continu2}).

If $\lambda=\beta$ then there is a solution $\eta$ such that
\begin{equation}\label{eta}
\eta(x)=[\left(
\begin{array}{lll}
x\\
0
\end{array}
\right)+O(e^{-\gamma x})],
\end{equation} as $x\rightarrow
-\infty,$ where $\gamma>0$ is a constant.
\end{proposition}
\begin{proof}
We will only prove the existence of solutions $\psi_{1}(x,k),$
that of $\psi_{2}(x,k)$ is almost the same.

First we will prove the existence of a function
$\psi_{1}(\cdot,k)$ on the domain $x\in [R,+\infty),$ $\lambda\in
\Omega$ satisfying the following equation
$$
\begin{array}{lll}
\psi_{1}(x,k)&=&e^{ikx}\left(
\begin{array}{lll}
1\\
0
\end{array}
\right)-\int_{x}^{+\infty}\left(
\begin{array}{lll}
-\frac{\sin k(x-y)}{k}&0\\
0&-\frac{1}{2\mu}e^{\mu(x-y)}
\end{array}
\right)W(y)\psi_{1}(y,k)dy\\
&-&\int^{x}_{R}\left(
\begin{array}{lll}
0&0\\
0&-\frac{1}{2\mu}e^{-\mu(x-y)}
\end{array} \right)W(y)\psi_{1}(y,k)dy,
\end{array}
$$ where $R$ is a sufficiently large constant.

Define $\psi_{k}(x):=e^{-ikx}\psi_{1}(x,k).$ We could rewrite the
equation as $$\psi_{k}(x)=\left(
\begin{array}{lll}
1\\
0
\end{array}
\right)-A_{k1}\psi_{k}-A_{k2}\psi_{k},$$ where the operators
$A_{k1}$ and $A_{k2}: \mathcal{L}^{\infty}([R,\infty))\rightarrow
\mathcal{L}^{\infty}([R,\infty))$ are defined as:
$$(A_{k1}\psi)(x)=\int_{x}^{+\infty}\left(
\begin{array}{lll}
-\frac{\sin k(x-y)}{k}&0\\
0&-\frac{1}{2\mu}e^{\mu(x-y)}
\end{array}
\right)W(y)e^{-ik(x-y)}\psi(y)dy$$ and
$$(A_{k2}\psi)(x)=\int_{R}^{x}\left(
\begin{array}{lll}
0&0\\
0&-\frac{1}{2\mu}e^{-\mu(x-y)}
\end{array}
\right)W(y)e^{-ik(x-y)}\psi(y)dy.$$ We claim that if the constant
$R$ is sufficiently large, then
$$\|A_{k1}\|_{\mathcal{L}^{\infty}([R,\infty))\rightarrow
\mathcal{L}^{\infty}([R,\infty))}+\|A_{k2}\|_{\mathcal{L}^{\infty}([R,\infty))\rightarrow
\mathcal{L}^{\infty}([R,\infty))}<1$$ for any $k.$ Indeed, by the
properties of the domain $\Omega$ from (~\ref{omega1}) and that
$|W(x)|\leq e^{-\alpha|x|}$ we have
\begin{equation}\label{ak1}
|A_{k1}\psi(x)|\leq
c_{1}\int_{x}^{\infty}e^{-\frac{\alpha}{2}|x|}dx\|\psi\|_{\mathcal{L}^{\infty}}\leq
c_{1}e^{-\epsilon_{0}|x|}\|\psi\|_{\mathcal{L}^{\infty}},
\end{equation}
\begin{equation}\label{ak2}
|A_{k2}\psi(x)|\leq
c_{1}\int^{x}_{R}e^{-c_{2}|x-y|}e^{-\frac{\alpha}{2}|y|}dy\|\psi\|_{\mathcal{L}^{\infty}}\leq
c_{1}e^{-\epsilon_{0}|x|}\|\psi\|_{\mathcal{L}^{\infty}}.
\end{equation}
for some constants $c_{1},\ c_{2},\ \epsilon_{0}>0.$ Hence if $R$
is sufficient large,
$\|A_{k1}\psi\|_{\mathcal{L}^{\infty}([R,\infty))}+\|A_{k2}\psi\|_{\mathcal{L}^{\infty}([R,\infty))}\leq
1.$ By the contraction lemma we could get that $\psi_{k}$ exists
and $$\psi_{k}(x)=\left(
\begin{array}{lll}
1\\
0
\end{array}
\right)+\sum_{n=1}^{+\infty}(A_{k1}+A_{k2})^{n}\left(
\begin{array}{lll}
1\\
0
\end{array}
\right).$$ The estimate (~\ref{psi1R}) is from Estimates
(~\ref{ak1}) and (~\ref{ak2}).

When $x$ is not necessarily large we estimate $\psi_{1}(x,k)$ by
Lemma ~\ref{ODE}:  $\forall R_{4}>0,$ if $x\in [R_{3},-R_{4}]$ and
if $\lambda\in [\beta,\beta+1],$ then we have
$$|\frac{d^{n}}{dk^{n}}\psi_{1}(x,k)|\leq c$$ for some constant $c$,
and $n=0,1,2.$ Thus Estimate (~\ref{psi1k}) is proven.

To prove Claim (~\ref{continu2}), we only need to consider the
case $\lambda=\beta,$ thus it is easier to prove it.

By the similar strategy we can find a solution $\eta$ to
$(H-\beta)\eta=0$ such that it satisfies the estimate (~\ref{eta})
and the following equation:
$$
\begin{array}{lll}
\eta(x)&=&\left(
\begin{array}{lll}
x\\
0
\end{array}
\right)-\int_{x}^{+\infty}\left(
\begin{array}{lll}
-\frac{\sin k(x-y)}{k}&0\\
0&-\frac{1}{2\mu}e^{\mu(x-y)}
\end{array}
\right)W(y)\eta(y) dy\\
&-&\int^{x}_{R}\left(
\begin{array}{lll}
0&0\\
0&-\frac{1}{2\mu}e^{-\mu(x-y)}
\end{array} \right)W(y)\eta dy.
\end{array}
$$
By a direct calculation we could prove that
$(H-\lambda)\psi_{1}(\cdot,k)=0$ and $(H-\beta)\eta=0.$
\end{proof}
\begin{proposition}\label{solutionxi1}
Recall $\mu=\sqrt{\lambda+\beta}$. There exists a solution
$\xi_{1}(x,\mu)$ to Equation (~\ref{solutionspace}) which is
analytic in k and as $x\rightarrow +\infty$
\begin{equation}\label{decayxi}
\xi_{1}(x,\mu)= e^{\mu x}(\left(
\begin{array}{lll}
0\\
1
\end{array}
\right) +O(e^{-\epsilon(\lambda)x}))
\end{equation}
for some $\epsilon(\lambda)>0.$
\end{proposition}
\begin{proof}
We follow an idea from ~\cite{Buslaev}. We will prove that if the
constant $R$ is sufficiently large, then there exists a function
$\xi_{1}$ such that
$$
\begin{array}{lll}
\xi_{1}(x,\mu)&=&e^{\mu x}\left(
\begin{array}{lll}
0\\
1
\end{array}
\right)+\int_{x}^{\infty}\left(
\begin{array}{lll}
0&0\\
0&-\frac{1}{2\mu}e^{\mu(x-y)}
\end{array}
\right)W(y)\xi_{1}(y,\mu)dy\\
& &+\int_{R}^{x}\left(
\begin{array}{lll}
\frac{\sin k(x-y)}{k}&0\\
0&-\frac{1}{2\mu}e^{-\mu(x-y)}
\end{array} \right)
W(y)\xi_{1}(y,\mu)dy.
\end{array}
$$ If we could prove the existence, it is easy to prove that $$(H-\lambda)\xi_{1}(\cdot,\mu)=0.$$ Let $\psi=e^{-\mu x}\xi_{1}(x,\mu),$ then this equation can be rewritten as
$$
\psi=\left(
\begin{array}{lll}
0\\
1
\end{array}
\right)+A_{\lambda 1}\psi+A_{\lambda 2}\psi,
$$
$A_{\lambda 1}$ and $A_{\lambda 2}:
\mathcal{L}^{\infty}([R,\infty))\rightarrow
\mathcal{L}^{\infty}([R,\infty))$ are operators defined as:
$$(A_{\lambda 1}\psi)(x)=\int_{x}^{\infty}\left(
\begin{array}{lll}
0&0\\
0&-\frac{1}{2\mu}
\end{array}
\right)W(y)\psi(y)dy,$$ and
$$(A_{\lambda 2}\psi)(x)=\int_{R}^{x}\left(
\begin{array}{lll}
\frac{\sin k(x-y)}{k}&0\\
0&-\frac{1}{2\mu}e^{-\mu(x-y)}
\end{array}
\right)e^{-\mu(x-y)}W(y)\psi(y)dy.$$ As usual we want to find a
large number $R,$ s.t. $\|A_{\lambda 1}+A_{\lambda
2}\|_{\mathcal{L}^{\infty}([R,\infty))\rightarrow
\mathcal{L}^{\infty}([R,\infty))}<1.$ Then we can implement the
contraction argument.

For $A_{\lambda 1},$
\begin{equation}\label{A1}
|A_{\lambda 1}\psi(x)|\leq c\int_{x}^{\infty}e^{-\alpha
y}dy\|\psi\|_{\mathcal{L}^{\infty}([R,\infty))},
\end{equation}
thus if $R$ is sufficiently large, then
$$\|A_{\lambda 1}\|_{\mathcal{L}^{\infty}([R,\infty))\rightarrow \mathcal{L}^{\infty}([R,\infty))}<1.$$
For $A_{\lambda 2},$ since there exists some constant $\epsilon>0$
such that $Re(\mu-\pm i k)>\epsilon>0,$ we obtain
\begin{equation}\label{A2}
|(A_{\lambda 2}\psi)(x)|\leq
c(1+|x|)\int_{R}^{x}(1+|y|)e^{-\epsilon(\lambda)(x-y)}e^{-\alpha|y|}dy
\|\psi\|_{\mathcal{L}^{\infty}([R,\infty))}
\end{equation}
for some constant $c$ independent of $R.$ Thus if $R\rightarrow
+\infty,$ then
$$\max_{x\in [R,\infty)}\{
\int_{R}^{x}e^{-\epsilon(\lambda)(x-y)}e^{-\alpha|x|}dy\}\rightarrow
0,$$ i.e.
$$\|A_{\lambda 2}\|_{\mathcal{L}^{\infty}([R,\infty))\rightarrow
\mathcal{L}^{\infty}([R,\infty))}\rightarrow 0$$ as $R\rightarrow
+\infty.$

We choose a large constant $R$ such that
$$\|A_{\lambda 1}+A_{\lambda 2}\|_{\mathcal{L}^{\infty}([R,\infty))\rightarrow \mathcal{L}^{\infty}([R,\infty))}<1.$$
The existence of the solution follows by a standard contraction
argument.

Since each function $(A_{\lambda 1}+A_{\lambda 2})^{n}\left(
\begin{array}{lll}
0\\
1
\end{array}
\right)$ is analytic in $k$, then by Lemma ~\ref{analy}
$\xi_{1}(x,\cdot)=\sum_{n=0}^{\infty} (A_{\lambda 1}+A_{\lambda
2})^{n}\left(
\begin{array}{lll}
0\\
1
\end{array}
\right)$ is analytic.

Estimate (~\ref{decayxi}) can be proved by Estimates (~\ref{A1})
and (~\ref{A2}).
\end{proof}
We define another solution to (~\ref{solutionspace}) by
$$\xi_{2}(x,\mu) := \xi_{1}(-x,\mu).$$
\subsection{Generalized Wronskian}
A generalized Wronskian function is defined in the next lemma,
whose proof is straightforward and is omitted here:
\begin{lemma}
If $X_{1}$ and $X_{2}$ satisfy $(H-\lambda)X_{i}=0,$ $(i=1,2)$
then
$$W(X_{1},X_{2}) := \partial_{x}X_{1}^{T}X_{2}-\partial_{x}X_{2}^{T}X_{1}=Const.$$
\end{lemma}
Define two $2\times 2$ matrices
\begin{equation}\label{f1f2}
\begin{array}{lll}
F_{1}(x,k)& := & \left[
\begin{array}{lll}
\psi_{1}(x,k),\phi_{1}(x,\mu)
\end{array}
\right],\\
F_{2}(x,k)& := & \left[
\begin{array}{lll}
\psi_{1}(-x,\mu),\phi_{2}(x,\mu)
\end{array}
\right].
\end{array}
\end{equation}
The $2\times 2$ matrix $D(k) := (\partial_{x}
F_{1}^{T})F_{2}-F_{1}^{T}(\partial_{x}F_{2})$ is independent of
$x$ because each entry is a Wronskian function. Observe that
$D(k)$ is a symmetric matrix. Let
\begin{equation}\label{Dk}
D(k)=:\left(
\begin{array}{lll}
D_{11}(k)&D_{12}(k)\\
D_{12}(k)&D_{22}(k)
\end{array}
\right).
\end{equation}
The entry $D_{22}(k)=W(\phi_{1}(\cdot,\mu),\phi_{2}(\cdot,\mu))$
will play an important role later.

Under an assumption that there are no eigenvalues embedded in the
essential spectrum one can prove, by strategies similar to that in
~\cite{Rauch,RSS1}, that $det D(k)\not=0$ and the operator
$1+(H_{0}-\lambda+i0)^{-1}W$ is invertible in some sense for any
$\lambda>\beta$. In this paper we approach these problems in a
different way and we do not use the assumption on the embeded
eigenvalues.

We have the following result:
\begin{theorem}\label{resonance}
$H$ has a resonance at the point $\beta$ if and only if
$det{D(0)}=0.$
\end{theorem}
\begin{proof}
First we prove the sufficient condition, i.e. assume $H$ has no
resonance at the point $\beta$. Since the vectors
$$\phi_{2}(\cdot,\sqrt{2\beta}),\ \psi_{2}(-\cdot,0),\ \eta,\
\xi_{2}(\cdot,\sqrt{2\beta})$$ form an basis in the solution space
$(H-\beta)\varphi=0,$ there exist $2\times 2$ matrices $A_{2}$ and
$B_{2}$, such that
\[ \left[
\psi_{1}(\cdot,0),\phi_{1}(\cdot,\sqrt{2\beta}) \right]= \left[
\psi_{2}(-\cdot,0),\phi_{2}(\cdot,\sqrt{2\beta})
\right]A_{2}+\left[\eta, \xi_{2}(\cdot,\sqrt{2\beta})
\right]B_{2}.
\]
We claim $det B_{2}\not=0.$

Indeed if $det B_{2}=0,$ we could choose an invertible matrix
$B_{1}$ so that
\[ B_{2}B_{1}=\left[
\begin{array}{lll}
\alpha_{1}&0\\
\alpha_{2}&0\\
\end{array}
 \right].
\]
Thus
$$
\begin{array}{lll}
[\gamma_{1},\gamma_{2}]& := &[\psi_{1}(\cdot,0),\phi_{1}(\cdot,\sqrt{2\beta})]B_{1}\\
&=&[\psi_{2}(-\cdot,0),\phi_{2}(\cdot,\sqrt{2\beta})]A_{2}B_{1}+[\eta_{3},0]
\end{array}
$$
for some function $\eta_{3}$, which implies the function
$\gamma_{2}$ is bounded at $-\infty$. Since we already know that
$\psi_{1}(\cdot,0)$ and $\phi_{1}(\cdot,\sqrt{2\beta})$ are
bounded at $+\infty$, $\gamma_{2}$ is a resonance at $\beta$. This
contradicts to the fact that there are no resonances at $\beta$.
Therefore $B_{2}$ is invertible.

Re-compute $D(0)$ to prove that it is invertible:
$$
\begin{array}{lll}
\begin{array}{lll}
D(0)&=&[\frac{d\psi_{1}(x,0)}{dx},\frac{d\phi_{1}(x,\sqrt{2\beta})}{dx}]^{T}[\psi_{2}(-x,0),\phi_{2}(x,\sqrt{2\beta})]\\
& &-[\psi_{1}(x,0),\phi_{1}(x,\sqrt{2\beta})]^{T}[\frac{d\psi_{2}(x,0)}{dx},\frac{d\phi_{2}(-x,\sqrt{2\beta})}{dx}]\\
&=
&B_{2}^{T}\{[\frac{d\eta}{dx}, \frac{d\xi_{2}(x,\sqrt{2\beta})}{dx}]^{T}[\psi_{2}(-x,0),\phi_{2}(x,\sqrt{2\beta})]\\
& &-[\eta,
\xi_{2}(x,\sqrt{2\beta})]^{T}[\frac{d\psi_{2}(-x,0)}{dx},\frac{d\phi_{2}(x,\sqrt{2\beta})}{dx}]
\}\\
&=&B_{2}^{T} \left[
\begin{array}{lll}
1&*\\
0&2\sqrt{2\beta}\\
\end{array}
\right],
\end{array}
\end{array}
$$
where $*$ is an unimportant constant. Therefore $det D(0)\not=0.$

Now, we prove the necessary condition: Suppose $detD(0)\not=0.$
Since the vectors
$$\phi_{1}(\cdot,\sqrt{2\beta}),\ \psi_{1}(\cdot,0),\ \eta(-\cdot),\
\xi_{2}(-\cdot,\sqrt{2\beta})$$ form a basis to the solution space
for the equation
$$(H-\beta)\varphi=0,$$ we only need to consider the linear
combination of these vectors when we look for a resonance. First
we exclude the vectors having $\eta(-\cdot),\
\xi_{2}(-\cdot,\sqrt{2\beta})$ components because at $\infty$ the
first one blows up exponentially fast, and the second blows up at
the rate of $x$, then any linear combination with them is
unbounded, i.e. is not an resonance.

$\phi_{1}(\cdot,\sqrt{2\beta})$ could not be a resonance otherwise
$D_{22}(0)=D_{12}(0)=0$ which implies $det D(0)=0.$ We claim that
for any scalar $z,$
$\psi_{1}(\cdot,0)+z\phi_{1}(\cdot,\sqrt{2\beta})$ could not be a
resonance: indeed, if it is a resonance, then let
$$G_{1}=F_{1}\left(
\begin{array}{lll}
1&z\\
0&1
\end{array}
\right),\ G_{2}=F_{2}\left(
\begin{array}{lll}
1&z\\
0&1
\end{array}
\right),$$ where, recall $F_{1}, F_{2}$ from Equation
(~\ref{f1f2}). Let
$$\left(
\begin{array}{lll}
a_{1}&a_{2}\\
a_{3}&a_{4}
\end{array}
\right):=W(G_{1},G_{2})=\left(
\begin{array}{lll}
1&0\\
z&1
\end{array}
\right)D(0)\left(
\begin{array}{lll}
1&z\\
0&1
\end{array}
\right),$$ where
$$a_{3}=W(\psi_{1}(\cdot,0)+z\phi_{1}(\cdot,\sqrt{2\beta}),\phi_{2}(\cdot,\sqrt{2\beta}))=0$$
and
$$a_{1}=W(\psi_{1}(\cdot,0)+z\phi_{1}(\cdot,\sqrt{2\beta}),\psi_{1}(-\cdot,0)+z\phi_{2}(\cdot,\sqrt{2\beta}))=0$$
which implies that $det D(0)=0.$
\end{proof}
Recall that the space
$L^{\gamma}$ defined in Equation (~\ref{weightedl2space}). The
following lemma explains the choice of the space $L^{-\alpha/2}$
in the next section.
\begin{lemma}\label{choosespace}
If a function $\phi\in L^{-\alpha/2}$ satisfies
$(H-\lambda)\phi=0$ and if $\lambda>\beta,$ then
\begin{equation}\label{wronskian0}
W(\phi,\phi_{1}(\pm\cdot,\mu))=0,
\end{equation} and
\begin{equation}\label{whychoosingspace}
\begin{array}{lll}
\phi&=&b_{+1}\phi_{1}(\cdot,\mu)+b_{+
2}\psi_{1}(\cdot,k)+b_{+3}\psi_{2}(\cdot,k)\\
&=&b_{-1}\phi_{1}(-\cdot,\mu)+b_{-2}\psi_{1}(-\cdot,k)+b_{-3}\psi_{2}(-\cdot,k)
\end{array}
\end{equation} for some constants $b_{\pm 1},\ b_{\pm 2},\ b_{\pm3}.$
\end{lemma}
\begin{proof}
The vectors
$\{\phi_{1}(\cdot,\mu),\psi_{1}(\cdot,k),\psi_{2}(\cdot,k),\xi_{1}(\cdot,\mu)\}$
form a basis to the solution space of $(H-\lambda)\phi=0.$
Moreover
$\phi_{1}(\cdot,\mu),\psi_{1}(\cdot,k),\psi_{2}(\cdot,k)\in
L^{-\alpha/2}([0,+\infty))$ while $\xi_{1}(\cdot,\mu)\not\in
L^{-\alpha/2}([0,+\infty))$ by the fact that $Re \mu>\alpha,$
which follows from the assumption that $\alpha<\beta$ made after
Equation (~\ref{alpha}). This implies the $+$ part of Equation
(~\ref{whychoosingspace}).

The $+$ part of Equation (~\ref{wronskian0}) follows from Equation
(~\ref{whychoosingspace}) and the following results:
$$W(\phi_{1}(\cdot,\mu),\phi_{1}(\cdot,\mu))=W(\psi_{1}(\cdot,k),\phi_{1}(\cdot,\mu))=W(\psi_{2}(\cdot,k),\phi_{1}(\cdot,\mu))=0$$
while
$$W(\xi_{1}(\cdot,\mu),\phi_{1}(\cdot,\mu))\not=0.$$

The $-$ part of the lemma is proven similarly.
\end{proof}
\subsection{Generalized Eigenfunction $e(x,k)$} In this subsection we prove that the function $e(x,k)$ in Equation (~\ref{definexk}) is well defined.
Recall the definition of the domain $\Omega$ from Equation
(~\ref{omega1}). For any $\lambda\in \Omega$ we define the
operator $\mathcal{R}^{+}(\lambda): L_{\alpha}\rightarrow
L_{-\alpha}$ by its integral kernel
\begin{equation}\label{kernel}
G^{+}_{k}(x,y)=\left(
\begin{array}{lll}
\frac{e^{ik|x-y|}}{2k}&0\\
0&-\frac{e^{-\mu|x-y|}}{2i\mu}
\end{array}
\right),
\end{equation}
where, recall that $k=\sqrt{\lambda-\beta}.$ The operator
$\mathcal{R}^{+}(\lambda)$ is continuation to the resolvent
$(H_{0}-\lambda)^{-1}, \ \lambda\in \Omega\cap \mathbb{C}^{+}$ in
the following sense. Observe that
$\sigma(H_{0})=\sigma_{ess}(H_{0})=(-\infty,-\beta]\cup
[\beta,\infty)$. For any functions $f,\ g\in L^{\alpha}$ and
$\lambda\in \mathbb{C}^{+}\cap \Omega$, the quadratic form
$\langle f, \mathcal{R}^{+}(\lambda)g\rangle$, $\lambda\in\Omega$,
is an analytic continuation of the quadratic form $\langle f,
(H_{0}-\lambda)^{-1}g\rangle$ from $\Omega\cap \mathbb{C}^{+}$ to
$\Omega.$

Similarly we define $\mathcal{R}^{-}(\lambda)$ using the integral
$$G^{-}_{k}(x,y)=\left(
\begin{array}{lll}
-\frac{e^{-ik|x-y|}}{2k}&0\\
0&-\frac{e^{-\mu|x-y|}}{2i\mu}
\end{array}
\right).$$ It is the analytic continuation of the resolvent
$(H_{0}-\lambda)^{-1}$ from $\lambda\in \Omega\bigcap
\mathbb{C}^{-}$ to $\Omega$.

Equation (~\ref{kernel}) and Inequality (~\ref{alpha}) imply that
if $\lambda\in \Omega$, then $\mathcal{R}^{\pm}(\lambda)W: \
L^{-\alpha/2}\rightarrow L^{-\alpha/2}$ are compact operators (in
fact, trace class operators, see ~\cite{RSI}).

The following theorem is the main result of this subsection:
\begin{theorem}\label{a31}
If $\lambda>\beta$ is not an eigenvalues of $H$ embedded in the
essential spectrum, then the operators
\begin{equation}\label{translation}
1+\mathcal{R}^{+}(\lambda)W:\ L^{-\alpha/2}\rightarrow
L^{-\alpha/2}
\end{equation} are invertible, and the functions
$e(\cdot,k)$ in Equation (~\ref{definexk}) are well defined and
can be written as
\begin{equation}\label{concreteexk}
e(x,k)=-i\frac{2D_{22}(k)k}{detD(k)}\eta(x,k),
\end{equation} where
$\eta(x,k)=\psi_{1}(x,k)-\frac{D_{12}(k)}{D_{22}(k)}\phi_{1}(x,\mu)$.
\end{theorem}
The proof of Theorem ~\ref{a31} will be after Lemma
~\ref{exlambda}, and will use the results from Lemma
~\ref{exkatinfty}, Propositions ~\ref{exkform}, ~\ref{zeromode}
and Lemma ~\ref{exlambda}.

The following simple lemma whose proof is obvious is important in
this subsection:
\begin{lemma}\label{goldrule}
Let $C$ be the operator of complex conjugating, and
$$\sigma_{3}:=\left(
\begin{array}{lll}
1&0\\
0&-1
\end{array} \right).$$ If $\lambda>\beta$, then
$$\sigma_{3}\mathcal{R}^{\pm}(\lambda)\sigma_{3}=\mathcal{R}^{\pm}(\lambda),\ \sigma_{3}W\sigma_{3}=W^{*},$$ and therefore
$$C\sigma_{3}(1+\mathcal{R}^{\pm}(\lambda)W)C\sigma_{3}=1+\mathcal{R}^{\mp}(\lambda)W,$$
$$\sigma_{3}(1+\mathcal{R}^{\pm}(\lambda)W^{*})\sigma_{3}=1+\mathcal{R}^{\pm}(\lambda)W.$$
\end{lemma}
\begin{corollary}\label{phicon} If $\lambda>\beta$,
then $\phi_{1}(x,\mu)=-\sigma_{3}\overline{\phi}_{1}(x,\mu).$
\end{corollary}
We start with studying the analytic function $D(k)$ introduced in
Equation (~\ref{Dk}).
\begin{lemma}\label{d22}
If there exists some $\lambda_{1}>\beta$ such that
$1+\mathcal{R}^{+}(\lambda_{1})W$ is not invertible, then either
$\lambda_{1}$ is an eigenvalue of $H$, or
$D_{22}(\sqrt{\lambda_{1}-\beta})\not=0.$
\end{lemma}
\begin{proof}
Assume $\lambda_{1}$ is not an eigenvalue of $H$ and assume by
contradiction that $D_{22}(\sqrt{\lambda_{1}-\beta})=0.$

By the definition of $D_{22}(k)$ and Lemma ~\ref{choosespace} we
can get that $\phi_{1}(x):=\phi_{1}(x,\sqrt{\beta+\lambda_{1}})$
is a bounded function, i.e.
$$\phi_{1}(x)=c_{1}\phi_{1}(-x)+c_{2}\psi_{2}(-x,\sqrt{\lambda_{1}-\beta})+c_{3}\psi_{1}(-x,\sqrt{\lambda_{1}-\beta})$$
for some constants $c_{i}, \ i=1,2,3$. Furthermore, since
$1+\mathcal{R}(\lambda_{1})W$ is not invertible, there exists a
function $g\in L^{-\alpha/2}\backslash \mathcal{L}^{2}$ such that
$(1+\mathcal{R}(\lambda_{1})W)g=0.$ By elementary calculations we
get that
$$g(x)=c_{4}\phi_{1}(x)+c_{5}\psi_{1}(x,\sqrt{\lambda_{1}-\beta})=c_{6}\phi_{1}(-x)+c_{7}\phi_{1}(-x,\sqrt{\lambda_{1}-\beta})$$
for some constants $c_{n},\ n=4,5,6,7.$

Since $g$ and $\phi_{1}$ satisfy the equation
$(H-\lambda_{1})\psi=0$, $W(\phi_{1},g)$ is independent of $x$.
Therefore
$$0=W(\phi_{1},g)=2c_{7}c_{3}\sqrt{\lambda_{1}-\beta}i,$$
then either $c_{7}=0$ or $c_{3}=0$. Similarly by calculating
$W(\phi_{1},g(-\cdot))$, we can get that either $c_{5}=0$ or
$c_{2}=0$. Hence by Corollary ~\ref{phicon} either $g\in
\mathcal{L}^{2}$ and is therefore an eigenfunction of $H$ or
$\phi_{1}\in \mathcal{L}^{2}$ and is therefore an eigenfunction of
$H$. By the assumption of the lemma,
$D_{22}(\sqrt{\lambda_{1}-\beta})\not=0.$
\end{proof}
Since $\mathcal{R}^{+}(\lambda)W$ is analytic in a neighborhood of
the semi-axis $[\beta,\infty)$ and since
$\|\mathcal{R}^{+}(\lambda)W\|_{L^{-\alpha/2}\rightarrow
L^{\alpha/2}}\rightarrow 0\rightarrow 0$ as $\lambda\rightarrow
\infty$ the operator $1+\mathcal{R}^{+}(\lambda)$ are not
invertible for at most finite number of points $\lambda\in
[\beta,\infty).$

Assume now that some point $\lambda_{1}>\beta$ is not an
eigenvalue of $H,$ and $1+\mathcal{R}^{+}(\lambda_{1})W$ is not
invertible, then $D_{22}(\sqrt{\lambda_{1}-\beta})\not=0.$ Since
$D_{22}$ is an analytic function of $k$, there exists a small
neighborhood $\Omega_{1}$ of $\lambda_{1}$ such that $\forall
\lambda\in \Omega_{1},$ $D_{22}(k)\not=0,$ and $\forall \lambda\in
\Omega_{1}\backslash \{\lambda_{1}\},$
$1+\mathcal{R}^{+}(\lambda)W$ is invertible as an operator from
$L^{-\alpha/2}\rightarrow L^{-\alpha/2}$. Hence we have
\begin{lemma}
The function $e(x,k):=(1+\mathcal{R}^{+}(\lambda)W)^{-1}\left(
\begin{array}{lll}
e^{ikx}\\
0
\end{array}
\right)$ is well defined for all
$\Omega_{1}\backslash\{\lambda_{1}\}$ and belongs to
$L^{-\alpha/2}$. Moreover it satisfies the equation
$(H-\lambda)e(\cdot,k)=0.$
\end{lemma}
Next using the lemma above we derive additional properties of the
function $e(x,k)$ for $\lambda\in
\Omega\backslash\{\lambda_{1}\}.$
\begin{lemma}\label{exkatinfty}
If at some point $\lambda\in \Omega$ the operator
$1+\mathcal{R}^{+}(\lambda)W$ defined in Equation
(~\ref{translation}) is invertible, then there are functions $s,\
a$ of $k,$ such that
$$e(x,k)=s(k)\psi_{1}(x,k)+a(k)\phi_{1}(x,\mu).$$ Especially,
$s(k)\not=0$ if $\lambda$ is sufficiently large.
\end{lemma}
\begin{proof}
Since for $\lambda\in \Omega_{1}\backslash\{\lambda_{1}\},$
$(H-\lambda)e(\cdot,k)=0$ and $e(\cdot,k)\in L^{-\alpha/2}$ we
have by Lemma ~\ref{choosespace} that
$$e(\cdot,k)=s(k)\psi_{1}(\cdot,k)+a(k)\phi_{2}(\cdot,\mu)+b(k)\psi_{2}(\cdot,k)$$
for some functions $s,\ a,\ b.$ Thus we only need to prove that
$b=0.$

From the properties of $\psi_{1},\psi_{2},\phi_{1}$ we can get
that
\begin{equation}\label{firstestimate}
e(\cdot,k)=s(k)\left(
\begin{array}{lll}
e^{ikx}\\
0
\end{array}
\right)+b(k)\left(
\begin{array}{lll}
e^{-ikx}\\
0
\end{array}
\right)+O(e^{-\frac{\alpha}{2} x})
\end{equation} as $x\rightarrow +\infty.$ From the definition of
the domains $\Omega$ and $\Omega_{1},$ we see that
$|Imk|<\alpha/2.$

On the other hand
\begin{equation}\label{secondestimate}
\begin{array}{lll}
\left(
\begin{array}{lll}
e^{ikx}\\
0
\end{array}
\right)&=&e(x,k)+\mathcal{R}^{+}(\lambda)We(\cdot,k)\\
&=&e(x,k)+e^{ikx}\left(
\begin{array}{lll}
a_{1}(k)\\
0
\end{array}
\right)+O(e^{-\frac{\alpha}{2} x})
\end{array}
\end{equation} for some constant $a_{1}(k)$
as $x\rightarrow +\infty.$ Comparing these two equations
(~\ref{firstestimate}), (\ref{secondestimate}), we find that
$b=0.$

By the fact that $$\lim_{\lambda\rightarrow
\infty}\|\mathcal{R}^{+}(\lambda)We(\cdot,k)\|_{\mathcal{L}^{\infty}}=0$$
and Equation (~\ref{secondestimate}), we can get that $s(k)$ could
not be zero if $\lambda$ is large.
\end{proof}
\begin{lemma}\label{d22notzero}
The analytic function $D_{22}(k)$ can be zero at only a discrete
subset of $\Omega$.
\end{lemma}
\begin{proof}
We prove by contradiction. Suppose not, then $D_{22}(k)=0$
globally. By Lemma ~\ref{choosespace}
$$W(e(\cdot,k),\phi_{1}(-\cdot,\mu))=0$$ for any $\lambda$ provided
that $e(x,k)$ is well defined. We proved in Lemma
~\ref{exkatinfty} that for large $\lambda,$ $e(x,k)$ is well
defined and $e(x,k)=s(k)\psi_{1}(x,k)+a(k)\phi_{1}(x,u)$ for some
constants $s(k)\not=0$ and $a(k)$. Since
$$D_{22}(k):=W(\phi_{1}(x,\mu),\phi_{1}(-x,\mu))=0,$$ we have
$$D_{12}(k)=D_{21}(k):=W(\psi_{1}(x,k),\phi_{1}(-x,k))=0$$ for any
large $k$, where, recall the definition of $D(k)$ in Equation
(~\ref{Dk}). Thus $Det D(k)=0$ globally which contradicts to the
fact that $detD(0)\not=0.$ Thus $D_{22}(k)=0$ only at a discrte
subset.
\end{proof}
\begin{proposition}\label{exkform}
Equation (~\ref{concreteexk}) holds for any $\lambda\in \Omega$ if
the opeator $1+\mathcal{R}^{+}(\lambda)W$ defined in Equation
(~\ref{translation}) is invertible.
\end{proposition}
\begin{proof}
Define $\Omega_{2}$ be the subset of $\Omega$ such that if
$\lambda\in \Omega_{3}$ then the operator
$1+\mathcal{R}^{+}(\lambda)W$ is invertible and
$D_{22}(\sqrt{\lambda-\beta})\not=0.$ By Lemma ~\ref{d22notzero}
we can see that $\Omega\backslash\Omega_{2}$ is a discrete subset
of $\Omega$, and $\Omega_{1}\subset\Omega_{2}.$ Therefore $e(x,k)$
is well defined if $k^{2}+\beta=\lambda\in\Omega_{2}$. By the fact
of $e(\cdot,k)\in L^{-\alpha/2}$ and Lemmas ~\ref{choosespace},
~\ref{exkatinfty} there exist functions $s$, $a,$ $k_{1},\ k_{2}$
of the variable $k$ such that
\begin{equation}\label{formulaexk}
\begin{array}{lll}
e(x,k)&=&s(k)\psi_{1}(x,k)+a(k)\phi_{1}(x,\mu)\\
&=&\psi_{2}(-x,k)+k_{1}(k)\psi_{1}(-x,k)+k_{2}(k)\phi_{2}(x,k).
\end{array}
\end{equation}
If $\lambda\in \Omega_{2},$ then the fact $D_{22}(k)\not=0$
implies that $s(k)\not=0.$ Define $s_{1}(k) := \frac{1}{s(k)}$ and
$k_{3}(k):=\frac{a(k)}{s(k)}.$ Then
$$e(x,k)=\frac{1}{s_{1}(k)}[\psi_{1}(x,k)+k_{3}(k)\phi_{1}(x,\mu)].$$
We claim $k_{3}(k)=-\frac{D_{12}(k)}{D_{22}(k)}.$ Indeed, consider
the matrix
$$
\begin{array}{lll}
(s_{1}(k)e(x,k),\phi_{1}(x,\mu))&=&(\psi_{1}(x,k),\phi_{1}(x,\mu))
\left(
\begin{array}{lll}
1&0\\
k_{3}(k)&1
\end{array}
\right)\\
&=&F_{1}(x,k)\left(
\begin{array}{lll}
1&0\\
k_{3}(k)&1
\end{array}
\right).
\end{array}
$$
Recall $F_{1}$ and $F_{2}$ from Equation (~\ref{f1f2}). In the
following computation we use the fact
$W(e(\cdot,k),\phi_{2}(\cdot,\mu))=0$:
$$
\begin{array}{lll}
& &\left(
\begin{array}{lll}
D_{11}(k)&D_{12}(k)\\
D_{12}(k)&D_{22}(k)
\end{array}
\right)=\partial_{x}F_{1}^{T}F_{2}-F_{1}^{T}\partial_{x}F_{2}\\
&=&\left(
\begin{array}{lll}
1&-k_{3}(k)\\
0&1
\end{array}
\right)\left(
\begin{array}{lll}
D_{11}(k)-\frac{D_{12}^{2}(k)}{D_{22}(k)}&0\\
0&D_{22}(k)
\end{array}
\right)\left(
\begin{array}{lll}
1&0\\
-k_{3}(k)&1
\end{array}
\right).
\end{array}
$$
This equality implies that $D_{12}(k)+k_{3}(k)D_{22}(k)=0$ or
equivalently $k_{3}(k)=-\frac{D_{12}(k)}{D_{22}(k)}.$ By Equation
(~\ref{formulaexk}),
$$
\begin{array}{lll}
&
&\psi_{1}(x,k)-\frac{D_{12}(k)}{D_{22}(k)}\phi_{1}(x,\mu)\\
&=&s_{1}(k)\psi_{2}(-x,k)+s_{1}(k)k_{1}(k)\psi_{1}(-x,k)+s_{1}(k)k_{2}(k)\phi_{2}(x,\mu).
\end{array}
$$
We use a Wronskian function to derive an expression for
$s_{1}(k):$
$$2ik s_{1}(k)=W(\psi_{1}(\cdot,k)-\frac{D_{12}(k)}{D_{22}(k)}\phi_{1}(\cdot,\mu),\psi_{1}(-\cdot,k))=D_{11}(k)-\frac{D_{12}^{2}(k)}{D_{22}(k)}.$$
Since the right hand side is analytic, $s_{1}(k)$ is meromorphic.

Therefore we proved Equation ~\ref{concreteexk} if $\lambda\in
\Omega_{2}.$ Since $D_{22}(k)=0$ only at a discrete subset of
$\Omega$ and $\eta(x,k)$ and $s_{1}(k)$ are meromorphic functions,
Equation ~\ref{concreteexk} holds if the function $e(x,k)$ is well
defined.
\end{proof}
Since $1+\mathcal{R}^{+}(\lambda)W$ is not invertible at
$\lambda_{1},$ we expect that
$[1+\mathcal{R}^{+}(\lambda)W]^{-1}\left(
\begin{array}{lll}
e^{ikx}\\
0
\end{array}
\right)$ blows up in some sense at $\lambda=\lambda_{1}$. We want
to determine the nature of this blow up.
\begin{proposition}\label{zeromode}
If $1+\mathcal{R}^{+}(\lambda)W$ is not invertible at some point
$\lambda_{1}>\beta,$ then:
$$s_{1}(\sqrt{\lambda_{1}-\beta})=
0;$$
\begin{equation}\label{etasolution}
[1+\mathcal{R}^{+}(\lambda_{1})W]\eta(\cdot,\sqrt{\lambda_{1}-\beta})=0;
\end{equation}
and $$\eta(\cdot,\sqrt{\lambda_{1}-\beta})\in L^{-\alpha/2},$$
where $\alpha$ is given in Equation (~\ref{alpha}) and
$L^{\gamma}:=e^{\gamma/4|x|}\mathcal{L}^{2}$
\end{proposition}
\begin{proof}
\begin{enumerate}
\item[(1)] The proof of $\eta(\cdot,\sqrt{\lambda_{1}-\beta})\in
L^{-\alpha/2}$ is easy: the analytic function
$$W(\eta(\cdot,k),\phi_{2}(\cdot,\mu))=0$$ for $\lambda\in \Omega_{1}\backslash \{\lambda_{1}\}$, therefore for any
$\lambda\in \Omega$. By Lemma ~\ref{choosespace} this implies that
$$\eta(\cdot,k)=k_{1}\psi_{1}(-\cdot,k)+k_{2}\psi_{2}(-\cdot,k)+k_{3}\phi_{2}(\cdot,\mu)$$
for some $k_{n},\ n=1,2,3.$ Thus if $\lambda>\beta$, then $\eta$
is bounded at $-\infty$. By the definition of $\eta(\cdot,k)$ it
is bounded at $+\infty$. Therefore $\eta(\cdot,k)\in
\mathcal{L}^{\infty}\subset L^{-\alpha/2}$ if $\lambda>\beta.$

\item[(2)] To prove
$(1+\mathcal{R}^{+}(\lambda_{1})W)\eta(\cdot,\sqrt{\lambda_{1}-\beta})=0,$
we use that $1+\mathcal{R}^{+}(\lambda_{1})W$ is not invertible,
and therefore there exists a function $g\in L^{-\alpha/2}$ such
that $(1+\mathcal{R}^{+}(\lambda_{1})W)g=0$ which implies that
$$g=z_{1}\psi_{1}(\cdot,\sqrt{\lambda_{1}-\beta})+z_{2}\phi_{1}$$
for some constants $z_{1}$ and $z_{2}$ by a similar argument as in
Equation (~\ref{secondestimate}). Since
$D_{22}(\sqrt{\lambda_{1}-\beta})\not=0,$ $z_{1}\not=0.$ Thus
without loss of generality we assume $z_{1}=1.$ We claim
\begin{equation}\label{geta}
z_{2}=-\frac{D_{12}(\sqrt{\lambda_{1}-\beta})}{D_{22}(\sqrt{\lambda_{1}-\beta})}
\end{equation}
or equivalently $g=\eta(\cdot,\sqrt{\lambda_{1}-\beta})$ (see
Equation (~\ref{defineta})). Indeed, using
$$\eta(\cdot,\sqrt{\lambda_{1}-\beta})-g=[z_{2}+\frac{D_{12}(\sqrt{\lambda_{1}-\beta})}{D_{22}(\sqrt{\lambda_{1}-\beta})}]\phi_{1}$$ and using the Wronskian function to calculate
$z_{2}+\frac{D_{12}(k)}{D_{22}(k)}$, we obtain
$$[z_{2}+\frac{D_{12}(\sqrt{\lambda_{1}-\beta})}{D_{22}(\sqrt{\lambda_{1}-\beta})}]D_{22}(\sqrt{\lambda_{1}-\beta})=W(\eta(\cdot,\sqrt{\lambda_{1}-\beta})-g,\phi_{2}(\cdot,\sqrt{\beta+\lambda_{1}}))=0.$$
Since $D_{22}(\sqrt{\lambda_{1}-\beta})\not=0$ as proven in Lemma
~\ref{d22}, we have Equation (~\ref{geta}). Therefore we have
$g=\eta(\cdot,\sqrt{\lambda_{1}-\beta}).$

\item[(3)] The equation $s_{1}(\sqrt{\lambda_{1}-\beta})=0$
follows from the following three facts
$$(1+\mathcal{R}^{+}(\lambda)W)\eta(\cdot,k)=s_{1}(k)\left(
\begin{array}{lll}
e^{ikx}\\
0
\end{array}
\right)$$ which follows from Equations (~\ref{definexk}) and
(~\ref{exketa}),
$$\lim_{k\rightarrow
\sqrt{\lambda_{1}-\beta}}\eta(\cdot,k)=\eta(\cdot,\sqrt{\lambda_{1}-\beta})\
\text{in}\ L^{-\alpha/2}$$ which can be proved by Lemma ~\ref{ODE}
and the Dominated Convergence Theorem, and Equation
(~\ref{etasolution}).
\end{enumerate}
\end{proof}
In the following lemma we prove that $s_{1}(k)$ could not be zero:
\begin{lemma}\label{exlambda}
There exist functions $s,\ a,\ b,\ \gamma$ such that
$$
\begin{array}{lll}
e(x,k)&=&s(k)\psi_{1}(x,k)+a(k)\phi_{1}(x,\mu)\\
&=&b(k)\phi_{2}(x,\mu)+\psi_{2}(-x,k)+r(k)\psi_{1}(-x,k),\\
\end{array}
$$
where $|s(k)|^{2}+|r(k)|^{2}=1$, and $\bar{s}r+\bar{r}s=0$.
\end{lemma}
\begin{proof}
By Lemmas ~\ref{choosespace}, ~\ref{exkatinfty} there exist
functions $s,\ a,\ b,\ c$ and $r$ such that
$$
\begin{array}{lll}
e(x,k)&=&s(k)\psi_{1}(x,k)+a(k)\phi_{1}(x,\mu)\\
&=&b(k)\phi_{2}(x,\mu)+c(k)\psi_{2}(-x,k)+r(k)\psi_{1}(-x,k).
\end{array}
$$ One can show that $c=1$ by a similar expansion as
in Equation (~\ref{secondestimate}) at $-\infty.$

We divide the proof into two cases: $s(k)=0$ and $s(k)\not=0.$
\begin{enumerate}
\item[(1)] If $s(k)=0:$ then $a(k)\not=0,$ thus
$$\frac{e(x,k)}{a(k)}=\phi_{1}(x,k).$$ By Corollary ~\ref{phicon},
$$-\sigma_{3}\overline{\frac{e(x,k)}{a(k)}}=\frac{e(x,k)}{a(k)}.$$ Thus
$$-\frac{1}{\bar{a}(k)}=\frac{r(k)}{a(k)},$$
which implies $|r(k)|=1.$ This proves the lemma when $s(k)=0.$

\item[(2)] If $s(k)\not=0:$ It is easy to get that $$
\begin{array}{lll}
e(-x,k)&=&\psi_{2}(x,k)+r(k)\psi_{1}(x,k)+b(k)\phi_{1}(x,\mu)\\
&=&s(k)\psi_{1}(-x,k)+a(k)\phi_{2}(x,\mu)
\end{array}
$$ is a solution to $H-\lambda.$
There exist $b_{1}, b_{2}$ such that $$
\begin{array}{lll}
\sigma_{3}\bar{e}(-x,k)
&=&\psi_{1}(x,k)+\bar{r}(k)\psi_{2}(x,k)+b_{1}(k)\phi_{1}(x,\mu)\\
&=&\bar{s}(k)\psi_{2}(-x,k)+b_{2}(k)\phi_{2}(x,\mu)
\end{array}
$$ satisfies $(H_{0}-\lambda+W)\sigma_{3}\bar{e}(\cdot,k)=0.$

Therefore
\begin{equation}\label{exk-infty}
e(x,k)=\frac{1}{\bar{s}(k)}\sigma_{3}\bar{e}(-x,k)+\frac{r(k)}{s(k)}e(-x,k)+d(k)\phi_{1}(x,\mu)
\end{equation}
for some $d(k).$ We claim that if $d(k)\not=0$ then $D_{22}(k)=0.$
Indeed, we already know that $e(\cdot,k)\in L^{-\alpha/2}.$ So
$\phi_{1}(\pm\cdot,\mu)\in L^{-\alpha/2}.$ Thus by Lemma
~\ref{choosespace},
$D_{22}(k)=W(\phi_{1}(\cdot,\mu),\phi_{1}(-\cdot,\mu))=0.$

Therefore if $D_{22}(k)\not=0,$ then $d(k)=0.$ Then Equation
(~\ref{exk-infty}) implies
$$\frac{1}{\bar{s}(k)}+\frac{r^{2}(k)}{s(k)}=s(k),$$
and $$ \frac{r(k)}{s(k)}=-\frac{\bar{r}(k)}{\bar{s}(k)}.
$$
Therefore $$|s(k)|^{2}+|r(k)|^{2}=1,\ \text{and}\
s(k)\bar{r}(k)+\bar{s}(k)r(k)=0.$$ Since $s$ and $r$ are
meromorphic functions of $k$, and since $D_{22}(k)=0$ only at
discrete points, the formula works for all $k.$
\end{enumerate}
\end{proof}
\text{Proof of Theorem ~\ref{a31}:} By Lemma ~\ref{exlambda} and
the proof of Proposition ~\ref{exkform}, one can obtain:
\begin{equation}\label{ska1k}
s(k)=\frac{1}{s_{1}(k)}=i\frac{2D_{22}(k)k}{detD(k)},
\end{equation}
and
\begin{equation}\label{noblowup}
|s(k)|^{2}=|\frac{1}{s_{1}(k)}|^{2}\leq 1.
\end{equation}
  And by
Equation (~\ref{noblowup}) we have that
$s_{1}(\sqrt{\lambda_{1}-\beta})\not=0.$ Hence the operator
$1+\mathcal{R}^{+}(\lambda)W$ must be invertible at the point
$\lambda_{1}>\beta,$ otherwise there is a contradiction by
Proposition ~\ref{zeromode}.
\begin{flushright}
$\square$
\end{flushright}
Also by Lemma ~\ref{exlambda} and the proof of Proposition
~\ref{exkform} we have that
\begin{equation}\label{ska1k2}
a(k)=-s(k)\frac{D_{12}(k)}{D_{22}(k)}=-\frac{2ikD_{12}(k)}{detD(k)}.
\end{equation} Moreover
$s,\ a,\ b,\ r$ are meromorphic functions of $k$. Since
$detD(0)\not=0$ they are analytic functions of $k$ in a
neighborhood of $0$.
\begin{proposition}
If $H$ has no resonance at $\beta$, then $$e(\cdot,0)=0.$$
\end{proposition}
\begin{proof}
By Theorem ~\ref{resonance} $detD(0)\not=0,$ and by Lemma
~\ref{exlambda} and Equations (~\ref{ska1k}) (~\ref{ska1k2})
$$e(x,k)=\frac{2ikD_{22}(k)}{detD(k)}\psi_{1}(x,k)-\frac{2ikD_{12}(k)}{detD(k)}\phi_{1}(x,k).$$
Hence $e(x,0)=0.$
\end{proof}
\subsection{Estimates on $e(x,k)$}
In this subsection we estimate the eigenfunctions
$$e(x,k)=[1+\mathcal{R}^{+}(\lambda)W]^{-1}\left(
\begin{array}{lll}
e^{ikx}\\
0
\end{array}
\right)$$ for all $\lambda>\beta$, which are well defined as
proved in the last subsection.
\begin{theorem}\label{exk}
If $\lambda\geq \beta$, then
$$\|\frac{d^{n}}{dk^{n}}(e(x,k)/k)\|_{\mathcal{L}^{2}(dk)}\leq c(1+|x|)^{n+1};$$
$$\|\frac{d^{n}}{dk^{n}}(e(x,k)/k)\|_{\mathcal{L}^{\infty}(dk)}\leq c(1+|x|)^{n+1};$$
$$\|f^{\#}\|_{\mathcal{H}^{2}}\leq c\|\rho_{-2}f\|_{2};$$
$$|\frac{d^{n}}{dk^{n}}f^{\#}|\leq c\|\rho_{-n}f\|_{1},$$
where $n=0,1,2,$ the constant $c$ is independent of $x,$ and,
recall, $\rho_{\nu}=(1+|x|)^{-\nu}.$
\end{theorem}
The estimates in Theorem ~\ref{exk} will be proved in Propositions
~\ref{roughestimate} and ~\ref{thirdestimate} and Corollary
~\ref{corothird}.
\begin{proposition}\label{roughestimate}
$$\|\frac{d^{n}}{dk^{n}}e(x,\cdot)\|_{\mathcal{L}^{\infty}(dk)}\leq c (1+|x|)^{n},$$
$$\|\frac{d^{n}}{dk^{n}}(e(x,\cdot)/k)\|_{\mathcal{L}^{\infty}(dk)}\leq c (1+|x|)^{n+1},$$
where $c$ is a constant independent of $x$, $n$ and $\lambda$.
$n=0,1,2,3$ and $\lambda>\beta>0.$
\end{proposition}
\begin{proof}
Since we proved $e(x,0)=0,$
$\|\frac{d^{n}}{dk^{n}}(e(x,k)/k)\|_{\mathcal{L}^{\infty}(dk)}$
can be estimated by
$\|\frac{d^{n+1}}{dk^{n+1}}e(x,k)\|_{\mathcal{L}^{\infty}(dk)}$.

We divide the proof into two cases: $\lambda>\beta+\epsilon_{0}$
and $\beta\leq \lambda\leq \beta+\epsilon_{0}$, where
$\epsilon_{0}$ is a small positive number to be specified later.
\begin{enumerate}
\item[(1)]If $\lambda>\beta+\epsilon_{0},$ then
$$
\begin{array}{lll}
e(x,k)&=&\left(
\begin{array}{lll}
e^{ikx}\\
0
\end{array}
\right)-\mathcal{R}^{+}(\lambda)We(x,k)\\
&=&\left(
\begin{array}{lll}
e^{ikx}\\
0
\end{array}
\right)-\int_{-\infty}^{\infty} \left(
\begin{array}{lll}
\frac{e^{ik|x-y|}}{2k}&0\\
0&-\frac{e^{-\mu |x-y|}}{2i\mu}\\
\end{array}
\right)We(y,k)dy.
\end{array}
$$
We estimate $\|e(x,k)\|_{\mathcal{L}^{\infty}(dx)}$ by
$\|e(\cdot,k)\|_{L^{-\alpha/2}}:$
$$\|e(\cdot,k)\|_{\mathcal{L}^{\infty}(dx)}\leq 1+\frac{c}{|k|}(\int_{-\infty}^{\infty}|W(y)e^{\frac{\alpha}{2}|y|}|^{2}dy)^{1/2}\|e(\cdot,k)\|_{L^{-\alpha/2}(dx)},$$
where the constant $c$ is independent of $\lambda.$

$\forall\ \epsilon_{0}>0,$ there exists a constant
$c(\epsilon_{0})>0$, such that if $\lambda>\beta+\epsilon_{0}$,
then
$$\|(1+\mathcal{R}^{+}(\lambda)W)^{-1}\|_{L^{-\alpha/2}\rightarrow
L^{-\alpha/2}}\leq c(\epsilon_{0}).$$ Thus if
$\lambda>\beta+\epsilon_{0}$ then we have
$\|e^{-\alpha/2|\cdot|}e(\cdot,k)\|_{\mathcal{L}^{2}(dx)}\leq
c(\epsilon_{0})$, hence
$\|e(\cdot,k)\|_{\mathcal{L}^{\infty}(dx)}\leq c(\epsilon_{0}).$

For $\frac{d}{dk}e(x,k),$ we need Fubini's Theorem to justify the
following computation. Since it is tedious and not hard, we do not
want to do it.
$$
\begin{array}{lll}
\frac{d}{dk}e(x,k)&=&ix\left(
\begin{array}{lll}
e^{ikx}\\
0
\end{array}
\right)
-\int_{-\infty}^{\infty}A(x,y,k)We(y,k)dy\\
&-&\int_{-\infty}^{\infty}\left(
\begin{array}{lll}
\frac{e^{ik|x-y|}}{2k}&0\\
0&\frac{e^{-\mu|x-y|}}{2\mu}
\end{array}
\right)W\frac{d}{dk}e(x,k)dy,
\end{array}
$$
where $$A(x,y,k):=\left(
\begin{array}{lll}
\frac{ik|x-y|e^{ik|x-y|}-ke^{ik|x-y|}}{2k}&0\\
0&-\frac{ik|x-y|e^{-\mu |x-y|}-ike^{-\mu|x-y|}}{2i\mu^{3}}
\end{array}
\right).$$ Similar reasoning proves that if
$\lambda>\beta+\epsilon_{0}$, then
$$\|\frac{d^{n}}{dk^{n}}e(x,\cdot)\|_{\mathcal{L}^{\infty}(dx)}\leq c(\epsilon_{0})(1+|x|)^{n},$$
where the constant $c$ is independent of $x$, $n=0,1,2,3$.

\item[(2)] After finishing the estimates of $e(x,k)$ when
$\lambda\geq \beta+\epsilon_{0},$ we consider the cases
$\beta\leq\lambda\leq\beta+\epsilon_{0}.$ We choose $\epsilon_{0}$
so small such that if $\lambda-\beta\leq \epsilon_{0}$, then
$detD(k)\not=0.$ When we estimate the functions $\phi_{1},\
\phi_{2},\ \psi_{1},\ \psi_{2},$ we always divide the domain
$(-\infty,\infty)$ into two parts: $(-\infty,0],\ [0,\infty).$ We
will use the same strategy to estimate $e(x,k)$ when $k$ is small.

In Lemma ~\ref{exlambda}, we prove that if $x\in [0,+\infty),$
$$e(x,k)=\frac{-2ikD_{22}(k)}{detD(k)}\psi_{1}(x,k)+\frac{2ikD_{12}(k)}{detD(k)}\phi_{1}(x,\mu).$$
If $x\in [0,+\infty)$, by Theorem ~\ref{mainsolution} we have
$$\|\frac{d^{n}}{dk^{n}}e(x,k)\|_{\mathcal{L}^{\infty}(dk)}\leq
c(1+|x|)^{n}$$ for $n=0,1,2,3.$ Similarly if $x\in (-\infty,0],$
then
$$\|\frac{d^{n}}{dk^{n}}e(x,k)\|_{\mathcal{L}^{\infty}(dk)}\leq c(1+|x|)^{n}$$
for $n=0,1,2.$
\end{enumerate}
Conclusion: there exists a constant $c>0$, such that if
$\lambda\geq \beta$, then
$$\|\frac{d^{n}}{dk^{n}}e(x,k)\|_{\mathcal{L}^{\infty}(dk)}\leq c(1+|x|)^{n}$$
where $n=0,1,2,3.$
\end{proof}
In the following lemma we decompose $e(x,k)$ into several parts
which are easier to understand.
\begin{lemma}\label{disekx}
If $x\geq 0$ and $\lambda>\beta$, then there exists function
$s_{2}$ such that
\begin{equation}\label{asympto}
|\frac{d^{n}}{dk^{n}}[e(x,k)-s_{2}(k)\left(
\begin{array}{lll}
e^{ikx}\\
0
\end{array}
\right)]|\leq c \frac{1}{1+|k|}e^{-\epsilon_{0}|x|}.
\end{equation}
If $x\leq 0$ and $\lambda>\beta,$ then there exists a function
$\gamma_{2}(k)$ such that
$$|\frac{d^{n}}{dk^{n}}[e(x,k)-\left(
\begin{array}{lll}
e^{ikx}\\
0
\end{array}
\right)-\gamma_{2}(k)\left(
\begin{array}{lll}
e^{-ikx}\\
0
\end{array}
\right)]|\leq c\frac{1}{1+|k|}e^{-\epsilon_{0}|x|}.$$ Also
$|\frac{d^{n}}{dk^{n}}s_{2}(k)|,\ \
|\frac{d^{n}}{dk^{n}}\gamma_{2}(k)|\leq c.$

All $c,\ \epsilon_{0}$ used do not depend on $x$ and $k$; and
$n=0,1,2.$
\end{lemma}
\begin{proof}
We only prove Estimate (~\ref{asympto}), the proof of the second
estimate is similar. As in the proof of Proposition
~\ref{roughestimate}, we divide the proof into two parts,
$\lambda>\beta+\epsilon_{0}$ and $\beta\leq \lambda\leq
\beta+\epsilon_{0}.$
\begin{enumerate}
\item[(1)] When $\lambda>\beta+\epsilon_{0},$ we start with the
definition of $e(x,k)$:
$$
\begin{array}{lll}
& &e(x,k)-\left(
\begin{array}{lll}
e^{ikx}\\
0
\end{array}
\right)\\
&=&-\frac{1}{2k}\int_{-\infty}^{\infty}\left(
\begin{array}{lll}
e^{ik|x-y|}&0\\
0&-\frac{ke^{-\mu|x-y|}}{\mu}
\end{array}
\right)W(y)e(y,k)dy\\
&=&-\frac{1}{2k}\int_{-\infty}^{\infty}\left(
\begin{array}{lll}
0&0\\
0&-\frac{ke^{-\mu|x-y|}}{\mu}
\end{array}
\right)W(y)e(y,k)dy\\
& &-\frac{1}{2k}e^{ikx}\int_{-\infty}^{\infty}\left(
\begin{array}{lll}
e^{-iky}&0\\
0&0
\end{array} \right)W(y)e(y,k)dy\\
& &-\frac{1}{2k}\int_{x}^{\infty}\left(
\begin{array}{lll}
2\sin k(x-y)&0\\
0&0
\end{array}
\right)W(y)e(y,k)dy
\end{array}
$$
All the three terms on the right hand side are nice functions, so we could use Fubini's Theorem to make the following calculations:\\
For the first term, if $x\in [0,+\infty)$ we have that
$$|\frac{d^{n}}{dk^{n}}\int_{-\infty}^{\infty}\left(
\begin{array}{lll}
0&0\\
0&-\frac{ke^{-\mu|x-y|}}{\mu}
\end{array}
\right)W(y)e(y,k)dy|\leq ce^{-\frac{\alpha}{4}|x|}$$ by
Proposition ~\ref{roughestimate}; for the second term:
$$-\frac{e^{ikx}}{2k}\int_{-\infty}^{\infty}\left(
\begin{array}{lll}
e^{-iky}&0\\
0&0
\end{array}
\right)W(y)e(y,k)dy=\left(
\begin{array}{lll}
s_{2}(k)\\
0
\end{array}
\right)e^{ikx},$$ where $s_{2}$ have the estimate
$$|\frac{d^{n}}{dk^{n}}s_{2}(k)|\leq c(\epsilon_{0})$$ for all
$\lambda>\beta+\epsilon_{0};$\\
and the third term:
$$|\frac{d^{n}}{dk^{n}}\int_{x}^{\infty}\left(
\begin{array}{lll}
2\sin k(x-y)&0\\
0&0
\end{array}
\right)W(y)e(y,k)dy|\leq ce^{-\frac{\alpha}{4}|x|}.$$ All the
constants $c$ used above are independent of $k,$ $x$ and $n$,
where $n=0,1,2.$

\item[(2)] We consider the case $\beta\leq \lambda\leq
\beta+\epsilon_{0}.$

Since $detD(0)\not=0,$ there exist $\epsilon_{0},\ \delta>0$ such
that if $|\lambda-\beta|\leq \epsilon_{0},$ then $|detD(k)|\geq
\delta$.

If $\beta\leq \lambda\leq \beta+\epsilon_{0},$ then Estimate
(~\ref{asympto}) can be proven by Lemma ~\ref{exlambda} in which
the fact that $s,\ a_{1},\ a_{2},\ \gamma$ are analytic functions
of $k$ is proved and the following fact: if $x>0,$ by Lemma
~\ref{ODE} and Estimate (~\ref{psi1R}) we have
$$|\frac{d^{n}}{dk^{n}}(\psi_{1}(x,k)-\left(
\begin{array}{lll}
e^{ikx}\\
0
\end{array}
\right))|\leq ce^{-\epsilon_{0}x},$$ for some constant $c$
independent of $x$ and $\lambda$, $n=0,1,2.$ Similar estimates can
be gotten for $\psi_{2}(\pm\cdot,k).$
\end{enumerate}
\end{proof}
We can prove the third estimate of Theorem ~\ref{exk} by using the
estimates made in Lemma ~\ref{disekx}.
\begin{proposition}\label{thirdestimate}
$$\|f^{\#}\|_{\mathcal{H}^{2}}\leq c\|(1+|\cdot|)^{2}f\|_{2},$$
where, recall that $$f^{\#}(k) = \int_{-\infty}^{\infty}
\bar{e}^{\%}(x,k)\cdot f(x) dx.$$
\end{proposition}
\begin{proof}
Decompose $f^{\#}$ into two parts,
$$
\begin{array}{lll}
&
&|\frac{d^{2}}{dk^{2}}\int_{-\infty}^{\infty}\bar{e}^{\%}(x,k)\cdot
f(x) dx|\\
&\leq&|\frac{d^{2}}{dk^{2}}\int_{0}^{\infty}\bar{e}^{\%}(x,k)\cdot
f(x)dx|+|\frac{d^{2}}{dk^{2}}\int_{-\infty}^{0}\bar{e}^{\%}(x,k)\cdot
f(x)dx|.
\end{array}
$$
By Lemma ~\ref{disekx},
$$
\begin{array}{lll}
& &|\frac{d^{2}}{dk^{2}}\int_{0}^{\infty}\bar{e}^{\%}(x,k)\cdot
f(x)dx|\\
&\leq&|\frac{d^{2}}{dk^{2}}\int_{0}^{\infty}(\bar{e}^{\%}(x,k)-\bar{s}_{2}(k)\left(
\begin{array}{lll}
e^{-ikx}\\
0
\end{array}
\right))\cdot f(x)dx|\\
& &+|\frac{d^{2}}{dk^{2}}\int_{0}^{\infty}\bar{s}_{2}(k)\left(
\begin{array}{lll}
e^{-ikx}\\
0
\end{array}
\right)\cdot f(x)dx|\\
&\leq&c\int_{0}^{\infty}\frac{e^{-\epsilon_{0}|x|}}{1+|k|}|f(x)|dx+c|\int_{0}^{+\infty}\left(
\begin{array}{lll}
e^{-ikx}\\
0
\end{array}
\right)\cdot (1+|x|)^{2}f(x)dx|\\
&\leq &c\frac{1}{1+|k|}\|f\|_{2}+c|\int_{0}^{+\infty}\left(
\begin{array}{lll}
e^{-ikx}\\
0
\end{array}
\right)\cdot (1+|x|)^{2}f(x)dx|.
\end{array}
$$
Similarly
$$
\begin{array}{lll}
& &|\frac{d^{2}}{dk^{2}}\int_{-\infty}^{0}\bar{e}^{\%}(x,k)\cdot
f(x)dx|\\
&\leq& c
\|f\|_{2}\frac{1}{1+|k|}+c|\int_{-\infty}^{0}\left(
\begin{array}{lll}
e^{-ikx}\\
0
\end{array}
\right)\cdot (1+|x|)^{2}f(x)dx|\\
& &+c|\int_{-\infty}^{0}\left(
\begin{array}{lll}
e^{ikx}\\
0
\end{array}
\right)\cdot (1+|x|)^{2}f(x)dx|.
\end{array}
$$
Combine the two parts together:
$$
\begin{array}{lll}
&
&|\frac{d^{2}}{dk^{2}}\int_{-\infty}^{\infty}\bar{e}^{\%}(x,k)\cdot
f(x)dx|\\
&\leq&
c\|f\|_{2}\frac{1}{1+|k|}+c|\int_{-\infty}^{\infty}\left(
\begin{array}{lll}
e^{-ikx}\\
0
\end{array}
\right)\cdot (1+|x|)^{2}f(x)dx|\\
& &+c|\int_{-\infty}^{0}\left(
\begin{array}{lll}
e^{ikx}\\
0
\end{array}
\right)\cdot (1+|x|)^{2}f(x)dx|.
\end{array}
$$
We conclude that $\|\frac{d^{2}}{dk^{2}}f^{\#}\|_{2}\leq
c\|(1+|\cdot|)^{2}f\|_{2}.$

It is easier to prove $\|f^{\#}\|_{2}\leq c\|f\|_{2},$ thus
$$\|f^{\#}\|_{\mathcal{H}^{2}}\leq c\|(1+|x|)^{2}f\|_{2}.$$
\end{proof}
\begin{corollary}\label{corothird}
$$\|\frac{d^{n}}{dk^{n}}(e(x,k)/k)\|_{\mathcal{L}^{2}(dk)}\leq c(1+|x|)^{n+1},$$
$$|\frac{d^{n}}{dk^{n}}f^{\#}|\leq c\|(1+|x|)^{n}f\|_{1},$$
where $n=0,1,2.$
\end{corollary}
\begin{proof}
When $|k|\leq 1,$ by Proposition ~\ref{roughestimate} we have
$$|\frac{d^{n}}{dk^{n}}(e(x,k)/k)|\leq c(1+|x|)^{n+1}.$$
When $k>1,$ by Proposition ~\ref{roughestimate} we have
$$|\frac{d^{n}}{dk^{n}}(e(x,k)/k)|\leq c\frac{1}{1+|k|}(1+|x|)^{n}.$$
Therefore
$$\|\frac{d^{n}}{dk^{n}}(e(x,k)/k)\|_{\mathcal{L}^{2}(dk)}\leq c(1+|x|)^{n+1}.$$
Recall $f^{\#}(k)=\langle e^{\%}(\cdot,k),f\rangle.$ By
Proposition ~\ref{roughestimate},
$$
|\frac{d^{n}}{dk^{n}}f^{\#}(k)| \leq c\int
|\frac{d^{n}}{dk^{n}}e^{\%}(x,k)||f(x)|dx \leq
c\int(1+|x|)^{n}|f(x)|dx.$$
\end{proof}
This completes the proof of Theorem ~\ref{exk}.

\section{Proof of Statements (A) and (B) of Proposition
~\ref{spectral}}\label{proofassu} In this appendix we prove
Proposition ~\ref{spectral} translated in the context of the
operator $H,$ i.e. for the family of operators $H(W):=H_{0}+W$
where the operators $H_{0},$ $W$ are defined in Equation
(~\ref{h0w}).\\
\textit{Proof of (A)}. Suppose for some $W_{0}$ Statement (SB) and
(SC) are satisfied. We use the notations and estimates from
Subsection ~\ref{sectionexk}. The Wronskian depends on the
potential $W$ and we display this dependence explicitly by writing
$D(k,W)$ for $D(k).$ By Definitions (~\ref{f1f2}) (~\ref{Dk}) we
can see that $det D(0,W)$ is a continuous functional of the
functions $\frac{d^{n}}{dx^{n}}\phi_{1}(x,\sqrt{2\beta},
W)|_{x=0}$ and $\frac{d^{n}}{dx^{n}}\psi_{1}(x,0,W)|_{x=0}$ where
$n=0,1.$ By Estimates (~\ref{continu2}), the last two functions
are continuous in variable $W.$ Thus if $det D(0,W_{0})\not=0$ for
some $W_{0},$ then there exists a constant $\epsilon>0$ such that
if the function $W$ satisfies
$\|e^{\alpha|x|}(W-W_{0})\|_{\mathcal{L}^{\infty}}\leq \epsilon,$
then $det D(0,W)\not=0$. By Theorem ~\ref{resonance} $H(W)$ has
no resonance at the point $\beta$. This completes the proof of (A).\\
\textit{Proof of (B)}. We fix the function $W$. It is not hard to
prove that the functions
$\frac{d^{n}}{dx^{n}}\phi_{1}(x,\sqrt{2\beta},sW),\
\frac{d^{n}}{dx^{n}}\psi_{1}(x,0,sW)\ (n=0,1)$ are analytic in the
variable $s\in \mathbb{C}.$ Therefore the function $det D(0,sW)$
is analytic in $s$ as well. Thus $det D(0,sW)$ is either
identically zero or vanishes at most at a discrete set of $s$. It
is left to prove the first case does not occur if
$\int_{-\infty}^{\infty}V_{3}\not=0$, where, recall
$V_{3}=V_{1}+V_{2}$ from Transformation (~\ref{trans}).

The proof is based on the following facts valid for sufficiently
small $s$:
\begin{itemize}
 \item[(I)]The Wronskian function
 $W(\phi_{1}(\cdot,\sqrt{2\beta},sW),\phi_{1}(-\cdot,\sqrt{2\beta},sW))\not=0$
 which will be proved in Lemma $\bold{B.2}$ below;
 \item[(II)] As
shown in Lemma $\bold{B.1}$ below there exists a solution
$\varphi_{1}(x,sW)$ with the following properties:
\begin{itemize}
\item[(DI)] for each $s$ there exist constants $c_{1}(s),
c_{2}(s)$ such that $c_{1}(s)\not=0$ if
$\int_{-\infty}^{\infty}V_{3}\not=0$ and
$$
\varphi_{1}(x,sW)-c_{1}(s)\left(
\begin{array}{lll}
x\\
0
\end{array} \right)-c_{2}(s)\left(
\begin{array}{lll}
1\\
0
\end{array} \right)
$$ decays exponentially fast at $-\infty;$
\item[(DII)] for any $x\in \mathbb{R}$ we have
$$|\varphi_{1}(x,sW)|\leq c(1+|x|).$$ The function $$\varphi_{1}(x,sW)-\left(
\begin{array}{lll}
1\\
0
\end{array}
\right)$$ decays exponentially fast at $+\infty.$
\end{itemize}
\end{itemize}
Given these facts we see that
$$W(\varphi_{1}(\cdot,sW),\varphi_{1}(-\cdot,sW))\not=0,
W(\phi_{1}(\cdot,\sqrt{2\beta},sW),\phi_{1}(-\cdot,\sqrt{2\beta},sW))\not=0,$$
$$W(\varphi_{1}(\cdot,sW),\phi_{1}(\pm\cdot,\sqrt{2\beta},sW))=0.$$
Since
$\varphi_{1}(x,s)=\psi_{1}(x,0,sW)+c_{3}(s)\phi_{1}(x,\sqrt{2\beta},sW)$
for some constant $c_{3}(s)$ and recalling the definition of
$D(0)$ hence $D(0,sW)$ from Equation (~\ref{Dk}) we can get that
$detD(0,sW)\not=0.$ This implies that the operator $H(sW)$ has
resonance only at discrete values of $s.$ The statement (B) is
proved (assuming Lemmas $\bold{B.1}$ and $\bold{B.2}$ below).
\begin{lemmaB}\label{last}
There exists a solution $\varphi_{1}(\cdot,sW)$ of the equation
$$[H(sW)-\beta]\varphi_{1}(\cdot,sW)=0$$ with the
properties stated in (DI) and (DII).
\end{lemmaB}
\begin{proof}
Recall the definition $$H(sW)=H_{0}+sW$$ with
$$H_{0}=\left(
\begin{array}{lll}
-\frac{d^{2}}{dx^{2}}+\beta&0\\
0&\frac{d^{2}}{dx^{2}}-\beta
\end{array} \right),\ W=1/2\left(
\begin{array}{lll}
V_{3}&-iV_{4}\\
-iV_{4}&-V_{3}
\end{array}
\right),$$ the functions $V_{3},V_{4}$ are smooth, even, and decay
exponentially fast at $\infty.$

We could rewrite the equation for
$$\varphi_{1}(\cdot,sW)=:\left(
\begin{array}{lll}
\varphi_{11}(\cdot,sW)\\
\varphi_{12}(\cdot,sW)
\end{array}
\right)$$ as
$$
\left(
\begin{array}{lll}
\varphi_{11}(\cdot,sW)\\
\varphi_{12}(\cdot,sW)
\end{array}
\right)=\left(
\begin{array}{lll}
1\\
0
\end{array}
\right)+s/2\left(
\begin{array}{lll}
\int_{x}^{\infty}\int_{y}^{\infty}V_{3}(t)\varphi_{11}(t,sW)-iV_{4}(t)\varphi_{12}(t,sW)dt\\
(-\frac{d^{2}}{dx^{2}}+2\beta)^{-1}(iV_{4}\varphi_{11}(\cdot,sW)+V_{3}\varphi_{12}(\cdot,sW))
\end{array}
\right);
$$
The proof of the existence of $\varphi_{1}(\cdot,sW)$ and the fact
that $|\varphi_{1}(x,sW)|\leq c(1+|x|)$ is easy because when $s$
is small we could use the contraction lemma. We will not go into
the details because we solve similar problems many times.

Since the Wronskian function
$W(\varphi_{1}(x,sW),\varphi_{1}(-x,sW))$ is independent of $x$
and analytic in $s$, it can be expanded in the variable $s$. We
only need to compute the first two terms of $\varphi_{1}(x,sW)$ in
terms of $s$ to prove (DII):
$$\varphi_{1}(x,sW)=\left(
\begin{array}{lll}
1\\
0
\end{array} \right)+s/2\left(
\begin{array}{lll}
\int_{x}^{\infty}\int_{y}^{\infty}V_{3}(t)dtdy\\
i(-\frac{d^{2}}{dx^{2}}+2\beta)^{-1}V_{4}
\end{array}
\right)+O(s^{2}).$$ Thus
$$
\begin{array}{lll}
& &W(\varphi_{1}(\cdot,sW),\varphi_{1}(-\cdot,sW))\\
&=&\frac{d}{dx}\varphi_{1}^{T}(x,sW)\varphi_{1}(-x,sW)-\varphi_{1}^{T}(x,sW)\frac{d}{dx}\varphi_{1}(-x,sW)\\
&=&-s\int_{x}^{\infty}V_{3}(t)dt-s\int_{-x}^{\infty}V_{3}(t)dt+O(s^{2})\\
&=&-s\int_{x}^{\infty}V_{3}(t)dt-s\int_{-\infty}^{x}V_{3}(-t)dt+O(s^{2})\\
&=&-s\int_{-\infty}^{+\infty}V_{3}+o(s^{2})+O(s^{2}).
\end{array}
$$
\end{proof}
\begin{lemmaB2}\label{blowup}
If $s$ is sufficiently small, then
$$W(\phi_{1}(\cdot,\sqrt{2\beta},sW),\phi_{1}(-\cdot,\sqrt{2\beta},sW))\not=0.$$
\end{lemmaB2}
\begin{proof}
For $n=0,1,$ as $s\rightarrow 0,$
$$\frac{d^{n}}{dx^{n}}\phi_{1}(\cdot,\sqrt{2\beta},sW)\rightarrow
\frac{d^{n}}{dx^{n}}\phi_{1}(\cdot,\sqrt{2\beta},0)$$ in the
$\mathcal{L}^{\infty}([0,\infty))$ norm. By Proposition
~\ref{phi1} we could get easily that
$$\phi_{1}(x,\sqrt{2\beta},0)=\left(
\begin{array}{lll}
0\\
e^{-\sqrt{2\beta}x}
\end{array}
\right).$$ Next we use the Wronskian function again: as
$s\rightarrow 0,$
$$
\begin{array}{lll}
&
&W(\phi_{1}(\cdot,\sqrt{2\beta},s),\phi_{1}(-\cdot,\sqrt{2\beta},s))\\
&=&\frac{d}{dx}\phi^{T}_{1}(x,\sqrt{2\beta},s)\phi_{1}(-x,\sqrt{2\beta},s)-\phi_{1}^{T}(\cdot,\sqrt{2\beta},s)\frac{d}{dx}\phi_{1}(-\cdot,\sqrt{2\beta},s)|_{x=0}\\
&\rightarrow &-2\sqrt{2\beta}.
\end{array}
$$
Thus we proved the lemma.
\end{proof}
\section{Proof of Proposition ~\ref{projection}}\label{appendixB}
In this appendix we will prove Proposition ~\ref{projection} in a
more general setting. We base our arguments on a general form of
the operator $L_{general}$ given in Subsection ~\ref{ap1}.
\begin{lemmaC}
For any constant $\lambda_{0}\not\in (-i\infty,-i\beta]\cup
[i\beta,i\infty)$, the operator-valued function
$(L_{general}-\lambda_{0}+z)^{-1}$ is an analytic function of $z$
in a small neighborhood of $0.$ Furthermore
$$(L(\lambda)-\lambda_{0}+z)^{-1}=\sum_{n=m_{0}}^{+\infty}z^{n}K_{n},$$
where $m_{0}>-\infty$ is an integer and $K_{n}$'s are operators.
\end{lemmaC}
\begin{proof}
Recall $L_{general}=L_{0}+U,$ where
$$L_{0}:=\left(
\begin{array}{lll}
0&-\frac{d^{2}}{dx^{2}}+\beta\\
\frac{d^{2}}{dx^{2}}-\beta&0
\end{array}
\right),\ U:=\left(
\begin{array}{lll}
0&V_{1}\\
-V_{2}&0
\end{array}
\right),$$ $\lambda$ is a positive constant, $V_{1}$ and $V_{2}$
are smooth functions decaying exponentially fast at $\infty.$
Since it is hard to get the Laurent series of
$(L_{general}-\lambda_{0}-z)^{-1}$ directly we make a
transformation:
$$(L(\lambda)-\lambda_{0}+z)^{-1}=(1+(L_{0}-\lambda_{0}+z)^{-1}U)^{-1}(L_{0}-\lambda_{0}+z)^{-1}.$$
We make expansion on each term: The operators
$(L_{0}-\lambda_{0}+z)^{-1}$ have no singularity when $z$ is
sufficiently small, so there exist operators $K_{1,n}\
(n=0,1\cdot\cdot\cdot)$ such that
$$(L_{0}-\lambda_{0}+z)^{-1}=\sum_{n=0}^{\infty}z^{n}K_{1,n}.$$
$(L_{0}-\lambda_{0}+z)^{-1}U$ are trace class operators, thus by
~\cite{RSI}, Theorem VI.14
$$(1+(L_{0}-\lambda_{0}+z)^{-1}U)^{-1}=\sum_{n=m_{0}}^{\infty}z^{n}K_{2,n}$$
where $K_{2,n}\ (n=m_{0},m_{0}+1\cdot\cdot\cdot)$ are operators
and $m_{0}>-\infty$ is an integer. The lemma is proved.
\end{proof}

The following is the main theorem of this section.
\begin{propositionC}
Suppose $A$ is an operator having a complex number $\theta$ as an
isolated eigenvalue and
$$(A-\theta-z)^{-1}=\sum_{n=m}^{+\infty} A_{n}z^{n},$$ where
$m>-\infty$ is an integer and $A_{n}$ are operators. Then we have
the following three results:
\begin{itemize}
 \item[(EnA)]
 $A_{-1}$ is a projection operator:
 $$P_{\theta}^{A}:=\frac{1}{2i\pi}\oint_{|x-\theta|=\epsilon}(A-x)^{-1}dx=A_{-1},$$
 where $\epsilon>0$ is a sufficiently small constant.
 $Range P_{\theta}^{A}$ is the space of eigenvectors and
 associated eigenvectors of $A$ with the eigenvalue $\theta,$ i.e.
 $$Range A_{-1}=\{x|(A-\theta)^{k}x=0\ \text{for some positive integer}\ k \}.$$

 Assume, if the operator
 $P_{\theta}^{A}$ is finite dimensional: $$Range
 P_{\theta}^{A}=\{\xi_{1},\cdot\cdot\cdot,\xi_{n}\},$$ then the operator $A^{*}$ has $n$ independent eigenvectors and
 associated eigenvectors $\eta_{1},\cdot\cdot\cdot, \eta_{n}$ with
 the
 eigenvalue $\bar{\theta};$
 \item[(EnB)] The $n\times n$ matrix $T=[T_{ij}]$
 where $T_{ij}:=\langle \eta_{i},\xi_{j}\rangle$ is invertible,
 \item[(EnC)] The operator $P_{\theta}^{A}$ is of the
 form:
 \begin{equation}\label{concretematrix}
 P_{\theta}^{A}f=(\xi_{1},\cdot\cdot\cdot,\xi_{n})T^{-1}\left(
 \begin{array}{lll}
 \langle \eta_{1},f\rangle\\
 \ \ \cdot\\
 \ \ \cdot\\
 \ \ \cdot\\
 \langle \eta_{n},f\rangle
 \end{array}
 \right).
 \end{equation}
\end{itemize}
\end{propositionC}
\begin{proof}
The proof of (EnA) is well known (see, e.g. ~\cite{RSIV,Kato}). To
prove (EnB), assume $det T=0.$ Then there exist constants
$b_{1},\cdot\cdot\cdot, b_{n}$ such that
$$b_{1}\eta_{1}+\cdot\cdot\cdot+b_{n}\eta_{n}\perp
\xi_{1},\cdot\cdot\cdot,\xi_{n}$$ and at least one of the
constants $b_{1},\cdot\cdot\cdot, b_{n}$ is not zero. Since
$P_{\theta}^{A}f\in
\text{Span}\{\xi_{1},\cdot\cdot\cdot,\xi_{n}\}$ for any vector
$f,$ we have
$$
\begin{array}{lll}
0&=&\langle
b_{1}\eta_{1}+b_{2}\eta_{2}+\cdot\cdot\cdot+b_{n}\eta_{n},P_{\theta}^{A}f\rangle\\
&=&\langle
P_{\bar{\theta}}^{A^{*}}(b_{1}\eta_{1}+b_{2}\eta_{2}+\cdot\cdot\cdot+b_{n}\eta_{n}),f\rangle\\
&=&\langle
b_{1}\eta_{1}+b_{2}\eta_{2}+\cdot\cdot\cdot+b_{n}\eta_{n},f\rangle.
\end{array}
$$ Thus
$$b_{1}\eta_{1}+\cdot\cdot\cdot+b_{n}\eta_{n}=0.$$ This
contradicts to the fact that the vectors
$\eta_{1},\cdot\cdot\cdot,\eta_{n}$ are linearly independent.
Therefore we proved that the matrix $T$ is invertible. For (EnC).
For any vector $f,$ $P_{\theta}^{A}f\in
\text{Span}\{\xi_{1},\cdot\cdot\cdot,\xi_{n}\}.$ Therefore they is
a $n\times 1$ scalar matrix $\left(
 \begin{array}{lll}
 a_{1},\cdot\cdot\cdot, a_{n}\\
 \end{array}
 \right)^{T}$ such that $$P_{\theta}^{A}f=(\xi_{1},\cdot\cdot\cdot,\xi_{n})\left(
 \begin{array}{lll}
 a_{1},\cdot\cdot\cdot,a_{n}\\
 \end{array}
 \right)^{T}.$$ What is left is to compute a concrete form of
 $a_{i}$'s.
We have the following formula $$\langle \eta_{i},f\rangle=\langle
P_{\bar{\theta}}^{A^{*}}\eta_{i},f\rangle=\langle
\eta_{i},P_{\theta}^{A}f\rangle,$$ while
$$\langle
\eta_{i},P_{\theta}^{A}f\rangle=(\langle
\eta_{i},\xi_{1}\rangle,\cdot\cdot\cdot,\langle
\eta_{i},\xi_{n}\rangle)\left(
 \begin{array}{lll}
 a_{1}\\
 \ \cdot\\
 \ \cdot\\
 \ \cdot\\
 a_{n}
 \end{array}
 \right)$$ which implies Formula (~\ref{concretematrix}).
\end{proof}

\end{document}